\documentclass{article}
\usepackage{amsfonts}
\usepackage{amsmath}
\usepackage{amssymb}
\usepackage{graphicx}
\usepackage{setspace}
\usepackage{CJK}
\usepackage{moresize}
\usepackage[utf8]{inputenc}
\usepackage[english]{babel}

\usepackage{pdflscape}

\usepackage{cite}
\usepackage{float}

\usepackage{authblk}
\usepackage{titlesec}
\usepackage{geometry}
\usepackage{lineno}
\usepackage{longtable}
\usepackage{array}
\usepackage{booktabs}

\usepackage[nottoc,numbib]{tocbibind}

\newcommand{\begM}{\begin{multline}}
\newcommand{\eM}{
\end{multline}}

\def\){\Big)}
\def\({\Big(}

\begin{document}

\title{Noether's theorems for first and second order field theories in flat spacetimes}
\author{Mark Robert Baker\thanks{mbaker66@uwo.ca}}
\affil{Department of Physics and Astronomy, Western University, London, ON, Canada}
\date{July 26, 2026}

\maketitle

\begin{abstract}
    Technical derivations of the Noether identity (that is, showing and explaining all steps using methods of the calculus of variations), as well as of Noether's first and second theorems, are typically presented in explicit detail in the mathematics literature for scalar functions with first order derivatives in the action. This presentation has drawbacks for physicists wishing to learn these techniques to apply to physical field theories that have higher rank tensor fields (potentials) or higher orders of derivatives in the action. Motivated by this, and heavily influenced by the calculus of variations presentation of Gelfand and Fomin, we detail complete derivations of the Noether identity and Noether's theorems for first and second order field theories with tensor fields of arbitrary rank in flat (e.g., Minkowski) spacetimes, showing all steps explicitly. We then give some physical examples of how to apply the first and second theorems to field theories in physics, and detail the origin of conventional spacetime symmetries from the Killing and conformal Killing equations. We conclude by briefly reviewing the many areas of physics where the theorems have been discussed and applied.
\end{abstract}

\vspace{1.0cm}

\textit{Author's note: The following content was primarily completed between 2016 and 2018, in order to gain a better understanding of the origin of the Noether identity and Noether's theorems \cite{Noether1918}, as part of graduate work which led to the PhD thesis \cite{Baker2021PHD}. It very closely follows the presentation of Gelfand and Fomin in \cite{GelfandFomin1963}, albeit for physical fields and higher orders (Gelfand and Fomin primarily focus on the first order case for scalar functions, more similar to Noether's presentation). While this is likely the most explicit and complete presentation of how to derive the Noether identity for first and second order field theories in flat spacetimes, we do not claim any novel results related to Noether's theorems in this article. In common sources, many of these details/ steps are skipped, making it difficult for the first time learner. This is why we felt compelled to share this, so that physicists looking to learn these expanded details have a source specific to application for physical fields, instead of trying to infer these results from more typical mathematics literature presentations. We upload this article to the arXiv on the 108$^{th}$ anniversary of the presentation of Noether's paper \cite{Noether1918}. Please reach out to the author by email if you have any suggestions for changes, details to add to the document, etc.}

\tableofcontents
\section{Introduction}

In this article, we provide explicit calculation of the Noether identity, as well as the statement of Noether's first and second theorems, for field theories of any rank of potential $\Phi_A$, and any dimension. In Noether's paper \cite{Noether1918}, and other common presentations of the calculus of variations (such as Gelfand and Fomin \cite{GelfandFomin1963} or Logan \cite{Logan1977}), this calculation is typically carried out for scalar functions, often without much explicit detail of the intermediate steps. Furthermore, this calculation is typically done for first order theories (e.g., Lagrangian terms of the form $\partial \Phi \partial \Phi$), largely because common physical Lagrangians (kinetic energy, Klein-Gordon, electrodynamics, etc.) have terms quadratic in first derivatives. However, if one considers Lagrangians with higher derivatives (such as $\Phi \partial \partial \Phi$ or $\partial \partial \Phi \partial \partial \Phi$), the entire Noether identity must be modified. That is why we present the derivation for both the first and second order cases in this document. There is much repetition between the two (we copied and pasted the first approach as the starting point for the second so that a reader can focus on one of the two sections independently, depending on their interest). The major differences are that in the first order section we discuss variations of higher ranked fields under coordinate transformations (a result independent of derivative order thus not repeated for the second order section), and in the second order section we discuss a commutation relation for the total variation and derivatives which can be used to derive the Noether identity at any order.\\

The goal is to write out every step explicitly, to obtain the Noether identity and both theorems, in both first and second order field theories. The article is organized as follows: the first two sections derive the Noether identity for first order and second order field theories respectively, with articulation of Noether's first and second theorems at the end. Next, in the origin of spacetime symmetries section, the Killing and conformal Killing vectors that serve as inputs to the identity are derived. After this in the applications section, the Noether identity is used for electrodynamics \cite{BesselHagen1921}, linearized Gauss-Bonnet gravity \cite{BK2019} and general relativity \cite{Noether1918,Klein1918}, recovering the known energy-momentum tensors and differential identities from the first and second theorems. The final section discusses the historical context and reviews applications of Noether's theorems in the physics literature.\\

In this introduction section, we will first go over some basic background knowledge, and intricate points, that are important for understanding the contents of this document:

\subsection{What is the Noether identity?}

The Noether identity is the core relationship from which Noether's first and second theorems are stated. It is a statement associated to invariance of the action, which requires careful application of the calculus of variations to an action under simultaneous transformation of (in the case of field theories used in physics) coordinates and fields. However, this cannot be computed by naive variations often used in e.g., Lagrangian mechanics to derive the Euler-Lagrange equation. For example, for a Lagrangian $L(x,v)$ in 1D, for the action condition $\delta S = 0$ it is often presented that one can simply act on the Lagrangian with the variation, based on the definition of the total differential of a multivariable function and rearrange as,

\begin{equation}
\delta L = \frac{\partial L}{\partial x} \delta x + \frac{\partial L}{\partial v} \delta v = \left[ \frac{\partial L}{\partial x}   - \frac{d}{dt} \frac{\partial L}{\partial v} \right] \delta x + \frac{d}{dt}[\frac{\partial L}{\partial v} \delta x]
\end{equation}

where $\delta v = \frac{d}{dt}\delta x$ was used above. This, set to zero under the time integral of the action as $\delta S = 0$, isolates the familiar Euler-Lagrange equation, and some additional ``boundary term'' $\frac{d}{dt}[\frac{\partial L}{\partial v} \delta x]$ which is often discarded. In a nutshell, in the Noether approach, this boundary term is kept and related to various conservation laws of the theory, depending on input symmetries through the $\delta x$ and $\delta t$ which are present under the total derivative. However, the above naive approach misses an essential term, and hides the fact that different ``total'' and ``substantial'' variations are present in the Noether calculation. In the above derivation, derivatives and variations are assumed to commute as $\delta \frac{d}{dt} = \frac{d}{dt} \delta$, a property of the ``substantial'' variation, which we refer to in this document as $\bar{\delta}$. The essential missing term comes from the transformation of the volume element, which is sometimes introduced in an ad-hoc fashion, allowing the naive approach to effectively yield the desired result. Perhaps these details are okay to overlook in introductory sources on the Euler-Lagrange equation, but omitting them can add confusion to those trying to learn about Noether's theorems in the context of field theories for the first time.

\subsection{Calculus of Variations, brief overview}

We refer the reader to the book by Gelfand and Fomin \cite{GelfandFomin1963} on the calculus of variations for technical details on this subject; in our opinion, this is one of the best sources (if not the best) on this topic. It was what we closely followed in our own calculations of the Noether identity in this document. Here, we will overview the very basic ideas required for understanding of Noether's theorems.\\

If we have an action principle based on the change in an action that depends on coordinates and fields, we write the action change as $\Delta S = S[\Phi_A^* (X^{*\alpha})] - S[\Phi_A (X^\alpha)]$, where * represents the changed quantities. By defining these quantities (fields $\Phi$ and coordinates $X$) as depending on some small parameter $\epsilon$, they can be Taylor expanded about $\epsilon = 0$, and all higher order $\mathcal{O}(\epsilon^2)$ terms can be neglected since multiplying the small parameter by itself yields a negligible quantity. This keeps only linear terms in $\epsilon$ which dramatically simplifies calculations and allows the two actions $S[\Phi_A^* (X^{*\alpha})] - S[\Phi_A (X^\alpha)]$ to be combined in terms of the original coordinates and fields. This is the basic theme throughout the calculus of variations, to convert large changes $\Delta$ to small changes $\delta$ by keeping only the first order term in an expansion based on small parameter $\epsilon$.\\

Once the change in the action is transformed and only linear $\epsilon$ are kept, we have what is defined as the total variation $\delta S$ of the action, from which the resulting terms can be recombined as the familiar Euler-Lagrange equation, and some terms under a total divergence; the combination of these is what is referred to as the Noether identity. This total divergence then depends explicitly on the variation of coordinates and fields which are by definition ``infinitesimal'' transformations as only first order $\epsilon$ has been kept at this stage. All of these quantities come from direct Taylor expansions that will be performed in this document.

\subsubsection*{Total versus Substantial Variations}

It is important to make a distinction between ``total'' and ``substantial'' variations. Symbolically, in this document we will use:

\begin{itemize}
    \item $\delta$ for total variation
    \item $\bar{\delta}$ for substantial variation
\end{itemize}

The difference in these two quantities is whether or not the coordinates are also being changed in the transformed object:

\begin{itemize}
    \item $\delta \Phi_A = \Phi_A^*(X^{*\alpha}) - \Phi_A(X^\alpha)$ is the total variation
    \item $\bar{\delta} \Phi_A = \Phi_A^*(X^{\alpha}) - \Phi_A(X^\alpha)$ is the substantial variation
\end{itemize}

As shown above, the total variation considers the change between the transformed and non transformed field, where the transformed field is a function of the transformed coordinates. The substantial variation has both the transformed and non transformed fields functions of the non transformed coordinates. Both of these will appear throughout the derivations in this document. They have an important distinction with regards to their commutation with partial derivatives:

\begin{itemize}
    \item  $\delta \partial \Phi_A \neq \partial \delta \Phi_A$ --- total variation does not commute
    \item  $\bar{\delta} \partial \Phi_A = \partial \bar{\delta} \Phi_A$ --- substantial variation does commute
\end{itemize}

Much of the work in the following document is defining both the total and substantial variations, such that the Noether identity can have substantial variations on the fields, allowing derivatives to be commuted and the Euler-Lagrange equation be extracted to yield the Noether identity in its conventional form.

\subsection{Notation table}

The final subsection includes a notation table, summarizing the notation used in this article compared to other common notations used in the physics literature. We follow Gelfand and Fomin \cite{GelfandFomin1963}, whose conventions differ cosmetically (but not substantively) from those common in the physics literature.  All variations are implied to be first order in the parameter $\epsilon$.\\

We note that in the article, the overall sign of the Lie derivative relations depends on the active versus passive convention for the transformations, as discussed in the subsection on variations of higher ranked $\Phi_A$. We also note the $g$ here for transformations is not to be confused with the metric $g_{\mu\nu}$ (which for the most part we won't use as this article focuses on flat e.g., Minkowski spacetimes). Furthermore, a third use, the $g$ in $\delta_g$ is used to imply gauge transformation.\\

The notation table is presented below. It also includes some notes on the particular quantity (it is not exhaustive, other quantities are introduced throughout):

\begin{table}[H]
\centering
\renewcommand{\arraystretch}{1.6}
\small
\begin{tabular}{|p{3.0cm}|p{1.5cm}|p{3.0cm}|p{6.0cm}|}
\hline
\textbf{Quantity} 
& \textbf{Present article} 
& \textbf{Other common physics notation} 
& \textbf{Notes} 
\\
\hline
Coordinates &
$X^\alpha$ &
$x^\mu$ &
$n$ components; volume element $dX$ \\
\hline
Fields (potential) &
$\Phi_A$ &
$\phi,\ A_\mu,\ h_{\mu\nu},\dots$ &
$m$ components; $A$ represents arbitrary rank of tensor \\
\hline
Transformation of coordinates, fields &
$G^\alpha,\ H_A$ &
Usually implicit &
In this article $X^{*\alpha} = G^\alpha(X,\Phi,\partial\Phi;\epsilon)$, $\Phi_A^* = H_A(X,\Phi,\partial\Phi;\epsilon)$ 
\\
\hline
Coordinate generator &
$g^\alpha$ &
$\xi^\alpha$ (i.e., symmetry / Killing vector) &
In this article $\delta X^\alpha = X^{*\alpha} - X^\alpha = \epsilon  g^\alpha$ 
\\
\hline
Total field generator &
$h_A$ &
Usually implicit &
In this article $\delta \Phi_A = \Phi_A^*(X^{*}) - \Phi_A(X) = \epsilon  h_A$
\\
\hline
Substantial field generator &
$\bar{h}_A$ &
The ``characteristic'' $Q$ &
In this article $\bar{\delta} \Phi_A = \Phi_A^*(X) - \Phi_A(X) = \epsilon  \bar{h}_A$ 
\\
\hline
Total variation &
$\delta$ &
$\delta$ or $\Delta$ &
Compares new field at new point to old field at old point; $\delta \partial \neq \partial \delta$ 
\\
\hline
Substantial variation &
$\bar{\delta}$ &
$\delta_0$ or $\bar{\delta}$, a.k.a. ``local'' or ``form'' variation &
Compares both fields at the same point; $\bar{\delta} \partial = \partial \bar{\delta}$ 
\\
\hline
Euler--Lagrange expression &
$E^A$ &
$\dfrac{\delta S}{\delta \Phi_A}$ &
Collected and multiplied by $\bar{\delta}\Phi_A$ in the Noether identity 
\\
\hline
Noether current &
$J^\rho$ &
$j^\rho$ &
Quantity under the total divergence in the Noether identity
\\
\hline
\end{tabular}
\caption{This table includes relations between the Gelfand and Fomin conventions used in this article and other common physics notation. The quantities $g^\alpha$, $h_A$, $\bar{h}_A$ are functions of $(X, \Phi, \partial\Phi)$ (with $\partial \partial \Phi$ added in the second order theory), often abbreviated as functions of $X$ alone since the fields are themselves functions of $X$.}
\end{table}

Some other notes used later in this article are as follows: First, the physics-literature symbol $\xi^\alpha$ for the coordinate generator corresponds to our $g^\alpha$; in the origin of spacetime symmetries section the Killing and conformal Killing vectors are denoted $v^\alpha$ (some arbitrary vector being solved for, ultimately the coordinate symmetries used in the Noether identity). Second, the symbol $\mathcal{L}$ denotes the Lagrangian density throughout; the Lie derivative along a vector $v$ is written $\mathsterling_v$. Finally, the number of coordinates is denoted $n$ in the general calculations, while in the spacetime applications the dimension is denoted $D$; thus $D = n$ throughout.

\section{Noether Identity for First Order Field Theories}

The following section highlights calculations related to the Noether identity, Noether's theorem stemming primarily from section 37 and 38 in Gelfand and Fomin, for first order actions with fields $\Phi_A$ of arbitrary rank.

\subsection{Variation of the action that depends on varied coordinates and fields}

The basic idea of this section is to consider the variation of a functional that depends on the variation of coordinates i.e. $X^\alpha = \langle x^1, x^2, x^3, x^4 \rangle$, fields i.e. $\Phi_A$ and derivatives of fields i.e. $\partial_\nu \Phi_A$. The volume element can be expressed as $dX = dx^1 dx^2 dx^3 dx^4$. The bounds of integration are the entire region $R_X$ in terms of the $X$ coordinates. The action functional then reads,

\begin{equation}
S[\Phi_A (X^\alpha)] = \int \int \int \int_{R_X} \mathcal{L} (X^\alpha, \Phi_A, \partial_\nu \Phi_A) dx^1 dx^2 dx^3 dx^4
\end{equation}

This above action is for 4D theories (common in physics), but can be in more general form expressed for $n$ coordinates, $m$ fields, and the corresponding derivatives of these fields with respect to the coordinates. If we consider $X^\alpha$ to be $n$ dimensional, then we can write the action compactly using 1 integral that is assumed to be $n$ integrals $(\int_{R_X} = \int \dots \int_{R_X}), (dX = dx^1 dx^2 \dots dx^n)$, 

\begin{equation}
S[\Phi_A (X^\alpha)] = \int_{R_X} \mathcal{L} (X^\alpha, \Phi_A, \partial_\nu \Phi_A) dX
\end{equation}

\subsubsection*{Transformation of coordinates and fields}

Now suppose we change both our coordinates and our fields based on some transformation that depends on a parameter $\epsilon$. We can define new coordinates and fields using the $*$ symbol,

\begin{equation}
X^{*\alpha} = G^\alpha (X, \Phi, \partial \Phi; \epsilon)
\end{equation}

\begin{equation}
\Phi_A^* = H_A (X, \Phi, \partial \Phi; \epsilon)
\end{equation}

the functions $G^\alpha$ and $H_A$ depend on the parameter $\epsilon$ and are differentiable with respect to $\epsilon$. The function $G^\alpha$ has $n$ components just like $X^\alpha$ does, and the function $H_A$ has $m$ components just like $\Phi_A$ does (for vector field theories $n = m$ such as in electrodynamics). They are defined such that at $\epsilon = 0$ the original coordinates and field are recovered,

\begin{equation}
X^\alpha = G^\alpha (X, \Phi, \partial \Phi; \epsilon = 0)
\end{equation}

\begin{equation}
\Phi_A = H_A (X, \Phi, \partial \Phi; \epsilon = 0)
\end{equation}

By Taylor expanding the new coordinates and fields about $\epsilon = 0$, we can relate the new coordinates ($*$) to the old coordinates,

\begin{equation}
X^{*\alpha} = G^\alpha (X, \Phi, \partial \Phi; \epsilon) = G^\alpha (X, \Phi, \partial \Phi; \epsilon = 0) + \epsilon \frac{\partial G^\alpha (X, \Phi, \partial \Phi; \epsilon)}{\partial \epsilon} |_{\epsilon = 0} + \mathcal{O}(\epsilon^2)
\end{equation}

\begin{equation}
\Phi_A^* = H_A (X, \Phi, \partial \Phi; \epsilon) = H_A (X, \Phi, \partial \Phi; \epsilon = 0) + \epsilon \frac{\partial H_A (X, \Phi, \partial \Phi; \epsilon)}{\partial \epsilon} |_{\epsilon = 0} + \mathcal{O}(\epsilon^2)
\end{equation}

the $\mathcal{O}(\epsilon^2)$ term represents all terms which are nonlinear in $\epsilon$. Since we are only interested in the linear part for the variation, we compactly write them as such (as we will eventually drop them). The first term in each Taylor expansion is exactly the original coordinates $X^\alpha$ and fields $\Phi_A$! We can compactly write the second term using the functions $g$ and $h$ since these terms will be frequently used in the following calculation,

\begin{equation}
g^\alpha (X, \Phi, \partial \Phi) = \frac{\partial G^\alpha (X, \Phi, \partial \Phi; \epsilon)}{\partial \epsilon} |_{\epsilon = 0}
\end{equation}

\begin{equation}
h_A (X, \Phi, \partial \Phi) = \frac{\partial H_A (X, \Phi, \partial \Phi; \epsilon)}{\partial \epsilon} |_{\epsilon = 0} 
\end{equation}

therefore we can rewrite the Taylor expansions to relate the two coordinates and fields,

\begin{equation}
X^{*\alpha} = X^\alpha + \epsilon g^\alpha (X, \Phi, \partial \Phi) + \mathcal{O}(\epsilon^2)
\end{equation}

\begin{equation}
\Phi_A^*(X^*) = \Phi_A(X) + \epsilon h_A (X, \Phi, \partial \Phi) + \mathcal{O}(\epsilon^2)
\end{equation}

where above we relate the transformed field at the transformed coordinate to the original field and coordinate.

\subsubsection*{Transformation of the action into new coordinates and fields}

If we use the new coordinate transformations on the action, we find that the bounds of integration, dependent variables on the Lagrangian, the volume element, are all now expressed in terms of $*$. In other words, the action is expressed completely in terms of a new set of coordinates and fields,

\begin{equation}
S[\Phi_A^* (X^{*\alpha})] = \int_{R_{X^*}} \mathcal{L} (X^{*\alpha}, \Phi_A^*, \partial_\nu^* \Phi_A^*) dX^*
\end{equation}

even the divergence is expressed in terms of the new coordinates. This is the whole point of the desired variation. The variation with respect to changing coordinates and fields is defined as the linear part of the difference between the two functionals when the coordinates and fields are changed. While $\Delta S$ is the full difference, 

\begin{equation}
\Delta S = S[\Phi_A^* (X^{*\alpha})] - S[\Phi_A (X^\alpha)]
\end{equation}

after keeping only the linear part we will find $\delta S$ (the linear part of $\Delta S= S[\Phi_A^* (X^{*\alpha})] - S[\Phi_A (X^\alpha)]$); but first we need to calculate $\Delta S$ in order to drop the nonlinear part (nonlinear in $\epsilon$). A transformation is a symmetry of the action if and only if,

\begin{equation}
S[\Phi_A^* (X^{*\alpha})] = S[\Phi_A (X^\alpha)]
\end{equation}

this invariance condition is behind both of Noether's theorems. It is distinct from the principle of stationary action, in which $\delta S = 0$ holds at linear order for solutions of the equation of motion. To derive the Noether identity we need to compute the difference in these action functionals $\Delta S = S[\Phi_A^* (X^{*\alpha})] - S[\Phi_A (X^\alpha)]$. The difference in the two action functionals can be found by,

\begin{equation}
\Delta S = \int_{R_{X^*}} \mathcal{L} (X^{*\alpha}, \Phi_A^*, \partial_\nu^* \Phi_A^*) dX^* - \int_{R_X} \mathcal{L} (X^\alpha, \Phi_A, \partial_\nu \Phi_A) dX
\end{equation}

\subsubsection*{Using the Jacobian to combine the bounds and region of integration}

in order to convert the first term into the same region of integration and volume element, we need to perform a change of variables using the Jacobian,

\begin{equation}
dX^* = \frac{\partial (X^{*1}, X^{*2}, \dots, X^{*n})}{\partial (X^1, X^2, \dots, X^n)} dX
\end{equation}

thus we can combine the two integrals,

\begin{equation}
\Delta S = \int_{R_{X}} \mathcal{L} (X^{*\alpha}, \Phi_A^*, \partial_\nu^* \Phi_A^*)  \frac{\partial (X^{*1}, X^{*2}, \dots, X^{*n})}{\partial (X^1, X^2, \dots, X^n)} dX - \int_{R_X} \mathcal{L} (X^\alpha, \Phi_A, \partial_\nu \Phi_A) dX
\end{equation}

\begin{equation}
\Delta S = \int_{R_{X}} [ \mathcal{L} (X^{*\alpha}, \Phi_A^*, \partial_\nu^* \Phi_A^*)  \frac{\partial (X^{*1}, X^{*2}, \dots, X^{*n})}{\partial (X^1, X^2, \dots, X^n)}  -  \mathcal{L} (X^\alpha, \Phi_A, \partial_\nu \Phi_A) ] dX
\end{equation}

the Jacobian is the determinant of the following matrix,

\begin{equation}
 \frac{\partial (X^{*1}, X^{*2}, \dots, X^{*n})}{\partial (X^1, X^2, \dots, X^n)} =
\begin{bmatrix}
\frac{\partial X^{*1}}{\partial X^1} & \frac{\partial X^{*2}}{\partial X^1} & \dots & \frac{\partial X^{*n}}{\partial X^1}\\
\frac{\partial X^{*1}}{\partial X^2} & \frac{\partial X^{*2}}{\partial X^2} & \dots & \frac{\partial X^{*n}}{\partial X^2}\\
\vdots & \vdots & \ddots & \vdots\\
\frac{\partial X^{*1}}{\partial X^n} & \frac{\partial X^{*2}}{\partial X^n} & \dots & \frac{\partial X^{*n}}{\partial X^n}
\end{bmatrix}_{n\times n}
\end{equation}

where we write this matrix with rows indexed by the original coordinates, the transpose of the convention in some references; the determinant (our desired calculation) is unaffected by this choice. To determine each of these components we need to recall the relationship between the coordinates,

\begin{equation}
X^{*\alpha} = X^\alpha + \epsilon g^\alpha (X, \Phi, \partial \Phi) + \mathcal{O}(\epsilon^2)
\end{equation}

we can write each element in the Jacobian in general as $\frac{\partial X^{*\alpha}}{\partial X^\beta}$, which we can find by differentiating the above expression,

\begin{equation}
\frac{\partial X^{*\alpha}}{\partial X^\beta} = \frac{\partial X^\alpha}{\partial X^\beta} + \epsilon \frac{\partial g^\alpha (X, \Phi, \partial \Phi)}{\partial X^\beta} + \frac{\partial \mathcal{O}(\epsilon^2)}{\partial X^\beta}
\end{equation}

since after taking the determinant the $\mathcal{O}(\epsilon^2)$ terms will be even higher order, and we only want terms linear in $\epsilon$ for the variation, we will drop them here. We can abbreviate the second term as just $g^\alpha$. Since $\Phi$ and $\partial \Phi$ themselves depend on $X$ ($\Phi(X)$ and $\partial \Phi(X)$), the function $g^\alpha (X, \Phi, \partial \Phi)$ is in the end a function of $X$ alone, which we write as $g^\alpha (X)$. Derivatives like $\frac{\partial g^\alpha}{\partial X^\beta}$ here and in the rest of the calculation are taken with this in mind (through the $X$ dependence of $\Phi$ and $\partial \Phi$ as well). The same goes for $h_A$ and $\bar{h}_A$.  Finally the first term is just the Kronecker delta,

\begin{equation}
\frac{\partial X^{*\alpha}}{\partial X^\beta} \approx \delta^\alpha_\beta + \epsilon \frac{\partial g^\alpha}{\partial X^\beta}
\end{equation}

inserting this into the Jacobian we have,

\begin{equation}
 \frac{\partial (X^{*1}, X^{*2}, \dots, X^{*n})}{\partial (X^1, X^2, \dots, X^n)} =
\begin{bmatrix}
1 + \epsilon \frac{\partial g^1}{\partial X^1} & \epsilon \frac{\partial g^2}{\partial X^1} & \dots & \epsilon \frac{\partial g^n}{\partial X^1}\\
\epsilon \frac{\partial g^1}{\partial X^2} & 1 + \epsilon \frac{\partial g^2}{\partial X^2} & \dots & \epsilon \frac{\partial g^n}{\partial X^2}\\
\vdots & \vdots & \ddots & \vdots\\
\epsilon \frac{\partial g^1}{\partial X^n} & \epsilon \frac{\partial g^2}{\partial X^n} & \dots & 1 + \epsilon \frac{\partial g^n}{\partial X^n}
\end{bmatrix}_{n\times n}
\end{equation}

to find this determinant we have to perform a cofactor expansion and keep only terms which are at most linear in epsilon. Based on the cofactor expansion all terms not on the diagonal will have submatrices at least $\epsilon^2$, thus they can all be neglected. What is left is the multiplication of all elements on the diagonal,

\begin{equation}
 \frac{\partial (X^{*1}, X^{*2}, \dots, X^{*n})}{\partial (X^1, X^2, \dots, X^n)} \approx (1 + \epsilon \frac{\partial g^1}{\partial X^1})(1 + \epsilon \frac{\partial g^2}{\partial X^2}) \dots (1 + \epsilon \frac{\partial g^n}{\partial X^n})
\end{equation}

this expands to, in order linear in epsilon,

\begin{equation}
 \frac{\partial (X^{*1}, X^{*2}, \dots, X^{*n})}{\partial (X^1, X^2, \dots, X^n)} \approx 1 + \epsilon \frac{\partial g^1}{\partial X^1} +  \epsilon \frac{\partial g^2}{\partial X^2} + \dots +  \epsilon \frac{\partial g^n}{\partial X^n}
\end{equation}

which can be expressed compactly in Einstein summation notation as,

\begin{equation}
 \frac{\partial (X^{*1}, X^{*2}, \dots, X^{*n})}{\partial (X^1, X^2, \dots, X^n)} \approx 1 + \epsilon \frac{\partial g^\alpha}{\partial X^\alpha} 
\end{equation}

returning to our difference in actions,

\begin{equation}
\Delta S = \int_{R_{X}} [ \mathcal{L} (X^{*\alpha}, \Phi_A^*, \partial_\nu^* \Phi_A^*)  \frac{\partial (X^{*1}, X^{*2}, \dots, X^{*n})}{\partial (X^1, X^2, \dots, X^n)}  -  \mathcal{L} (X^\alpha, \Phi_A, \partial_\nu \Phi_A) ] dX
\end{equation}

we can insert the evaluated Jacobian,

\begin{equation}
\Delta S = \int_{R_{X}} [ \mathcal{L} (X^{*\alpha}, \Phi_A^*, \partial_\nu^* \Phi_A^*) (1 + \epsilon \frac{\partial g^\alpha}{\partial X^\alpha} ) -  \mathcal{L} (X^\alpha, \Phi_A, \partial_\nu \Phi_A) ] dX
\end{equation}

\subsubsection*{Taylor expanding the Lagrangian from the new coordinates and fields}

Expanding out our integrand,

\begin{equation}
\Delta S = \int_{R_{X}} [ \mathcal{L} (X^{*\alpha}, \Phi_A^*, \partial_\nu^* \Phi_A^*)  + \epsilon   \mathcal{L} (X^{*\alpha}, \Phi_A^*, \partial_\nu^* \Phi_A^*) \frac{\partial g^\alpha}{\partial X^\alpha}  -  \mathcal{L} (X^\alpha, \Phi_A, \partial_\nu \Phi_A) ] dX
\end{equation}

we can Taylor expand the Lagrangian terms $\mathcal{L} (X^{*\alpha}, \Phi_A^*, \partial_\nu^* \Phi_A^*)$ which are expressed in terms of the new coordinates and fields, and keep only order 1 relative to $\epsilon$. Taylor expanding this multivariable function,

\begin{multline}
\mathcal{L} (X^{*\alpha}, \Phi_A^*, \partial_\nu^* \Phi_A^*) = \mathcal{L} (X^\alpha, \Phi_A, \partial_\nu \Phi_A) 
+ \frac{\partial \mathcal{L}(X^\alpha, \Phi_A, \partial_\nu \Phi_A)}{\partial X^\beta} (X^{*\beta} - X^\beta) \\
+ \frac{\partial \mathcal{L}(X^\alpha, \Phi_A, \partial_\nu \Phi_A)}{\partial \Phi_A} (\Phi_A^* - \Phi_A) 
+ \frac{\partial \mathcal{L}(X^\alpha, \Phi_A, \partial_\nu \Phi_A)}{\partial (\partial_\rho \Phi_A)} (\partial_\rho^* \Phi_A^* - \partial_\rho \Phi_A) + \dots
\end{multline}

the second part of the final three terms are just what we define the variations of these input variables to be (the linear change between the new and old systems),

\begin{equation}
\delta X^\beta = X^{*\beta} - X^\beta
\end{equation}

\begin{equation}
\delta \Phi_A = \Phi_A^*(X^{*\alpha}) - \Phi_A(X^\alpha)
\end{equation}

\begin{equation}
\delta (\partial_\rho \Phi_A) = \partial_\rho^* \Phi_A^*(X^{*\alpha}) - \partial_\rho \Phi_A(X^\alpha)
\end{equation}

where the fields are functions of their corresponding coordinates, thus the Taylor expansion reads,

\begin{multline}
\mathcal{L} (X^{*\alpha}, \Phi_A^*, \partial_\nu^* \Phi_A^*) = \mathcal{L} (X^\alpha, \Phi_A, \partial_\nu \Phi_A) 
+ \frac{\partial \mathcal{L}(X^\alpha, \Phi_A, \partial_\nu \Phi_A)}{\partial X^\beta} \delta X^\beta \\
+ \frac{\partial \mathcal{L}(X^\alpha, \Phi_A, \partial_\nu \Phi_A)}{\partial \Phi_A} \delta \Phi_A 
+ \frac{\partial \mathcal{L}(X^\alpha, \Phi_A, \partial_\nu \Phi_A)}{\partial (\partial_\rho \Phi_A)} \delta (\partial_\rho \Phi_A) + \dots
\end{multline}

each of these variations is proportional to first order epsilon so these are the only terms we will maintain, for example recalling the relationship between the coordinates,

\begin{equation}
X^{*\alpha} = X^\alpha + \epsilon g^\alpha (X, \Phi, \partial \Phi) + \mathcal{O}(\epsilon^2)
\end{equation}

the variation is at first order $\epsilon$,

\begin{equation}
\delta X^\alpha = X^{*\alpha} - X^\alpha = \epsilon g^\alpha (X, \Phi, \partial \Phi) 
\end{equation}

Recall the difference in actions, 

\begin{equation}
\Delta S = \int_{R_{X}} [ \mathcal{L} (X^{*\alpha}, \Phi_A^*, \partial_\nu^* \Phi_A^*)  + \epsilon   \mathcal{L} (X^{*\alpha}, \Phi_A^*, \partial_\nu^* \Phi_A^*) \frac{\partial g^\alpha}{\partial X^\alpha}  -  \mathcal{L} (X^\alpha, \Phi_A, \partial_\nu \Phi_A) ] dX
\end{equation}

\subsubsection*{The desired variation $\delta S$}

The first term above will have 4 terms up to order epsilon from the Taylor expansion, the second term will have one term order epsilon. Since we drop all nonlinear in epsilon terms at this point, we have recovered $\delta S$ from $\Delta S$, since the variation $\delta S$ is defined as the linear part of the difference in the two actions. We are left with,

\begin{multline}
\delta S = \int_{R_{X}} [ \mathcal{L} (X^\alpha, \Phi_A, \partial_\nu \Phi_A) 
+ \frac{\partial \mathcal{L}(X^\alpha, \Phi_A, \partial_\nu \Phi_A)}{\partial X^\beta} \delta X^\beta 
+ \frac{\partial \mathcal{L}(X^\alpha, \Phi_A, \partial_\nu \Phi_A)}{\partial \Phi_A} \delta \Phi_A \\
+ \frac{\partial \mathcal{L}(X^\alpha, \Phi_A, \partial_\nu \Phi_A)}{\partial (\partial_\rho \Phi_A)} \delta (\partial_\rho \Phi_A)  + \epsilon   \mathcal{L} (X^\alpha, \Phi_A, \partial_\nu \Phi_A) \frac{\partial g^\alpha}{\partial X^\alpha}  -  \mathcal{L} (X^\alpha, \Phi_A, \partial_\nu \Phi_A) ] dX
\end{multline}

the first and last terms exactly cancel, thus,

\begin{equation}
\delta S = \int_{R_{X}} [  \frac{\partial \mathcal{L}(X^\alpha, \Phi_A, \partial_\nu \Phi_A)}{\partial X^\beta} \delta X^\beta 
+ \frac{\partial \mathcal{L}(X^\alpha, \Phi_A, \partial_\nu \Phi_A)}{\partial \Phi_A} \delta \Phi_A 
+ \frac{\partial \mathcal{L}(X^\alpha, \Phi_A, \partial_\nu \Phi_A)}{\partial (\partial_\rho \Phi_A)} \delta (\partial_\rho \Phi_A)  + \epsilon   \mathcal{L} (X^\alpha, \Phi_A, \partial_\nu \Phi_A) \frac{\partial g^\alpha}{\partial X^\alpha} ] dX
\end{equation}

now that we have expressed every term in terms of the Lagrangian that depends on the original coordinates and fields, we can drop the notation, and from here we know that $\mathcal{L} = \mathcal{L}(X^\alpha, \Phi_A, \partial_\nu \Phi_A)$,

\begin{equation}
\delta S = \int_{R_{X}} [  \frac{\partial \mathcal{L}}{\partial X^\beta} \delta X^\beta 
+ \frac{\partial \mathcal{L}}{\partial \Phi_A} \delta \Phi_A 
+ \frac{\partial \mathcal{L}}{\partial (\partial_\rho \Phi_A)} \delta (\partial_\rho \Phi_A)  
+ \epsilon   \mathcal{L} \frac{\partial g^\alpha}{\partial X^\alpha} ] dX
\end{equation}

In order to continue we require explicit expressions for the variations $\delta X^\beta $, $ \delta \Phi_A$ and $\delta (\partial_\rho \Phi_A) $, which will be the topic of the next section.

\subsection{Determining expressions for the variations $\delta X^\beta $ and  $ \delta \Phi_A$ }

\subsubsection*{Calculating the variation $\delta X^\beta$}

We define the variation $\delta X^\beta$ as,

\begin{equation}
\delta X^\beta = X^{*\beta} - X^\beta
\end{equation}

based on the relationship between the two coordinates as previously defined,

\begin{equation}
X^{*\beta} = X^\beta + \epsilon g^\beta (X, \Phi, \partial \Phi) + \mathcal{O}(\epsilon^2)
\end{equation}

The difference between these two coordinates is,

\begin{equation}
\Delta X^\beta = X^{*\beta} - X^\beta = \epsilon g^\beta (X, \Phi, \partial \Phi) + \mathcal{O}(\epsilon^2)
\end{equation}

which is defined as a variation at first order,

\begin{equation}
\delta X^\beta = X^{*\beta} - X^\beta = \epsilon g^\beta (X, \Phi, \partial \Phi) 
\end{equation}

since $\Phi$ and $\partial \Phi$ depend on $X$, i.e. $\Phi(X)$ and $\partial \Phi (X)$, we can abbreviate $g^\beta (X, \Phi, \partial \Phi) = g^\beta (X)$, thus our variation reads,

\begin{equation}
\delta X^\beta = \epsilon g^\beta (X) 
\end{equation}

which gives us an expression for the linear part of the difference between the two coordinates.

\subsubsection*{Calculating the variation $ \delta \Phi_A$}

We define the variation $ \delta \Phi_A$ as,

\begin{equation}
\delta \Phi_A = \Phi_A^*(X^{*\alpha}) - \Phi_A(X^\alpha)
\end{equation}

where the variation is the linear part of the difference between the two fields as a function of their respective coordinates. Based on the relationship between the two fields as previously defined,

\begin{equation}
\Phi_A^*(X^{*\alpha}) = \Phi_A(X^\alpha) + \epsilon h_A (X, \Phi, \partial \Phi) + \mathcal{O}(\epsilon^2)
\end{equation}

The difference between these two fields is,

\begin{equation}
\Delta \Phi_A = \Phi_A^*(X^{*\alpha}) - \Phi_A(X^\alpha) =  \epsilon h_A (X, \Phi, \partial \Phi) + \mathcal{O}(\epsilon^2)
\end{equation}

which is defined as a variation at first order,

\begin{equation}
\delta \Phi_A = \Phi_A^*(X^{*\alpha}) - \Phi_A(X^\alpha) =  \epsilon h_A (X, \Phi, \partial \Phi) 
\end{equation}

since $\Phi$ and $\partial \Phi$ depend on $X$, i.e. $\Phi(X)$ and $\partial \Phi (X)$, we can abbreviate $h_A (X, \Phi, \partial \Phi) = h_A (X)$, thus our variation reads,

\begin{equation}
\delta \Phi_A = \epsilon h_A (X) 
\end{equation}

which gives us an expression for the linear part of the difference between the two fields. 

\subsubsection*{Calculating the variation $\bar{\delta} \Phi_A$}

In order to complete the calculation we must also consider the variation $\bar{\delta} \Phi_A$, which we will introduce the bar to define it as the change in the two fields when both are defined in terms of the same $X^\alpha$ coordinate,

\begin{equation}
\bar{\delta} \Phi_A = \Phi_A^*(X^\alpha) - \Phi_A(X^\alpha)
\end{equation}

in this case we have a new relationship between our fields,

\begin{equation}
\Phi_A^*(X^\alpha) = \Phi_A(X^\alpha) + \epsilon \bar{h}_A (X, \Phi, \partial \Phi) + \mathcal{O}(\epsilon^2)
\end{equation}

where $\bar{h}_A$ is associated with the difference between the two fields when we vary our fields but maintain the same coordinates for each. The difference between the two is,

\begin{equation}
\bar{\Delta} \Phi_A = \Phi_A^*(X^\alpha) - \Phi_A(X^\alpha) = \epsilon \bar{h}_A (X, \Phi, \partial \Phi) + \mathcal{O}(\epsilon^2)
\end{equation}

and the variation is defined at order epsilon, abbreviating to,

\begin{equation}
\bar{\delta} \Phi_A = \epsilon \bar{h}_A (X) 
\end{equation}

which gives us a relationship for the change in fields when the input coordinates are the same.

\subsubsection*{Calculating the relationship between the field variations $ \delta \Phi_A$ and $\bar{\delta} \Phi_A$}

We can determine a relationship between these two variations by returning to the difference,

\begin{equation}
\Delta \Phi_A = \Phi_A^*(X^{*\alpha}) - \Phi_A(X^\alpha) 
\end{equation}

if we add and subtract $\Phi_A^*(X^\alpha)$ in this expression we get,

\begin{equation}
\Delta \Phi_A = [ \Phi_A^*(X^{*\alpha}) - \Phi_A^*(X^\alpha) ] + [ \Phi_A^*(X^\alpha) - \Phi_A(X^\alpha) ]
\end{equation}

immediately we can recognize the second term from our bar variation calculation,

\begin{equation}
\Delta \Phi_A = [ \Phi_A^*(X^{*\alpha}) - \Phi_A^*(X^\alpha) ] + \bar{\Delta} \Phi_A
\end{equation}

if we Taylor expand the first part of the first term about $X^\alpha$,

\begin{equation}
\Phi_A^*(X^{*\alpha}) = \Phi_A^*(X^\alpha)  + \frac{\partial \Phi_A^*(X^\alpha)}{\partial X^\beta} (X^{*\beta} - X^\beta) + \mathcal{O}(\epsilon^2)
\end{equation}

recall that $\delta X^\beta = X^{*\beta} - X^\beta$, which is first order in $\epsilon$,

\begin{equation}
\Phi_A^*(X^{*\alpha}) = \Phi_A^*(X^\alpha)  + \frac{\partial \Phi_A^*(X^\alpha)}{\partial X^\beta} \delta X^\beta + \mathcal{O}(\epsilon^2)
\end{equation}

inserting this Taylor expansion into the change in fields relationship,

\begin{equation}
\Delta \Phi_A = [\Phi_A^*(X^\alpha)  + \frac{\partial \Phi_A^*(X^\alpha)}{\partial X^\beta} \delta X^\beta + \mathcal{O}(\epsilon^2) - \Phi_A^*(X^\alpha) ] + \bar{\Delta} \Phi_A
\end{equation}

the first and fourth terms in the square brackets cancel,

\begin{equation}
\Delta \Phi_A = [ \frac{\partial \Phi_A^*(X^\alpha)}{\partial X^\beta} \delta X^\beta + \mathcal{O}(\epsilon^2)  ] + \bar{\Delta} \Phi_A
\end{equation}

we can replace the derivative with one in the original coordinates by recalling,

\begin{equation}
\Phi_A^*(X^\alpha) = \Phi_A(X^\alpha) + \epsilon \bar{h}_A (X, \Phi, \partial \Phi) + \mathcal{O}(\epsilon^2)
\end{equation}

differentiating both sides,

\begin{equation}
\frac{\partial \Phi_A^*(X^\alpha)}{\partial X^\beta} = \frac{\partial \Phi_A(X^\alpha)}{\partial X^\beta} + \epsilon \frac{\partial }{\partial X^\beta} \bar{h}_A (X, \Phi, \partial \Phi) + \frac{\partial }{\partial X^\beta} \mathcal{O}(\epsilon^2)
\end{equation}

inserting this back into the change in fields relationship,

\begin{equation}
\Delta \Phi_A = [ (\frac{\partial \Phi_A(X^\alpha)}{\partial X^\beta} + \epsilon \frac{\partial }{\partial X^\beta} \bar{h}_A (X, \Phi, \partial \Phi) + \frac{\partial }{\partial X^\beta} \mathcal{O}(\epsilon^2)) \delta X^\beta + \mathcal{O}(\epsilon^2)  ] + \bar{\Delta} \Phi_A
\end{equation}

for variations $ \delta \Phi_A$ and $\bar{\delta} \Phi_A$ we require only the linear portion of this expression with respect to epsilon, therefore the equation reduces to (after neglecting second, third and fourth terms which are $\epsilon^2$ or greater),

\begin{equation}
\delta \Phi_A = \frac{\partial \Phi_A(X^\alpha)}{\partial X^\beta} \delta X^\beta  + \bar{\delta} \Phi_A
\end{equation}

this gives us a relationship between the two different variations of fields which we will commonly abbreviate as,

\begin{equation}
\delta \Phi_A = \frac{\partial \Phi_A}{\partial X^\beta} \delta X^\beta  + \bar{\delta} \Phi_A
\end{equation}

which is similar to the variation taken for Poincare translation in field theory. Typically we will write this more compactly as,

\begin{equation}
\delta \Phi_A = \partial_\beta \Phi_A \delta X^\beta  + \bar{\delta} \Phi_A
\end{equation}

\subsubsection*{A relationship between the two derivatives}

Suppose we want to determine the relationship between the derivatives,

\begin{equation}
 \frac{\partial }{\partial X^{*\rho}} \quad , \quad
 \frac{\partial }{\partial X^{\rho}}  
\end{equation}

we can start by noting the relationship between the two coordinates,

\begin{equation}
X^{*\alpha} = X^\alpha + \epsilon g^\alpha (X, \Phi, \partial \Phi) + \mathcal{O}(\epsilon^2)
\end{equation}

differentiating both sides of this expression and keeping only terms linear in $\epsilon$, and abbreviating $g$,

\begin{equation}
\frac{\partial X^{*\alpha}}{\partial X^\beta} \approx \frac{\partial X^\alpha}{\partial X^\beta} + \epsilon \frac{\partial g^\alpha (X)}{\partial X^\beta}
\end{equation}

where the first term is just the Kronecker delta,

\begin{equation}
\frac{\partial X^{*\alpha}}{\partial X^\beta} \approx \delta^\alpha_\beta + \epsilon \frac{\partial g^\alpha (X)}{\partial X^\beta}
\end{equation}

to change variable between the two derivatives we have,

\begin{equation}
\frac{\partial}{\partial X^\beta} =  \frac{\partial X^{*\alpha}}{\partial X^\beta} \frac{\partial}{\partial X^{*\alpha}}
\end{equation}

which we can introduce our expression for $\frac{\partial X^{*\alpha}}{\partial X^\beta} $ on the right hand side,

\begin{equation}
\frac{\partial}{\partial X^\beta} = (\delta^\alpha_\beta + \epsilon \frac{\partial g^\alpha (X)}{\partial X^\beta}) \frac{\partial}{\partial X^{*\alpha}}
\end{equation}

our derivative operator can contract with the delta, yielding,

\begin{equation}
\frac{\partial}{\partial X^\beta} =  \frac{\partial}{\partial X^{*\beta}} + \epsilon \frac{\partial g^\alpha (X)}{\partial X^\beta} \frac{\partial}{\partial X^{*\alpha}}
\end{equation}

therefore the difference in the two derivatives is,

\begin{equation}
\frac{\partial}{\partial X^\beta} -  \frac{\partial}{\partial X^{*\beta}} = \epsilon \frac{\partial g^\alpha (X)}{\partial X^\beta} \frac{\partial}{\partial X^{*\alpha}}
\end{equation}

Before turning to the variation of the derivatives of the fields, we will first specify the field variations under coordinate transformations at each rank.

\subsection{Variations of higher ranked $\Phi_A$ under coordinate transformations}

In this section we will talk specifically about the variation of field definitions under a coordinate transformation. Earlier we defined the relationship,

\begin{equation}
\Phi_A^*(X^*) = \Phi_A(X) + \epsilon h_A (X, \Phi, \partial \Phi) + \mathcal{O}(\epsilon^2)
\end{equation}

where $h_A$ is some arbitrary generator of a field transformation. How this behaves for specifically a coordinate transformation alone is an important question for physical applications; here we determine this transformation relationship explicitly. As we previously calculated, keeping the linear part of the above transformation we have (since the fields are uniquely determined by $X$),

\begin{equation}
\delta \Phi_A = \epsilon h_A (X) 
\end{equation}

this definition holds for fields of any rank; as we will show, the $h_A$ also will change depending on the rank of the field. See e.g., \cite{Banados2016} for additional discussion.

\subsubsection*{Scalar (rank 0 tensor) transformation}

Suppose we have a scalar $\Phi$. We know that from tensor transformation laws we are relating the transformed object to the untransformed object under a coordinate transformation. A scalar transforms under a coordinate transformation as,

\begin{equation}
    \Phi^*(X^{*\alpha}) = \Phi(X^\alpha)
\end{equation}

what this means is, 

\begin{equation}
    \Delta \Phi = \Phi^*(X^{*\alpha}) - \Phi(X^\alpha) = 0 
\end{equation}

Therefore for the previous result at linear order $\delta \Phi = \epsilon h (X) $, we have that $\delta \Phi = 0$ and thus $h = 0$ for a scalar (the total variation is zero for a scalar). This is directly related to why the Lagrangian is constructed as a scalar: the action built from it is then invariant under coordinate transformations, the condition from which the Noether identity is derived. For curved spacetimes, the scalar density must also include the $\sqrt{-g}$ Jacobian contributions (this is $1$ in flat spacetimes considered in this article).\\

For the scalar field therefore we have for the relationship to the substantial variation

\begin{equation}
\delta \Phi = \partial_\beta \Phi \delta X^\beta  + \bar{\delta} \Phi
\end{equation}

but since $\delta \Phi = 0$ we have,

\begin{equation}
\bar{\delta} \Phi = -\partial_\beta \Phi \delta X^\beta  
\end{equation}

which is the form of the so-called canonical variational symmetry \cite{BakerLinnemannSmeenk2021}.

\subsubsection*{Vector (rank 1 tensor) transformation}

However, for a vector, the (covariant) tensor transformation law for $\Phi_\mu$ is,

\begin{equation}
\Phi_\mu^*(X^{*\alpha}) = \frac{\partial X^{\beta}}{\partial X^{*\mu}} \Phi_\beta(X^\alpha)
\end{equation}

we now need to take into account the Jacobian for this under the infinitesimal coordinate transformation defined earlier as $X^{*\alpha} = X^\alpha + \epsilon g^\alpha(X) + \mathcal{O}(\epsilon^2)$ which gave,

\begin{equation}
\frac{\partial X^{*\alpha}}{\partial X^\beta} = \delta^\alpha_\beta
+ \epsilon \frac{\partial g^\alpha}{\partial X^\beta} + \mathcal{O}(\epsilon^2)
\end{equation}

but for the covariant vector we need the opposite differentiation which we can develop from,

\begin{equation}
X^\alpha =     X^{*\alpha} - \epsilon g^\alpha(X) - \mathcal{O}(\epsilon^2)
\end{equation}

thus differentiating we have,

\begin{equation}
\frac{\partial X^{\alpha}}{\partial X^{*\beta}} = \delta^\alpha_\beta
- \epsilon \frac{\partial g^\alpha}{\partial X^\beta} + \mathcal{O}(\epsilon^2)
\end{equation}

inserting this into the vector transformation law,

\begin{equation}
\Phi_\mu^*(X^{*\alpha}) 
= \Phi_\mu(X^\alpha)
- \epsilon \frac{\partial g^\beta}{\partial X^\mu} \Phi_\beta(X^\alpha)
+ \mathcal{O}(\epsilon^2)
\end{equation}

If we compare this to the previously defined relationship,

\begin{equation}
\Phi_A^*(X^*) = \Phi_A(X) + \epsilon h_\mu (X, \Phi, \partial \Phi) + \mathcal{O}(\epsilon^2)
\end{equation}

we have the result that,

\begin{equation}
    h_\mu = - \frac{\partial g^\beta}{\partial X^\mu} \Phi_\beta = - (\partial_\mu g^\beta)  \Phi_\beta
\end{equation}

this means we have the field variation,

\begin{equation}
    \delta \Phi_\mu = \epsilon h_\mu = - \epsilon (\partial_\mu g^\beta)  \Phi_\beta
\end{equation}

using $\delta X^\beta = \epsilon g^\beta$ we can write this as,

\begin{equation}
    \delta \Phi_\mu = -  (\partial_\mu \delta X^\beta)  \Phi_\beta
\end{equation}

recall before we determined the difference between the total and substantial variations to be, in the case of vector $A \to \mu$,

\begin{equation}
\delta \Phi_\mu = \partial_\beta \Phi_\mu \delta X^\beta  + \bar{\delta} \Phi_\mu
\end{equation}

thus the substantial variation can be expressed as,

\begin{equation}
\bar{\delta} \Phi_\mu = \delta \Phi_\mu - \partial_\beta \Phi_\mu \delta X^\beta  
\end{equation}

inserting $\delta \Phi_\mu$,

\begin{equation}
\bar{\delta} \Phi_\mu = -  (\partial_\mu \delta X^\beta)  \Phi_\beta - \partial_\beta \Phi_\mu \delta X^\beta  
\end{equation}

but what we have left can be expressed as the Lie derivative of a vector,

\begin{equation}
    \mathsterling_{\delta X} \Phi_\mu =   (\partial_\mu \delta X^\beta)  \Phi_\beta + \partial_\beta \Phi_\mu \delta X^\beta  
\end{equation}

leaving us with the substantial variation of the fields as,

\begin{equation}
\bar{\delta} \Phi_\mu = -      \mathsterling_{\delta X} \Phi_\mu
\end{equation}

commonly in physics we write $\delta X^\beta = \xi^\beta$ and this becomes $\bar{\delta} \Phi_\mu = -      \mathsterling_{\xi} \Phi_\mu$.\\

It is important to note that the negative sign here is due to the ``passive sign convention'' used in defining the coordinate transformation as $X^{*\alpha} = X^\alpha + \epsilon g^\alpha(X) $. If we had used $X^{*\alpha} = X^\alpha - \epsilon g^\alpha(X) $, known as ``active sign convention'', we would be left with $\bar{\delta} \Phi_\mu =       \mathsterling_{\xi} \Phi_\mu$ for the substantial variation.\\

\subsubsection*{Rank 2 tensor transformation}

This result, that the substantial variation is related to the Lie derivative, generalizes to all ranks. For example, a second rank covariant tensor,

\begin{equation}
\Phi_{\mu\nu}^*(X^{*\alpha}) = \frac{\partial X^{\beta}}{\partial X^{*\mu}} \frac{\partial X^{\gamma}}{\partial X^{*\nu}} \Phi_{\beta\gamma}(X^\alpha)
\end{equation}

we again need to use,

\begin{equation}
\frac{\partial X^{\beta}}{\partial X^{*\mu}} = \delta^\beta_\mu
- \epsilon \frac{\partial g^\beta}{\partial X^\mu} + \mathcal{O}(\epsilon^2)
\end{equation}

\begin{equation}
\frac{\partial X^{\gamma}}{\partial X^{*\nu}} = \delta^\gamma_\nu
- \epsilon \frac{\partial g^\gamma}{\partial X^\nu} + \mathcal{O}(\epsilon^2)
\end{equation}

inserting this,

\begin{equation}
    \Phi_{\mu\nu}^*(X^{*\alpha}) = (\delta^\beta_\mu
- \epsilon \frac{\partial g^\beta}{\partial X^\mu} + \mathcal{O}(\epsilon^2)) (\delta^\gamma_\nu
- \epsilon \frac{\partial g^\gamma}{\partial X^\nu} + \mathcal{O}(\epsilon^2)) \Phi_{\beta\gamma}(X^\alpha)
\end{equation}

expanding we are left with,

\begin{equation}
\Phi_{\mu\nu}^*(X^{*\alpha}) 
= \Phi_{\mu\nu}(X^\alpha)
- \epsilon \frac{\partial g^\beta}{\partial X^\mu} \Phi_{\beta\nu}(X^\alpha)
- \epsilon \frac{\partial g^\beta}{\partial X^\nu} \Phi_{\mu\beta}(X^\alpha)
+ \mathcal{O}(\epsilon^2)
\end{equation}

If we compare this to the previously defined relationship,

\begin{equation}
\Phi_A^*(X^*) = \Phi_A(X) + \epsilon h_{\mu\nu} (X, \Phi, \partial \Phi) + \mathcal{O}(\epsilon^2)
\end{equation}

we have the result that,

\begin{equation}
    h_{\mu\nu} = - \frac{\partial g^\beta}{\partial X^\mu} \Phi_{\beta\nu} - \frac{\partial g^\beta}{\partial X^\nu} \Phi_{\mu\beta} = - (\partial_\mu g^\beta)  \Phi_{\beta\nu} - (\partial_\nu g^\beta)  \Phi_{\mu\beta}
\end{equation}

this means we have the field variation,

\begin{equation}
    \delta \Phi_{\mu\nu} = \epsilon h_{\mu\nu} = - \epsilon (\partial_\mu g^\beta)  \Phi_{\beta\nu} - \epsilon (\partial_\nu g^\beta)  \Phi_{\mu\beta}
\end{equation}

using $\delta X^\beta = \epsilon g^\beta$ we can write this as,

\begin{equation}
    \delta \Phi_{\mu\nu} = -  (\partial_\mu \delta X^\beta)  \Phi_{\beta\nu} -  (\partial_\nu \delta X^\beta)  \Phi_{\mu\beta}
\end{equation}

recall before we determined the difference between the total and substantial variations to be, in the case of a second rank tensor $A \to \mu\nu$,

\begin{equation}
\delta \Phi_{\mu\nu} = \partial_\beta \Phi_{\mu\nu} \delta X^\beta  + \bar{\delta} \Phi_{\mu\nu}
\end{equation}

thus the substantial variation can be expressed as,

\begin{equation}
\bar{\delta} \Phi_{\mu\nu} = \delta \Phi_{\mu\nu} - \partial_\beta \Phi_{\mu\nu} \delta X^\beta  
\end{equation}

inserting $\delta \Phi_{\mu\nu}$,

\begin{equation}
\bar{\delta} \Phi_{\mu\nu} = -  (\partial_\mu \delta X^\beta)  \Phi_{\beta\nu} -  (\partial_\nu \delta X^\beta)  \Phi_{\mu\beta} - \partial_\beta \Phi_{\mu\nu} \delta X^\beta  
\end{equation}

but what we have left can be expressed as the Lie derivative of a second rank tensor,

\begin{equation}
    \mathsterling_{\delta X} \Phi_{\mu\nu} =   (\partial_\mu \delta X^\beta)  \Phi_{\beta\nu} +  (\partial_\nu \delta X^\beta)  \Phi_{\mu\beta} + \partial_\beta \Phi_{\mu\nu} \delta X^\beta  
\end{equation}

leaving us with the substantial variation of the fields as,

\begin{equation}
\bar{\delta} \Phi_{\mu\nu} = -      \mathsterling_{\delta X} \Phi_{\mu\nu}
\end{equation}

\subsubsection*{General substantial variation under a coordinate transformation}

Altogether, for any rank of tensor, each additional rank will contribute 1 additional term associated to the Lie derivative at the given rank. So the following general relationship for the substantial variation under a coordinate transformation is,

\begin{equation}
\bar{\delta} \Phi_A = -      \mathsterling_{\delta X} \Phi_A
\end{equation}

We note that other transformations (such as gauge transformations $\delta_g \Phi_A$) can be incorporated as well (not done here) that would cause the substantial variation to be of the form,

\begin{equation}
\bar{\delta} \Phi_A = -      \mathsterling_{\delta X} \Phi_A + \delta_g \Phi_A
\end{equation}

this is the basis for the Bessel-Hagen method which we will discuss briefly at the end of the document.

\subsection{Determining expressions for the variation $\delta (\partial_\rho \Phi_A) $}

\subsubsection*{Introducing the variation $\delta (\partial_\rho \Phi_A) $}

We define the difference $\Delta (\partial_\rho \Phi_A) $ as,

\begin{equation}
\Delta (\partial_\rho \Phi_A)  = \frac{\partial \Phi_A^*(X^{*\alpha})}{\partial X^{*\rho}} - \frac{\partial \Phi_A(X^\alpha)}{\partial X^\rho}
\end{equation}

where the variation is the linear part of the difference between the two derivatives of fields as a function of their respective coordinates (the linear part of the above expression). In order to evaluate this variation we need to add and subtract the following terms just like we did in the previous subsection,

\begin{equation}
\pm \frac{\partial \Phi_A(X^{*\alpha})}{\partial X^{*\rho}} \quad , \quad \pm \frac{\partial \Phi_A(X^{*\alpha})}{\partial X^{\rho}}
\end{equation}

introducing all of these terms,

\begin{equation}
\Delta (\partial_\rho \Phi_A)  = \frac{\partial \Phi_A^*(X^{*\alpha})}{\partial X^{*\rho}} - \frac{\partial \Phi_A(X^\alpha)}{\partial X^\rho} + \frac{\partial \Phi_A(X^{*\alpha})}{\partial X^{*\rho}} - \frac{\partial \Phi_A(X^{*\alpha})}{\partial X^{*\rho}} + \frac{\partial \Phi_A(X^{*\alpha})}{\partial X^{\rho}} - \frac{\partial \Phi_A(X^{*\alpha})}{\partial X^{\rho}}
\end{equation}

and rearranging/ combining,

\begin{equation}
\Delta (\partial_\rho \Phi_A)  = \frac{\partial \Phi_A^*(X^{*\alpha})}{\partial X^{*\rho}} - \frac{\partial \Phi_A(X^{*\alpha})}{\partial X^{*\rho}} 
+ \frac{\partial \Phi_A(X^{*\alpha})}{\partial X^{\rho}}   - \frac{\partial \Phi_A(X^\alpha)}{\partial X^\rho} 
+ \frac{\partial \Phi_A(X^{*\alpha})}{\partial X^{*\rho}} - \frac{\partial \Phi_A(X^{*\alpha})}{\partial X^{\rho}}
\end{equation}

\begin{equation}
\Delta (\partial_\rho \Phi_A)  = \frac{\partial [\Phi_A^*(X^{*\alpha}) - \Phi_A(X^{*\alpha})]}{\partial X^{*\rho}} 
+ \frac{\partial [\Phi_A(X^{*\alpha}) - \Phi_A(X^\alpha)]}{\partial X^{\rho}}  
+ [\frac{\partial }{\partial X^{*\rho}} - \frac{\partial }{\partial X^{\rho}}] \Phi_A(X^{*\alpha})
\end{equation}

each of the 3 terms above deserves separate attention because each involves a fairly lengthy calculation from here. We will start with the first term.

\subsubsection*{$\delta (\partial_\rho \Phi_A) $ First term calculation}

We start with the first term,

\begin{equation}
 \frac{\partial [\Phi_A^*(X^{*\alpha}) - \Phi_A(X^{*\alpha})]}{\partial X^{*\rho}} 
\end{equation}

recall the relationship between the derivatives for the two coordinates,

\begin{equation}
\frac{\partial}{\partial X^\beta} -  \frac{\partial}{\partial X^{*\beta}} = \epsilon \frac{\partial g^\alpha (X)}{\partial X^\beta} \frac{\partial}{\partial X^{*\alpha}}
\end{equation}

thus,

\begin{equation}
\frac{\partial}{\partial X^\rho} -  \frac{\partial}{\partial X^{*\rho}} = \epsilon \frac{\partial g^\alpha (X)}{\partial X^\rho} \frac{\partial}{\partial X^{*\alpha}}
\end{equation}

we can use this to rewrite the derivative,

\begin{equation}
 \frac{\partial [\Phi_A^*(X^{*\alpha}) - \Phi_A(X^{*\alpha})]}{\partial X^{*\rho}}  =  \frac{\partial [\Phi_A^*(X^{*\alpha}) - \Phi_A(X^{*\alpha})]}{\partial X^{\rho}}  -  \epsilon \frac{\partial g^\beta (X)}{\partial X^\rho} \frac{\partial [\Phi_A^*(X^{*\alpha}) - \Phi_A(X^{*\alpha})]}{\partial X^{*\beta}} 
\end{equation}

we can also note that the term in the derivative is related to the variation,

\begin{equation}
\bar{\delta} \Phi_A (X^\alpha) = \Phi_A^*(X^\alpha) - \Phi_A(X^\alpha)
\end{equation}

only here we have this variation as a function of the new coordinates,

\begin{equation}
\bar{\delta} \Phi_A (X^{*\alpha}) = \Phi_A^*(X^{*\alpha}) - \Phi_A(X^{*\alpha}) 
\end{equation}

therefore we can rewrite the first term as,

\begin{equation}
 \frac{\partial [\Phi_A^*(X^{*\alpha}) - \Phi_A(X^{*\alpha})]}{\partial X^{*\rho}}  =  \frac{\partial \bar{\delta} \Phi_A (X^{*\alpha})}{\partial X^{\rho}}  -  \epsilon \frac{\partial g^\beta (X)}{\partial X^\rho} \frac{\partial \bar{\delta} \Phi_A (X^{*\alpha})}{\partial X^{*\beta}} 
\end{equation}

recall that we defined in terms of the barred function,

\begin{equation}
\bar{\delta} \Phi_A(X^\alpha)  = \epsilon \bar{h}_A (X) 
\end{equation}

thus for this case in the new coordinates,

\begin{equation}
\bar{\delta} \Phi_A(X^{*\alpha})  = \epsilon \bar{h}_A (X^*) 
\end{equation}

the first term now reads,

\begin{equation}
 \frac{\partial [\Phi_A^*(X^{*\alpha}) - \Phi_A(X^{*\alpha})]}{\partial X^{*\rho}} =  \frac{\partial \epsilon \bar{h}_A (X^*) }{\partial X^{\rho}}  -  \epsilon \frac{\partial g^\beta (X)}{\partial X^\rho} \frac{\partial \epsilon \bar{h}_A (X^*) }{\partial X^{*\beta}} 
\end{equation}

the second term can be dropped since it is higher order in $\epsilon$, leaving us with only,

\begin{equation}
 \frac{\partial [\Phi_A^*(X^{*\alpha}) - \Phi_A(X^{*\alpha})]}{\partial X^{*\rho}} = \epsilon \frac{\partial  \bar{h}_A (X^*) }{\partial X^{\rho}}
\end{equation}

Taylor expanding $ \bar{h}_A (X^*) $ about $X$,

\begin{equation}
\epsilon \bar{h}_A (X^*) =\epsilon \bar{h}_A (X) + \mathcal{O}(\epsilon^2) 
\end{equation}

again allows only the first term in order $\epsilon$, thus finally we have for the first term,

\begin{equation}
 \frac{\partial [\Phi_A^*(X^{*\alpha}) - \Phi_A(X^{*\alpha})]}{\partial X^{*\rho}} = \epsilon \frac{\partial  \bar{h}_A (X) }{\partial X^{\rho}}
\end{equation}

\subsubsection*{$\delta (\partial_\rho \Phi_A) $ Second term calculation}

Now for the second term,

\begin{equation}
 \frac{\partial [\Phi_A(X^{*\alpha}) - \Phi_A(X^\alpha)]}{\partial X^{\rho}}  
\end{equation}

we can Taylor expand the $\Phi_A(X^{*\alpha})$,

\begin{equation}
\Phi_A(X^{*\alpha}) = \Phi_A(X^\alpha) + \frac{\partial \Phi_A(X^\alpha)}{\partial X^\beta} (X^{*\beta} - X^\beta) + \mathcal{O}(\epsilon^2)
\end{equation}

keeping only the first order in $\epsilon$ and recalling $\delta X^\beta = X^{*\beta} - X^\beta$,

\begin{equation}
\Phi_A(X^{*\alpha}) \approx \Phi_A(X^\alpha) + \frac{\partial \Phi_A(X^\alpha)}{\partial X^\beta} \delta X^\beta
\end{equation}

inserting this expansion into the second term,

\begin{equation}
  \frac{\partial [\Phi_A(X^{*\alpha}) - \Phi_A(X^\alpha)]}{\partial X^{\rho}}  = \frac{\partial [\Phi_A(X^\alpha) + \frac{\partial \Phi_A(X^\alpha)}{\partial X^\beta} \delta X^\beta - \Phi_A(X^\alpha)]}{\partial X^{\rho}}  
\end{equation}

the first and third terms cancel, leaving,

\begin{equation}
  \frac{\partial [\Phi_A(X^{*\alpha}) - \Phi_A(X^\alpha)]}{\partial X^{\rho}}  = \frac{\partial }{\partial X^{\rho}}  [\frac{\partial \Phi_A(X^\alpha)}{\partial X^\beta} \delta X^\beta ]
\end{equation}

we now recall the form of the variation of coordinates in terms of the function $g$,

\begin{equation}
\delta X^\beta = \epsilon g^\beta (X) 
\end{equation}

inserting this,

\begin{equation}
  \frac{\partial [\Phi_A(X^{*\alpha}) - \Phi_A(X^\alpha)]}{\partial X^{\rho}}  = \frac{\partial }{\partial X^{\rho}}  [\frac{\partial \Phi_A(X^\alpha)}{\partial X^\beta}  \epsilon g^\beta (X)  ]
\end{equation}

and using the product rule,

\begin{equation}
  \frac{\partial [\Phi_A(X^{*\alpha}) - \Phi_A(X^\alpha)]}{\partial X^{\rho}}  = \epsilon  [(  \frac{\partial }{\partial X^{\rho}} \frac{\partial \Phi_A(X^\alpha)}{\partial X^\beta})   g^\beta (X) + \frac{\partial \Phi_A(X^\alpha)}{\partial X^\beta}   \frac{\partial  g^\beta (X) }{\partial X^{\rho}} ]
\end{equation}

\begin{equation}
  \frac{\partial [\Phi_A(X^{*\alpha}) - \Phi_A(X^\alpha)]}{\partial X^{\rho}}  = \epsilon  (  \frac{\partial }{\partial X^{\rho}} \frac{\partial \Phi_A(X^\alpha)}{\partial X^\beta})   g^\beta (X) + \epsilon \frac{\partial \Phi_A(X^\alpha)}{\partial X^\beta}   \frac{\partial  g^\beta (X) }{\partial X^{\rho}} 
\end{equation}

we find a first order epsilon expression for the second desired term.

\subsubsection*{$\delta (\partial_\rho \Phi_A) $ Third term calculation}

Finally for the third term,

\begin{equation}
[\frac{\partial }{\partial X^{*\rho}} - \frac{\partial }{\partial X^{\rho}}] \Phi_A(X^{*\alpha})
\end{equation}

we found the relationship between these two derivatives earlier on as,

\begin{equation}
\frac{\partial}{\partial X^\rho} -  \frac{\partial}{\partial X^{*\rho}} = \epsilon \frac{\partial g^\beta (X)}{\partial X^\rho} \frac{\partial}{\partial X^{*\beta}}
\end{equation}

therefore,

\begin{equation}
\frac{\partial}{\partial X^{*\rho}} - \frac{\partial}{\partial X^\rho}   = - \epsilon \frac{\partial g^\beta (X)}{\partial X^\rho} \frac{\partial}{\partial X^{*\beta}}
\end{equation}

and the third term reads,

\begin{equation}
[\frac{\partial }{\partial X^{*\rho}} - \frac{\partial }{\partial X^{\rho}}] \Phi_A(X^{*\alpha}) = - \epsilon \frac{\partial g^\beta (X)}{\partial X^\rho} \frac{\partial \Phi_A(X^{*\alpha}) }{\partial X^{*\beta}} 
\end{equation}

which we can again Taylor expand $\Phi_A(X^{*\alpha})$ and drop higher order terms to basically replace it with $\Phi_A(X^\alpha)$,

\begin{equation}
[\frac{\partial }{\partial X^{*\rho}} - \frac{\partial }{\partial X^{\rho}}] \Phi_A(X^{*\alpha}) = - \epsilon \frac{\partial g^\beta (X)}{\partial X^\rho} \frac{\partial \Phi_A(X^\alpha) }{\partial X^{*\beta}} 
\end{equation}

we can also replace the $\beta$ derivative with,

\begin{equation}
\frac{\partial}{\partial X^\beta} -  \frac{\partial}{\partial X^{*\beta}} = \epsilon \frac{\partial g^\sigma (X)}{\partial X^\beta} \frac{\partial}{\partial X^{*\sigma}}
\end{equation}

thus the higher order term again drops and we just get the derivative in the original coordinates,

\begin{equation}
[\frac{\partial }{\partial X^{*\rho}} - \frac{\partial }{\partial X^{\rho}}] \Phi_A(X^{*\alpha}) = - \epsilon \frac{\partial g^\beta (X)}{\partial X^\rho} \frac{\partial \Phi_A(X^\alpha) }{\partial X^\beta} 
\end{equation}

we now have expressions for each of the terms in terms of first order epsilon. It is necessary to put them all together.

\subsubsection*{Combining the 3 terms}

Recall that we wished to determine each of the following 3 terms,

\begin{equation}
\Delta (\partial_\rho \Phi_A)  = \frac{\partial [\Phi_A^*(X^{*\alpha}) - \Phi_A(X^{*\alpha})]}{\partial X^{*\rho}} 
+ \frac{\partial [\Phi_A(X^{*\alpha}) - \Phi_A(X^\alpha)]}{\partial X^{\rho}}  
+ [\frac{\partial }{\partial X^{*\rho}} - \frac{\partial }{\partial X^{\rho}}] \Phi_A(X^{*\alpha})
\end{equation}

with respect to first order in epsilon. We found each of these 3 in the prior sections,

\begin{equation}
 \frac{\partial [\Phi_A^*(X^{*\alpha}) - \Phi_A(X^{*\alpha})]}{\partial X^{*\rho}} = \epsilon \frac{\partial  \bar{h}_A (X) }{\partial X^{\rho}}
\end{equation}

\begin{equation}
  \frac{\partial [\Phi_A(X^{*\alpha}) - \Phi_A(X^\alpha)]}{\partial X^{\rho}}  = \epsilon  (  \frac{\partial }{\partial X^{\rho}} \frac{\partial \Phi_A(X^\alpha)}{\partial X^\beta})   g^\beta (X) + \epsilon \frac{\partial \Phi_A(X^\alpha)}{\partial X^\beta}   \frac{\partial  g^\beta (X) }{\partial X^{\rho}} 
\end{equation}

\begin{equation}
[\frac{\partial }{\partial X^{*\rho}} - \frac{\partial }{\partial X^{\rho}}] \Phi_A(X^{*\alpha}) = - \epsilon \frac{\partial g^\beta (X)}{\partial X^\rho} \frac{\partial \Phi_A(X^\alpha) }{\partial X^\beta} 
\end{equation}

inserting each of them we find,

\begin{equation}
\Delta (\partial_\rho \Phi_A)  = \epsilon \frac{\partial  \bar{h}_A (X) }{\partial X^{\rho}}
+  \epsilon  (  \frac{\partial }{\partial X^{\rho}} \frac{\partial \Phi_A(X^\alpha)}{\partial X^\beta})   g^\beta (X) 
+ \epsilon \frac{\partial \Phi_A(X^\alpha)}{\partial X^\beta}   \frac{\partial  g^\beta (X) }{\partial X^{\rho}} 
- \epsilon \frac{\partial g^\beta (X)}{\partial X^\rho} \frac{\partial \Phi_A(X^\alpha) }{\partial X^\beta} 
\end{equation}

the third and fourth terms exactly cancel!

\begin{equation}
\Delta (\partial_\rho \Phi_A)  = \epsilon \frac{\partial  \bar{h}_A (X) }{\partial X^{\rho}}
+  \epsilon  (  \frac{\partial }{\partial X^{\rho}} \frac{\partial \Phi_A(X^\alpha)}{\partial X^\beta})   g^\beta (X) 
\end{equation}

recalling the variations,

\begin{equation}
\delta X^\beta = \epsilon g^\beta (X) 
\end{equation}

\begin{equation}
\bar{\delta} \Phi_A(X)  = \epsilon \bar{h}_A (X) 
\end{equation}

since we are left with only the linear part we have found the desired variation $\Delta (\partial_\rho \Phi_A)$,

\begin{equation}
\delta (\partial_\rho \Phi_A)  =  \frac{\partial \bar{\delta} \Phi_A(X)}{\partial X^{\rho}} 
+    (  \frac{\partial }{\partial X^{\rho}} \frac{\partial \Phi_A(X^\alpha)}{\partial X^\beta})   \delta X^\beta
\end{equation}

which is the derivative of the bar variation, and a second order contribution due to the fields,

\begin{equation}
\delta (\partial_\rho \Phi_A)  =  \frac{\partial  }{\partial X^{\rho}} \bar{\delta} \Phi_A(X)
+    [ \frac{\partial }{\partial X^{\rho}} \frac{\partial }{\partial X^\beta} \Phi_A(X^\alpha)]   \delta X^\beta
\end{equation}

this is the desired variation which we will be using to simplify the variation of the action from the prior subsection. Typically we will write this more compactly as,

\begin{equation}
\delta (\partial_\rho \Phi_A)  = \partial_\rho \bar{\delta} \Phi_A
+    [\partial_\rho \partial_\beta \Phi_A]   \delta X^\beta
\end{equation}

\subsection{Using the field variations on the variation of the action}

\subsubsection*{Inserting the variations}

Recall that we calculated the variation of the action to be,

\begin{equation}
\delta S = \int_{R_{X}} [  \frac{\partial \mathcal{L}}{\partial X^\beta} \delta X^\beta 
+ \frac{\partial \mathcal{L}}{\partial \Phi_A} \delta \Phi_A 
+ \frac{\partial \mathcal{L}}{\partial (\partial_\rho \Phi_A)} \delta (\partial_\rho \Phi_A)  
+ \epsilon   \mathcal{L} \frac{\partial g^\alpha}{\partial X^\alpha} ] dX
\end{equation}

we then found the variations,

\begin{equation}
\delta \Phi_A = \partial_\beta \Phi_A \delta X^\beta  + \bar{\delta} \Phi_A
\end{equation}

and,

\begin{equation}
\delta (\partial_\rho \Phi_A)  = \partial_\rho \bar{\delta} \Phi_A  +    [\partial_\rho \partial_\beta \Phi_A]   \delta X^\beta
\end{equation}

inserting these variations into the variation of the action,

\begin{equation}
\delta S = \int_{R_{X}} [  \frac{\partial \mathcal{L}}{\partial X^\beta} \delta X^\beta 
+ \frac{\partial \mathcal{L}}{\partial \Phi_A} \partial_\beta \Phi_A \delta X^\beta  +  \frac{\partial \mathcal{L}}{\partial \Phi_A} \bar{\delta} \Phi_A
+ \frac{\partial \mathcal{L}}{\partial (\partial_\rho \Phi_A)} \partial_\rho \bar{\delta} \Phi_A  +   \frac{\partial \mathcal{L}}{\partial (\partial_\rho \Phi_A)} [\partial_\rho \partial_\beta \Phi_A]   \delta X^\beta  
+ \epsilon   \mathcal{L} \frac{\partial g^\alpha}{\partial X^\alpha} ] dX
\end{equation}

we can rewrite the sixth term using,

\begin{equation}
\delta X^\beta = \epsilon g^\beta (X) 
\end{equation}

thus,

\begin{equation}
\delta S = \int_{R_{X}} [  \frac{\partial \mathcal{L}}{\partial X^\beta} \delta X^\beta 
+ \frac{\partial \mathcal{L}}{\partial \Phi_A} \partial_\beta \Phi_A \delta X^\beta  +  \frac{\partial \mathcal{L}}{\partial \Phi_A} \bar{\delta} \Phi_A
+ \frac{\partial \mathcal{L}}{\partial (\partial_\rho \Phi_A)} \partial_\rho \bar{\delta} \Phi_A  +   \frac{\partial \mathcal{L}}{\partial (\partial_\rho \Phi_A)} [\partial_\rho \partial_\beta \Phi_A]   \delta X^\beta  
+    \mathcal{L} \partial_\beta \delta X^\beta] dX
\end{equation}

\subsubsection*{Combining terms in the integrand}

Rearranging the terms,

\begin{equation}
\delta S = \int_{R_{X}} [  \frac{\partial \mathcal{L}}{\partial X^\beta} \delta X^\beta 
+ \frac{\partial \mathcal{L}}{\partial \Phi_A} \partial_\beta \Phi_A \delta X^\beta
 +   \frac{\partial \mathcal{L}}{\partial (\partial_\rho \Phi_A)} [\partial_\rho \partial_\beta \Phi_A]   \delta X^\beta  
+    \mathcal{L} \partial_\beta \delta X^\beta
 +  \frac{\partial \mathcal{L}}{\partial \Phi_A} \bar{\delta} \Phi_A
+ \frac{\partial \mathcal{L}}{\partial (\partial_\rho \Phi_A)} \partial_\rho \bar{\delta} \Phi_A  
] dX
\end{equation}

we find that everything on the first line can be expressed as a result of the product rule,

\begin{equation}
\partial_\beta (\mathcal{L} \delta X^\beta) = (\partial_\beta \mathcal{L}) \delta X^\beta + \mathcal{L} \partial_\beta \delta X^\beta
\end{equation}

in order to differentiate the multivariable function $\mathcal{L}$ we need to apply the chain rule for a multivariable function,

\begin{equation}
\partial_\beta \mathcal{L} =  \frac{\partial \mathcal{L}}{\partial X^\beta} 
+ \frac{\partial \mathcal{L}}{\partial \Phi_A} \partial_\beta \Phi_A 
 +   \frac{\partial \mathcal{L}}{\partial (\partial_\rho \Phi_A)} [\partial_\beta \partial_\rho \Phi_A]  
\end{equation}

inserting back into the product rule,

\begin{equation}
\partial_\beta (\mathcal{L} \delta X^\beta) = \frac{\partial \mathcal{L}}{\partial X^\beta} \delta X^\beta 
+ \frac{\partial \mathcal{L}}{\partial \Phi_A} \partial_\beta \Phi_A \delta X^\beta
 +   \frac{\partial \mathcal{L}}{\partial (\partial_\rho \Phi_A)} [\partial_\beta \partial_\rho \Phi_A]   \delta X^\beta  + \mathcal{L} \partial_\beta \delta X^\beta
\end{equation}

but this is exactly what we have in the variation of the action! Therefore, after inserting this back into the action,

\begin{equation}
\delta S = \int_{R_{X}} [  \partial_\beta (\mathcal{L} \delta X^\beta)
 +  \frac{\partial \mathcal{L}}{\partial \Phi_A} \bar{\delta} \Phi_A
+ \frac{\partial \mathcal{L}}{\partial (\partial_\rho \Phi_A)} \partial_\rho \bar{\delta} \Phi_A  
] dX
\end{equation}

We want to separate the terms into an EOM and a conservation law by writing everything for the EOM in terms of $\bar{\delta} \Phi_A$ and everything else under a divergence. Using the product rule (IBP) we can also expand the third term,

\begin{equation}
 \partial_\rho  [\frac{\partial \mathcal{L}}{\partial (\partial_\rho \Phi_A)}\bar{\delta} \Phi_A]   = [\partial_\rho  \frac{\partial \mathcal{L}}{\partial (\partial_\rho \Phi_A)}] \bar{\delta} \Phi_A   + \frac{\partial \mathcal{L}}{\partial (\partial_\rho \Phi_A)} [\partial_\rho \bar{\delta} \Phi_A  ]
\end{equation}

thus we can reexpress the desired term in our integrand as,

\begin{equation}
 \frac{\partial \mathcal{L}}{\partial (\partial_\rho \Phi_A)} [\partial_\rho \bar{\delta} \Phi_A  ] =  \partial_\rho  [\frac{\partial \mathcal{L}}{\partial (\partial_\rho \Phi_A)}\bar{\delta} \Phi_A] -  [\partial_\rho  \frac{\partial \mathcal{L}}{\partial (\partial_\rho \Phi_A)}] \bar{\delta} \Phi_A
\end{equation}

inserting this back into the action,

\begin{equation}
\delta S = \int_{R_{X}} [  \partial_\beta (\mathcal{L} \delta X^\beta)
 +  \frac{\partial \mathcal{L}}{\partial \Phi_A} \bar{\delta} \Phi_A
+ \partial_\rho  (\frac{\partial \mathcal{L}}{\partial (\partial_\rho \Phi_A)}\bar{\delta} \Phi_A) 
-  (\partial_\rho  \frac{\partial \mathcal{L}}{\partial (\partial_\rho \Phi_A)}) \bar{\delta} \Phi_A 
] dX
\end{equation}

we can separate the EOM from the conservation law,

\begin{equation}
\delta S = \int_{R_{X}} [ 
   \frac{\partial \mathcal{L}}{\partial \Phi_A} \bar{\delta} \Phi_A
-  (\partial_\rho  \frac{\partial \mathcal{L}}{\partial (\partial_\rho \Phi_A)}) \bar{\delta} \Phi_A 
+ \partial_\rho  (\frac{\partial \mathcal{L}}{\partial (\partial_\rho \Phi_A)}\bar{\delta} \Phi_A) 
+ \partial_\beta (\mathcal{L} \delta X^\beta)
] dX
\end{equation}

\begin{equation}
\delta S = \int_{R_{X}} [ 
 \left(  \frac{\partial \mathcal{L}}{\partial \Phi_A}
-  \partial_\rho  \frac{\partial \mathcal{L}}{\partial (\partial_\rho \Phi_A)} \right) \bar{\delta} \Phi_A 
+ \partial_\rho  (\frac{\partial \mathcal{L}}{\partial (\partial_\rho \Phi_A)}\bar{\delta} \Phi_A) 
+ \partial_\rho (\mathcal{L} \delta X^\rho)
] dX
\end{equation}

Factoring out the $\partial_\rho$,

\begin{equation}
\delta S = \int_{R_{X}} [ 
 \left(  \frac{\partial \mathcal{L}}{\partial \Phi_A}
-  \partial_\rho  \frac{\partial \mathcal{L}}{\partial (\partial_\rho \Phi_A)} \right) \bar{\delta} \Phi_A 
+ \partial_\rho  \left( \frac{\partial \mathcal{L}}{\partial (\partial_\rho \Phi_A)}\bar{\delta} \Phi_A 
+ \mathcal{L} \delta X^\rho \right)
] dX
\end{equation}

We now have an expression for the EOM and conservation law. The required variations for the conservation law are,

\begin{equation}
 \bar{\delta} \Phi_A = \delta \Phi_A -  \partial_\beta \Phi_A \delta X^\beta
\end{equation}

and,

\begin{equation}
\delta X^\beta = X^{*\beta} - X^\beta
\end{equation}

where for coordinate and gauge transformations we express the field transformations as,

\begin{equation}
\bar{\delta} \Phi_A = -      \mathsterling_{\delta X} \Phi_A + \delta_g \Phi_A
\end{equation}

We note that the Lagrangian $\mathcal{L}$ here is the Lagrangian in the original coordinates, dependent on the original coordinates, and the original fields.

\subsection{Noether's Theorems}

We have the result that the action variation (from the condition $S[\Phi_A (X^\alpha)] = S[\Phi_A^* (X^{*\alpha})]$),

\begin{equation}
\delta S = \int_{R_{X}} [ 
 \left(  \frac{\partial \mathcal{L}}{\partial \Phi_A}
-  \partial_\rho  \frac{\partial \mathcal{L}}{\partial (\partial_\rho \Phi_A)} \right) \bar{\delta} \Phi_A 
+ \partial_\rho  \left( \frac{\partial \mathcal{L}}{\partial (\partial_\rho \Phi_A)}\bar{\delta} \Phi_A 
+ \mathcal{L} \delta X^\rho \right)
] dX
\end{equation}

for a symmetry transformation the action is invariant, $\delta S = 0$, for any region of integration, thus the integrand must vanish and one therefore has the condition,

\begin{equation}
     \left(  \frac{\partial \mathcal{L}}{\partial \Phi_A}
-  \partial_\rho  \frac{\partial \mathcal{L}}{\partial (\partial_\rho \Phi_A)} \right) \bar{\delta} \Phi_A 
+ \partial_\rho  \left( \frac{\partial \mathcal{L}}{\partial (\partial_\rho \Phi_A)}\bar{\delta} \Phi_A 
+ \mathcal{L} \delta X^\rho \right) = 0
\end{equation}

This is known as the Noether identity. The first term includes the Euler-Lagrange equation,

\begin{equation}
   E^A =  \frac{\partial \mathcal{L}}{\partial \Phi_A}
-  \partial_\rho  \frac{\partial \mathcal{L}}{\partial (\partial_\rho \Phi_A)}  
\end{equation}

The second term has a total divergence of a quantity known as the Noether current,

\begin{equation}
    J^\rho =  \frac{\partial \mathcal{L}}{\partial (\partial_\rho \Phi_A)}\bar{\delta} \Phi_A 
+ \mathcal{L} \delta X^\rho
\end{equation}

Therefore the Noether identity can be expressed compactly as $E^A \bar{\delta} \Phi_A + \partial_\rho J^\rho = 0$. It is from this Noether current that conserved objects can be derived.\\

The Noether identity presented above is the starting point for Noether's two theorems.

\subsubsection*{Noether's first theorem}

Noether's first theorem states that every continuous (variational) symmetry of the action with a finite number of parameters yields linearly independent conserved currents, one per parameter. For physical field theories, this means that there is a one to one correspondence between the number of action symmetries (typically the finite-dimensional continuous group of symmetries associated to the spacetime metric, such as Lorentz, Poincar\'e, conformal symmetry groups) and the number of conservation laws derived from these symmetries.\\

Pragmatically, this involves direct use of the Noether current,

\begin{equation}
    J^\rho =  \frac{\partial \mathcal{L}}{\partial (\partial_\rho \Phi_A)}\bar{\delta} \Phi_A 
+ \mathcal{L} \delta X^\rho
\end{equation}

where one must identify the coordinate symmetries $\delta X^\rho$ (these are the ones associated to the spacetime symmetries, often solved for by solving the Killing equation) and the field symmetries $\bar{\delta} \Phi_A $ (defining these is more nuanced and different methods exist, we will discuss physical applications later in this article). The $\delta X^\rho$ are therefore the Killing vectors $v^\rho$ of a later section, in other words the spacetime symmetry generators, that are solved for in a given metric spacetime (we will give examples later). In applications the infinitesimal parameter $\epsilon$ is absorbed into the transformation parameters (e.g. $a_\beta$, $\omega_{\beta\alpha}$, $S$, $k_\mu$), so that $\delta X^\beta$ and $\bar{\delta} \Phi_A$ are written without an explicit $\epsilon$. We note that internal symmetries, for which $\delta X^\rho = 0$ and $\bar{\delta} \Phi_A \neq 0$, are covered by the same identity; the Killing vector description applies to the spacetime symmetries.\\

More formally, for Noether's first theorem, if $S[\Phi_A (X^\alpha)]$ is invariant, and the equation of motion is satisfied $E^A = 0$, then,

\begin{equation}
 \partial_\rho  \left( \frac{\partial \mathcal{L}}{\partial (\partial_\rho \Phi_A)}\bar{\delta} \Phi_A 
+ \mathcal{L} \delta X^\rho \right) = 0
\end{equation}

then for our transforms of coordinates and fields that leave the action invariant, the above quantity will define a conservation law (i.e., a conserved object exists). The currents obtained this way are conserved on shell, and are unique up to identically conserved (improper) currents.\\

\subsubsection*{Noether's second theorem}

Noether's second theorem is about continuous (infinite) groups of transformations which leave the action invariant. In physics, the common example is the gauge transformations in e.g., electrodynamics. More formally, if $S[\Phi_A (X^\alpha)]$ is invariant under transformations depending on $r$ arbitrary functions, there are $r$ identities connected to the Euler-Lagrange equations. For physics this is a statement that each gauge symmetry implies differential identities of the equations of motion.\\

More formally, Noether's second theorem involves considering that the field transformation depends on $r$ arbitrary functions $ \phi_i$ and their derivatives,

\begin{equation}
   \bar{\delta} \Phi_A =  a_{Ai} \phi_i + b_{Ai}^\mu \partial_\mu \phi_i + \dots
\end{equation}

where $i = 1,2,\dots,r$, and $\phi_i$ is, for example, a gauge parameter. If the total divergence in the Noether identity $E^A \bar{\delta} \Phi_A + \partial_\rho J^\rho = 0$ vanishes (after integrating in the action and using appropriate boundary conditions), we are left with the condition for the remaining action integral,

\begin{equation}
    \int E^A \bar{\delta} \Phi_A dX = 0
\end{equation}

which for the above transformation gives,

\begin{equation}
\int E^A (a_{Ai} \phi_i + b_{Ai}^\mu \partial_\mu \phi_i + \dots) dX = 0
\end{equation}

in the action integral we can then integrate by parts the derivative of the parameter in each term, the resulting total derivatives vanishing at each order (the parameter $\phi_i$ is taken to vanish at the boundary so all such terms vanish), allowing the arbitrary function (i.e., gauge parameter) to be factored out, yielding,

\begin{equation}
\int (E^A a_{Ai}  - \partial_\mu (E^A  b_{Ai}^\mu)  + \dots) \phi_i dX = 0
\end{equation}

where we note the signs of each term alternate above. Therefore since $\phi_i$ is arbitrary, we are left with a set of identities of the Euler-Lagrange expressions $E^A$ of the form,

\begin{equation}
    E^A a_{Ai} - \partial_\mu (E^A  b_{Ai}^\mu)  + \dots = 0
\end{equation}

where the above identity is what Noether has \cite{Noether1918} in her Eq.16. Examples include $\partial_\mu \partial_\nu F^{\mu\nu} = 0$ in electrodynamics and $\nabla_\mu G^{\mu\nu} = 0$ in general relativity; both include first order derivatives of the gauge parameter in their respective $\bar{\delta} \Phi_A$, which is why we left off at first order in the above calculations, but in principle this can be done for any order (e.g., some higher spin models). These results can be derived as above from the relevant Noether identity used for the first theorem, where boundary terms vanish as above, and integrating by parts the Euler-Lagrange equation term yields divergences of the equation of motion. We will give an example later in the article. In summary, for physical theories typically considered: restricting $\delta X^\mu$ to e.g., the Killing vectors yields the first theorem's conserved currents, while leaving the gauge parameter arbitrary (electrodynamics) or $\delta X^\mu$ arbitrary (GR) yields the second theorem's off-shell identities.

\section{Noether Identity for Second Order Field Theories}

Typically, presentations of Noether's theorems for physics involve first order field theories and not second order field theories (although presentation of the expected results can be found in the literature). Most mathematics presentations also focus on first order actions for some scalar functions. Therefore it is much harder to find explicit calculation details for second order field theories following rigorous methods of the calculus of variations. For this reason we will in this section repeat almost everything from scratch, so that there is a self contained presentation of this calculation. Relative to the first order calculation, four things change: the Taylor expansion gains one term, the chain rule gains one term, one additional double integration by parts appears, and the Euler--Lagrange expression and current gain their $\partial \partial$ terms; every other step (apart from computing the required $\delta (\partial_\omega \partial_\rho \Phi_A)$ variation) is identical.

\subsection{Variation of a multiple integral functional that depends on varied coordinates and fields - higher order}

The basic idea of this section is to consider the variation of a functional that depends on the variation of coordinates i.e. $X^\alpha = \langle x^1, x^2, x^3, x^4 \rangle$, fields i.e. $\Phi_A$, derivatives of fields i.e. $\partial_\nu \Phi_A$, and second derivatives of fields $\partial_\lambda \partial_\nu \Phi_A$. The volume element can be expressed as $dX = dx^1 dx^2 dx^3 dx^4$. The bounds of integration are the entire region $R_X$ in terms of the $X$ coordinates. The action functional then reads,

\begin{equation}
S[\Phi_A (X^\alpha)] = \int \int \int \int_{R_X} \mathcal{L} (X^\alpha, \Phi_A, \partial_\nu \Phi_A, \partial_\lambda \partial_\nu \Phi_A) dx^1 dx^2 dx^3 dx^4
\end{equation}

This above action is for 4D theories (common in physics), but can be in more general form expressed for $n$ coordinates, $m$ fields, and the corresponding derivatives of these fields with respect to the coordinates. If we consider $X^\alpha$ to be $n$ dimensional, then we can write the action compactly using 1 integral that is assumed to be $n$ integrals $(\int_{R_X} = \int \dots \int_{R_X}), (dX = dx^1 dx^2 \dots dx^n)$, 

\begin{equation}
S[\Phi_A (X^\alpha)] = \int_{R_X} \mathcal{L} (X^\alpha, \Phi_A, \partial_\nu \Phi_A, \partial_\lambda \partial_\nu \Phi_A) dX
\end{equation}

Since partial derivatives commute, $\partial_\omega \partial_\rho \Phi_A = \partial_\rho \partial_\omega \Phi_A$, the derivative $\frac{\partial \mathcal{L}}{\partial (\partial_\omega \partial_\rho \Phi_A)}$ appearing below is only defined up to its part symmetric in $(\omega, \rho)$; we take this object to be symmetric in these indices, a convention used explicitly when the total divergence is factored at the end of this section.

\subsubsection*{Transformation of coordinates and fields}

Now suppose we change both our coordinates and our fields based on some transformation that depends on a parameter $\epsilon$. We can define new coordinates and fields using the $*$ symbol,

\begin{equation}
X^{*\alpha} = G^\alpha (X, \Phi, \partial \Phi, \partial \partial \Phi; \epsilon)
\end{equation}

\begin{equation}
\Phi_A^* = H_A (X, \Phi, \partial \Phi, \partial \partial \Phi; \epsilon)
\end{equation}

the functions $G^\alpha$ and $H_A$ depend on the parameter $\epsilon$ and are differentiable with respect to $\epsilon$. The function $G^\alpha$ has $n$ components just like $X^\alpha$ does, and the function $H_A$ has $m$ components just like $\Phi_A$ does. They are defined such that at $\epsilon = 0$ the original coordinates and field are recovered,

\begin{equation}
X^\alpha = G^\alpha (X, \Phi, \partial \Phi, \partial \partial \Phi; \epsilon = 0)
\end{equation}

\begin{equation}
\Phi_A = H_A (X, \Phi, \partial \Phi, \partial \partial \Phi; \epsilon = 0)
\end{equation}

By Taylor expanding the new coordinates and fields about $\epsilon = 0$, we can relate the new coordinates ($*$) to the old coordinates,

\begin{equation}
X^{*\alpha} = G^\alpha (X, \Phi, \partial \Phi, \partial \partial \Phi; \epsilon) = G^\alpha (X, \Phi, \partial \Phi, \partial \partial \Phi; \epsilon = 0) + \epsilon \frac{\partial G^\alpha (X, \Phi, \partial \Phi, \partial \partial \Phi; \epsilon)}{\partial \epsilon} |_{\epsilon = 0} + \mathcal{O}(\epsilon^2)
\end{equation}

\begin{equation}
\Phi_A^* = H_A (X, \Phi, \partial \Phi, \partial \partial \Phi; \epsilon) = H_A (X, \Phi, \partial \Phi, \partial \partial \Phi; \epsilon = 0) + \epsilon \frac{\partial H_A (X, \Phi, \partial \Phi, \partial \partial \Phi; \epsilon)}{\partial \epsilon} |_{\epsilon = 0} + \mathcal{O}(\epsilon^2)
\end{equation}

the $\mathcal{O}(\epsilon^2)$ term represents all terms which are nonlinear in $\epsilon$. Since we are only interested in the linear part for the variation, we compactly write them as such (as we will eventually drop them). The first term in each Taylor expansion is exactly the original coordinates $X^\alpha$ and fields $\Phi_A$! We can compactly write the second term using the functions $g$ and $h$ since these terms will be frequently used in the following calculation,

\begin{equation}
g^\alpha (X, \Phi, \partial \Phi, \partial \partial \Phi) = \frac{\partial G^\alpha (X, \Phi, \partial \Phi, \partial \partial \Phi; \epsilon)}{\partial \epsilon} |_{\epsilon = 0}
\end{equation}

\begin{equation}
h_A (X, \Phi, \partial \Phi, \partial \partial \Phi) = \frac{\partial H_A (X, \Phi, \partial \Phi, \partial \partial \Phi; \epsilon)}{\partial \epsilon} |_{\epsilon = 0} 
\end{equation}

These definitions are completely general; for tensor fields under pure coordinate transformations the induced form of $\bar{h}_A$ is derived in the first order section's subsection on variations of higher ranked $\Phi_A$. Therefore we can rewrite the Taylor expansions to relate the two coordinates and fields,

\begin{equation}
X^{*\alpha} = X^\alpha + \epsilon g^\alpha (X, \Phi, \partial \Phi, \partial \partial \Phi) + \mathcal{O}(\epsilon^2)
\end{equation}

\begin{equation}
\Phi_A^* = \Phi_A + \epsilon h_A (X, \Phi, \partial \Phi, \partial \partial \Phi) + \mathcal{O}(\epsilon^2)
\end{equation}

\subsubsection*{Transformation of the action into new coordinates and fields}

If we use the new coordinate transformations on the action, we find that the bounds of integration, dependent variables on the Lagrangian, the volume element, are all now expressed in terms of $*$. In other words, the action is expressed completely in terms of a new set of coordinates and fields,

\begin{equation}
S[\Phi_A^* (X^{*\alpha})] = \int_{R_{X^*}} \mathcal{L} (X^{*\alpha}, \Phi_A^*, \partial_\nu^* \Phi_A^*, \partial^*_\lambda \partial^*_\nu \Phi^*_A) dX^*
\end{equation}

even the divergence is expressed in terms of the new coordinates. This is the whole point of the desired variation. The variation with respect to changing coordinates and fields is defined as the linear part of the difference between the two functionals when the coordinates and fields are changed. While $\Delta S$ is the full difference, 

\begin{equation}
\Delta S = S[\Phi_A^* (X^{*\alpha})] - S[\Phi_A (X^\alpha)]
\end{equation}

after keeping only the linear part we will find $\delta S$ (the linear part of $\Delta S = S[\Phi_A^* (X^{*\alpha})] - S[\Phi_A (X^\alpha)]$); but first we need to calculate $\Delta S$ in order to drop the nonlinear part (nonlinear in $\epsilon$).\\

The difference in the two action functionals can be found by,

\begin{equation}
\Delta S = \int_{R_{X^*}} \mathcal{L} (X^{*\alpha}, \Phi_A^*, \partial_\nu^* \Phi_A^*, \partial^*_\lambda \partial^*_\nu \Phi^*_A) dX^* - \int_{R_X} \mathcal{L} (X^\alpha, \Phi_A, \partial_\nu \Phi_A, \partial_\lambda \partial_\nu \Phi_A) dX
\end{equation}

\subsubsection*{Using the Jacobian to combine the bounds and region of integration}

in order to convert the first term into the same region of integration and volume element, we need to perform a change of variables using the Jacobian,

\begin{equation}
dX^* = \frac{\partial (X^{*1}, X^{*2}, \dots, X^{*n})}{\partial (X^1, X^2, \dots, X^n)} dX
\end{equation}

thus we can combine the two integrals,

\begin{equation}
\Delta S = \int_{R_{X}} \mathcal{L} (X^{*\alpha}, \Phi_A^*, \partial_\nu^* \Phi_A^*, \partial^*_\lambda \partial^*_\nu \Phi^*_A)  \frac{\partial (X^{*1}, X^{*2}, \dots, X^{*n})}{\partial (X^1, X^2, \dots, X^n)} dX - \int_{R_X} \mathcal{L} (X^\alpha, \Phi_A, \partial_\nu \Phi_A, \partial_\lambda \partial_\nu \Phi_A) dX
\end{equation}

\begin{equation}
\Delta S = \int_{R_{X}} [ \mathcal{L} (X^{*\alpha}, \Phi_A^*, \partial_\nu^* \Phi_A^*, \partial^*_\lambda \partial^*_\nu \Phi^*_A)  \frac{\partial (X^{*1}, X^{*2}, \dots, X^{*n})}{\partial (X^1, X^2, \dots, X^n)}  -  \mathcal{L} (X^\alpha, \Phi_A, \partial_\nu \Phi_A, \partial_\lambda \partial_\nu \Phi_A) ] dX
\end{equation}

the Jacobian is the determinant of the following matrix,

\begin{equation}
 \frac{\partial (X^{*1}, X^{*2}, \dots, X^{*n})}{\partial (X^1, X^2, \dots, X^n)} =
\begin{bmatrix}
\frac{\partial X^{*1}}{\partial X^1} & \frac{\partial X^{*2}}{\partial X^1} & \dots & \frac{\partial X^{*n}}{\partial X^1}\\
\frac{\partial X^{*1}}{\partial X^2} & \frac{\partial X^{*2}}{\partial X^2} & \dots & \frac{\partial X^{*n}}{\partial X^2}\\
\vdots & \vdots & \ddots & \vdots\\
\frac{\partial X^{*1}}{\partial X^n} & \frac{\partial X^{*2}}{\partial X^n} & \dots & \frac{\partial X^{*n}}{\partial X^n}
\end{bmatrix}_{n\times n}
\end{equation}

where we write this matrix with rows indexed by the original coordinates, the transpose of the convention in some references; the determinant (our desired calculation) is unaffected by this choice. To determine each of these components we need to recall the relationship between the coordinates,

\begin{equation}
X^{*\alpha} = X^\alpha + \epsilon g^\alpha (X, \Phi, \partial \Phi, \partial \partial \Phi) + \mathcal{O}(\epsilon^2)
\end{equation}

we can write each element in the Jacobian in general as $\frac{\partial X^{*\alpha}}{\partial X^\beta}$, which we can find by differentiating the above expression,

\begin{equation}
\frac{\partial X^{*\alpha}}{\partial X^\beta} = \frac{\partial X^\alpha}{\partial X^\beta} + \epsilon \frac{\partial g^\alpha (X, \Phi, \partial \Phi, \partial \partial \Phi)}{\partial X^\beta} + \frac{\partial \mathcal{O}(\epsilon^2)}{\partial X^\beta}
\end{equation}

since after taking the determinant the $\mathcal{O}(\epsilon^2)$ terms will be even higher order, and we only want terms linear in $\epsilon$ for the variation, we will drop them here.  We can abbreviate the second term as just $g^\alpha$. Since $\Phi$, $\partial \Phi$ and $\partial \partial \Phi$ themselves depend on $X$ ($\Phi(X)$, $\partial \Phi(X)$ and $\partial \partial \Phi(X)$), the function $g^\alpha (X, \Phi, \partial \Phi, \partial \partial \Phi)$ is in the end a function of $X$ alone, which we write as $g^\alpha (X)$. Derivatives like $\frac{\partial g^\alpha}{\partial X^\beta}$ here and in the rest of the calculation are taken with this in mind (through the $X$ dependence of $\Phi$, $\partial \Phi$ and $\partial \partial \Phi$ as well). The same goes for $h_A$ and $\bar{h}_A$.  Finally the first term is just the Kronecker delta,

\begin{equation}
\frac{\partial X^{*\alpha}}{\partial X^\beta} \approx \delta^\alpha_\beta + \epsilon \frac{\partial g^\alpha}{\partial X^\beta}
\end{equation}

inserting this into the Jacobian we have,

\begin{equation}
 \frac{\partial (X^{*1}, X^{*2}, \dots, X^{*n})}{\partial (X^1, X^2, \dots, X^n)} =
\begin{bmatrix}
1 + \epsilon \frac{\partial g^1}{\partial X^1} & \epsilon \frac{\partial g^2}{\partial X^1} & \dots & \epsilon \frac{\partial g^n}{\partial X^1}\\
\epsilon \frac{\partial g^1}{\partial X^2} & 1 + \epsilon \frac{\partial g^2}{\partial X^2} & \dots & \epsilon \frac{\partial g^n}{\partial X^2}\\
\vdots & \vdots & \ddots & \vdots\\
\epsilon \frac{\partial g^1}{\partial X^n} & \epsilon \frac{\partial g^2}{\partial X^n} & \dots & 1 + \epsilon \frac{\partial g^n}{\partial X^n}
\end{bmatrix}_{n\times n}
\end{equation}

to find this determinant we have to perform a cofactor expansion and keep only terms which are at most linear in epsilon. Based on the cofactor expansion all terms not on the diagonal will have submatrices at least $\epsilon^2$, thus they can all be neglected. What is left is the multiplication of all elements on the diagonal,

\begin{equation}
 \frac{\partial (X^{*1}, X^{*2}, \dots, X^{*n})}{\partial (X^1, X^2, \dots, X^n)} \approx (1 + \epsilon \frac{\partial g^1}{\partial X^1})(1 + \epsilon \frac{\partial g^2}{\partial X^2}) \dots (1 + \epsilon \frac{\partial g^n}{\partial X^n})
\end{equation}

this expands to, in order linear in epsilon,

\begin{equation}
 \frac{\partial (X^{*1}, X^{*2}, \dots, X^{*n})}{\partial (X^1, X^2, \dots, X^n)} \approx 1 + \epsilon \frac{\partial g^1}{\partial X^1} +  \epsilon \frac{\partial g^2}{\partial X^2} + \dots +  \epsilon \frac{\partial g^n}{\partial X^n}
\end{equation}

which can be expressed compactly in Einstein summation notation as,

\begin{equation}
 \frac{\partial (X^{*1}, X^{*2}, \dots, X^{*n})}{\partial (X^1, X^2, \dots, X^n)} \approx 1 + \epsilon \frac{\partial g^\alpha}{\partial X^\alpha} 
\end{equation}

returning to our difference in actions,

\begin{equation}
\Delta S = \int_{R_{X}} [ \mathcal{L} (X^{*\alpha}, \Phi_A^*, \partial_\nu^* \Phi_A^*, \partial^*_\lambda \partial^*_\nu \Phi^*_A)  \frac{\partial (X^{*1}, X^{*2}, \dots, X^{*n})}{\partial (X^1, X^2, \dots, X^n)}  -  \mathcal{L} (X^\alpha, \Phi_A, \partial_\nu \Phi_A, \partial_\lambda \partial_\nu \Phi_A) ] dX
\end{equation}

we can insert the evaluated Jacobian,

\begin{equation}
\Delta S = \int_{R_{X}} [ \mathcal{L} (X^{*\alpha}, \Phi_A^*, \partial_\nu^* \Phi_A^*, \partial^*_\lambda \partial^*_\nu \Phi^*_A) (1 + \epsilon \frac{\partial g^\alpha}{\partial X^\alpha} ) -  \mathcal{L} (X^\alpha, \Phi_A, \partial_\nu \Phi_A, \partial_\lambda \partial_\nu \Phi_A) ] dX
\end{equation}

\subsubsection*{Taylor expanding the Lagrangian from the new coordinates and fields}

Expanding out our integrand,

\begin{equation}
\Delta S = \int_{R_{X}} [ \mathcal{L} (X^{*\alpha}, \Phi_A^*, \partial_\nu^* \Phi_A^*, \partial^*_\lambda \partial^*_\nu \Phi^*_A)  + \epsilon   \mathcal{L} (X^{*\alpha}, \Phi_A^*, \partial_\nu^* \Phi_A^*, \partial^*_\lambda \partial^*_\nu \Phi^*_A) \frac{\partial g^\alpha}{\partial X^\alpha}  -  \mathcal{L} (X^\alpha, \Phi_A, \partial_\nu \Phi_A, \partial_\lambda \partial_\nu \Phi_A) ] dX
\end{equation}

we can Taylor expand the Lagrangian terms $\mathcal{L} (X^{*\alpha}, \Phi_A^*, \partial_\nu^* \Phi_A^*, \partial^*_\lambda \partial^*_\nu \Phi^*_A)$ which are expressed in terms of the new coordinates and fields, and keep only order 1 relative to $\epsilon$. Taylor expanding this multivariable function,

\begin{multline}
\mathcal{L} (X^{*\alpha}, \Phi_A^*, \partial_\nu^* \Phi_A^*, \partial^*_\lambda \partial^*_\nu \Phi^*_A) = \mathcal{L} (X^\alpha, \Phi_A, \partial_\nu \Phi_A, \partial_\lambda \partial_\nu \Phi_A) 
+ \frac{\partial \mathcal{L}(X^\alpha, \Phi_A, \partial_\nu \Phi_A, \partial_\lambda \partial_\nu \Phi_A)}{\partial X^\beta} (X^{*\beta} - X^\beta) \\
+ \frac{\partial \mathcal{L}(X^\alpha, \Phi_A, \partial_\nu \Phi_A, \partial_\lambda \partial_\nu \Phi_A)}{\partial \Phi_A} (\Phi_A^* - \Phi_A) 
+ \frac{\partial \mathcal{L}(X^\alpha, \Phi_A, \partial_\nu \Phi_A, \partial_\lambda \partial_\nu \Phi_A)}{\partial (\partial_\rho \Phi_A)} (\partial_\rho^* \Phi_A^* - \partial_\rho \Phi_A) 
\\
+  \frac{\partial \mathcal{L}(X^\alpha, \Phi_A, \partial_\nu \Phi_A, \partial_\lambda \partial_\nu \Phi_A)}{\partial (\partial_\omega \partial_\rho \Phi_A)} (\partial_\omega^* \partial_\rho^* \Phi_A^* - \partial_\omega \partial_\rho \Phi_A) 
+ \dots
\end{multline}

the second part of the final four terms are just what we define the variations of these input variables to be (the linear change between the new and old systems),

\begin{equation}
\delta X^\beta = X^{*\beta} - X^\beta
\end{equation}

\begin{equation}
\delta \Phi_A = \Phi_A^*(X^{*\alpha}) - \Phi_A(X^\alpha)
\end{equation}

\begin{equation}
\delta (\partial_\rho \Phi_A) = \partial_\rho^* \Phi_A^*(X^{*\alpha}) - \partial_\rho \Phi_A(X^\alpha)
\end{equation}

\begin{equation}
\delta (\partial_\omega \partial_\rho \Phi_A) = \partial_\omega^* \partial_\rho^* \Phi_A^*(X^{*\alpha}) - \partial_\omega \partial_\rho \Phi_A(X^\alpha)
\end{equation}

where the fields are functions of their corresponding coordinates, thus the Taylor expansion reads,

\begin{multline}
\mathcal{L} (X^{*\alpha}, \Phi_A^*, \partial_\nu^* \Phi_A^*, \partial^*_\lambda \partial^*_\nu \Phi^*_A) = \mathcal{L} (X^\alpha, \Phi_A, \partial_\nu \Phi_A, \partial_\lambda \partial_\nu \Phi_A) 
+ \frac{\partial \mathcal{L}(X^\alpha, \Phi_A, \partial_\nu \Phi_A, \partial_\lambda \partial_\nu \Phi_A)}{\partial X^\beta} 
\delta X^\beta \\
+ \frac{\partial \mathcal{L}(X^\alpha, \Phi_A, \partial_\nu \Phi_A, \partial_\lambda \partial_\nu \Phi_A)}{\partial \Phi_A} \delta \Phi_A
+ \frac{\partial \mathcal{L}(X^\alpha, \Phi_A, \partial_\nu \Phi_A, \partial_\lambda \partial_\nu \Phi_A)}{\partial (\partial_\rho \Phi_A)} \delta (\partial_\rho \Phi_A)
+  \frac{\partial \mathcal{L}(X^\alpha, \Phi_A, \partial_\nu \Phi_A, \partial_\lambda \partial_\nu \Phi_A)}{\partial (\partial_\omega \partial_\rho \Phi_A)} \delta (\partial_\omega \partial_\rho \Phi_A)
+ \dots
\end{multline}

each of these variations is proportional to first order epsilon so these are the only terms we will maintain, for example recalling the relationship between the coordinates,

\begin{equation}
X^{*\alpha} = X^\alpha + \epsilon g^\alpha (X, \Phi, \partial \Phi, \partial \partial \Phi) + \mathcal{O}(\epsilon^2)
\end{equation}

the variation is at first order $\epsilon$,

\begin{equation}
\delta X^\alpha = X^{*\alpha} - X^\alpha = \epsilon g^\alpha (X, \Phi, \partial \Phi, \partial \partial \Phi) 
\end{equation}

Recall the difference in actions, 

\begin{equation}
\Delta S = \int_{R_{X}} [ \mathcal{L} (X^{*\alpha}, \Phi_A^*, \partial_\nu^* \Phi_A^*, \partial^*_\lambda \partial^*_\nu \Phi^*_A)  + \epsilon   \mathcal{L} (X^{*\alpha}, \Phi_A^*, \partial_\nu^* \Phi_A^*, \partial^*_\lambda \partial^*_\nu \Phi^*_A) \frac{\partial g^\alpha}{\partial X^\alpha}  -  \mathcal{L} (X^\alpha, \Phi_A, \partial_\nu \Phi_A, \partial_\lambda \partial_\nu \Phi_A) ] dX
\end{equation}

\subsubsection*{The desired variation $\delta S$}

The first term above will have 5 terms up to order epsilon from the Taylor expansion, the second term will have one term order epsilon. Since we drop all nonlinear in epsilon terms at this point, we have recovered $\delta S$ from $\Delta S$, since the variation $\delta S$ is defined as the linear part of the difference in the two actions. We are left with,

\begin{multline}
\delta S =  \int_{R_{X}} [\mathcal{L} (X^\alpha, \Phi_A, \partial_\nu \Phi_A, \partial_\lambda \partial_\nu \Phi_A) 
+ \frac{\partial \mathcal{L}(X^\alpha, \Phi_A, \partial_\nu \Phi_A, \partial_\lambda \partial_\nu \Phi_A)}{\partial X^\beta} 
\delta X^\beta 
\\
+ \frac{\partial \mathcal{L}(X^\alpha, \Phi_A, \partial_\nu \Phi_A, \partial_\lambda \partial_\nu \Phi_A)}{\partial \Phi_A} \delta \Phi_A
+ \frac{\partial \mathcal{L}(X^\alpha, \Phi_A, \partial_\nu \Phi_A, \partial_\lambda \partial_\nu \Phi_A)}{\partial (\partial_\rho \Phi_A)} \delta (\partial_\rho \Phi_A)
\\
+  \frac{\partial \mathcal{L}(X^\alpha, \Phi_A, \partial_\nu \Phi_A, \partial_\lambda \partial_\nu \Phi_A)}{\partial (\partial_\omega \partial_\rho \Phi_A)} \delta (\partial_\omega \partial_\rho \Phi_A)  
+ \epsilon    \mathcal{L} (X^\alpha, \Phi_A, \partial_\nu \Phi_A, \partial_\lambda \partial_\nu \Phi_A) \frac{\partial g^\alpha}{\partial X^\alpha} 
 -  \mathcal{L} (X^\alpha, \Phi_A, \partial_\nu \Phi_A, \partial_\lambda \partial_\nu \Phi_A) ] dX
\end{multline}

the first and last terms exactly cancel, thus,

\begin{multline}
\delta S =  \int_{R_{X}} [
 \frac{\partial \mathcal{L}(X^\alpha, \Phi_A, \partial_\nu \Phi_A, \partial_\lambda \partial_\nu \Phi_A)}{\partial X^\beta} 
\delta X^\beta 
+ \frac{\partial \mathcal{L}(X^\alpha, \Phi_A, \partial_\nu \Phi_A, \partial_\lambda \partial_\nu \Phi_A)}{\partial \Phi_A} \delta \Phi_A
+ \frac{\partial \mathcal{L}(X^\alpha, \Phi_A, \partial_\nu \Phi_A, \partial_\lambda \partial_\nu \Phi_A)}{\partial (\partial_\rho \Phi_A)} \delta (\partial_\rho \Phi_A)
\\
+  \frac{\partial \mathcal{L}(X^\alpha, \Phi_A, \partial_\nu \Phi_A, \partial_\lambda \partial_\nu \Phi_A)}{\partial (\partial_\omega \partial_\rho \Phi_A)} \delta (\partial_\omega \partial_\rho \Phi_A)  
+ \epsilon    \mathcal{L} (X^\alpha, \Phi_A, \partial_\nu \Phi_A, \partial_\lambda \partial_\nu \Phi_A) \frac{\partial g^\alpha}{\partial X^\alpha}  ] dX
\end{multline}

now that we have expressed every term in terms of the Lagrangian that depends on the original coordinates and fields, we can drop the notation, and from here we know that $\mathcal{L} =  \mathcal{L} (X^\alpha, \Phi_A, \partial_\nu \Phi_A, \partial_\lambda \partial_\nu \Phi_A)$,

\begin{equation}
\delta S =  \int_{R_{X}} [
 \frac{\partial \mathcal{L}}{\partial X^\beta} \delta X^\beta 
+ \frac{\partial \mathcal{L}}{\partial \Phi_A} \delta \Phi_A
+ \frac{\partial \mathcal{L}}{\partial (\partial_\rho \Phi_A)} \delta (\partial_\rho \Phi_A)
+  \frac{\partial \mathcal{L}}{\partial (\partial_\omega \partial_\rho \Phi_A)} \delta (\partial_\omega \partial_\rho \Phi_A)  
+ \epsilon    \mathcal{L}  \frac{\partial g^\alpha}{\partial X^\alpha}  ] dX
\end{equation}

In order to continue we require explicit expressions for the variations $\delta X^\beta $, $ \delta \Phi_A$, $\delta (\partial_\rho \Phi_A) $ and $\delta (\partial_\omega \partial_\rho \Phi_A) $, which will be the topic of the next sections.

\subsection{Determining expressions for the variations $\delta X^\beta $ and  $ \delta \Phi_A$ }

\subsubsection*{Calculating the variation $\delta X^\beta$}

We define the variation $\delta X^\beta$ as,

\begin{equation}
\delta X^\beta = X^{*\beta} - X^\beta
\end{equation}

based on the relationship between the two coordinates as previously defined,

\begin{equation}
X^{*\beta} = X^\beta + \epsilon g^\beta (X, \Phi, \partial \Phi, \partial \partial \Phi) + \mathcal{O}(\epsilon^2)
\end{equation}

The difference between these two coordinates is,

\begin{equation}
\Delta X^\beta = X^{*\beta} - X^\beta = \epsilon g^\beta (X, \Phi, \partial \Phi, \partial \partial \Phi) + \mathcal{O}(\epsilon^2)
\end{equation}

which is defined as a variation at first order,

\begin{equation}
\delta X^\beta = X^{*\beta} - X^\beta = \epsilon g^\beta (X, \Phi, \partial \Phi, \partial \partial \Phi) 
\end{equation}

since $\Phi$, $\partial \Phi$ and $\partial \partial \Phi$ depend on $X$, i.e. $\Phi(X)$, $\partial \Phi (X)$ and $\partial \partial \Phi (X)$, we can abbreviate $g^\beta (X, \Phi, \partial \Phi, \partial \partial \Phi) = g^\beta (X)$, thus our variation reads,

\begin{equation}
\delta X^\beta = \epsilon g^\beta (X) 
\end{equation}

which gives us an expression for the linear part of the difference between the two coordinates.

\subsubsection*{Calculating the variation $ \delta \Phi_A$}

We define the variation $ \delta \Phi_A$ as,

\begin{equation}
\delta \Phi_A = \Phi_A^*(X^{*\alpha}) - \Phi_A(X^\alpha)
\end{equation}

where the variation is the linear part of the difference between the two fields as a function of their respective coordinates. Based on the relationship between the two fields as previously defined,

\begin{equation}
\Phi_A^*(X^{*\alpha}) = \Phi_A(X^\alpha) + \epsilon h_A (X, \Phi, \partial \Phi, \partial \partial \Phi) + \mathcal{O}(\epsilon^2)
\end{equation}

The difference between these two fields is,

\begin{equation}
\Delta \Phi_A = \Phi_A^*(X^{*\alpha}) - \Phi_A(X^\alpha) =  \epsilon h_A (X, \Phi, \partial \Phi, \partial \partial \Phi) + \mathcal{O}(\epsilon^2)
\end{equation}

which is defined as a variation at first order,

\begin{equation}
\delta \Phi_A = \Phi_A^*(X^{*\alpha}) - \Phi_A(X^\alpha) =  \epsilon h_A (X, \Phi, \partial \Phi, \partial \partial \Phi) 
\end{equation}

since $\Phi$, $\partial \Phi$ and $\partial \partial \Phi$ depend on $X$, i.e. $\Phi(X)$, $\partial \Phi (X)$ and $\partial \partial \Phi (X)$, we can abbreviate $h_A (X, \Phi, \partial \Phi, \partial \partial \Phi) = h_A (X)$, thus our variation reads,

\begin{equation}
\delta \Phi_A = \epsilon h_A (X) 
\end{equation}

which gives us an expression for the linear part of the difference between the two fields. 

\subsubsection*{Calculating the variation $\bar{\delta} \Phi_A$}

In order to complete the calculation we must also consider the variation $\bar{\delta} \Phi_A$, which we will introduce the bar to define it as the change in the two fields when both are defined in terms of the same $X^\alpha$ coordinate,

\begin{equation}
\bar{\delta} \Phi_A = \Phi_A^*(X^\alpha) - \Phi_A(X^\alpha)
\end{equation}

in this case we have a new relationship between our fields,

\begin{equation}
\Phi_A^*(X^\alpha) = \Phi_A(X^\alpha) + \epsilon \bar{h}_A (X, \Phi, \partial \Phi, \partial \partial \Phi) + \mathcal{O}(\epsilon^2)
\end{equation}

where $\bar{h}_A$ is associated with the difference between the two fields when we vary our fields but maintain the same coordinates for each. The difference between the two is,

\begin{equation}
\bar{\Delta} \Phi_A = \Phi_A^*(X^\alpha) - \Phi_A(X^\alpha) = \epsilon \bar{h}_A (X, \Phi, \partial \Phi, \partial \partial \Phi) + \mathcal{O}(\epsilon^2)
\end{equation}

and the variation is defined at order epsilon, abbreviating to,

\begin{equation}
\bar{\delta} \Phi_A = \epsilon \bar{h}_A (X) 
\end{equation}

which gives us a relationship for the change in fields when the input coordinates are the same.

\subsubsection*{Calculating the relationship between the field variations $ \delta \Phi_A$ and $\bar{\delta} \Phi_A$}

We can determine a relationship between these two variations by returning to the difference,

\begin{equation}
\Delta \Phi_A = \Phi_A^*(X^{*\alpha}) - \Phi_A(X^\alpha) 
\end{equation}

if we add and subtract $\Phi_A^*(X^\alpha)$ in this expression we get,

\begin{equation}
\Delta \Phi_A = [ \Phi_A^*(X^{*\alpha}) - \Phi_A^*(X^\alpha) ] + [ \Phi_A^*(X^\alpha) - \Phi_A(X^\alpha) ]
\end{equation}

immediately we can recognize the second term from our bar variation calculation,

\begin{equation}
\Delta \Phi_A = [ \Phi_A^*(X^{*\alpha}) - \Phi_A^*(X^\alpha) ] + \bar{\Delta} \Phi_A
\end{equation}

if we Taylor expand the first part of the first term about $X^\alpha$,

\begin{equation}
\Phi_A^*(X^{*\alpha}) = \Phi_A^*(X^\alpha)  + \frac{\partial \Phi_A^*(X^\alpha)}{\partial X^\beta} (X^{*\beta} - X^\beta) + \mathcal{O}(\epsilon^2)
\end{equation}

recall that $\delta X^\beta = X^{*\beta} - X^\beta$, which is first order in $\epsilon$,

\begin{equation}
\Phi_A^*(X^{*\alpha}) = \Phi_A^*(X^\alpha)  + \frac{\partial \Phi_A^*(X^\alpha)}{\partial X^\beta} \delta X^\beta + \mathcal{O}(\epsilon^2)
\end{equation}

inserting this Taylor expansion into the change in fields relationship,

\begin{equation}
\Delta \Phi_A = [\Phi_A^*(X^\alpha)  + \frac{\partial \Phi_A^*(X^\alpha)}{\partial X^\beta} \delta X^\beta + \mathcal{O}(\epsilon^2) - \Phi_A^*(X^\alpha) ] + \bar{\Delta} \Phi_A
\end{equation}

the first and fourth terms in the square brackets cancel,

\begin{equation}
\Delta \Phi_A = [ \frac{\partial \Phi_A^*(X^\alpha)}{\partial X^\beta} \delta X^\beta + \mathcal{O}(\epsilon^2)  ] + \bar{\Delta} \Phi_A
\end{equation}

we can replace the derivative with one in the original coordinates by recalling,

\begin{equation}
\Phi_A^*(X^\alpha) = \Phi_A(X^\alpha) + \epsilon \bar{h}_A (X, \Phi, \partial \Phi, \partial \partial \Phi) + \mathcal{O}(\epsilon^2)
\end{equation}

differentiating both sides,

\begin{equation}
\frac{\partial \Phi_A^*(X^\alpha)}{\partial X^\beta} = \frac{\partial \Phi_A(X^\alpha)}{\partial X^\beta} + \epsilon \frac{\partial }{\partial X^\beta} \bar{h}_A (X, \Phi, \partial \Phi, \partial \partial \Phi) + \frac{\partial }{\partial X^\beta} \mathcal{O}(\epsilon^2)
\end{equation}

inserting this back into the change in fields relationship,

\begin{equation}
\Delta \Phi_A = [ (\frac{\partial \Phi_A(X^\alpha)}{\partial X^\beta} + \epsilon \frac{\partial }{\partial X^\beta} \bar{h}_A (X, \Phi, \partial \Phi, \partial \partial \Phi) + \frac{\partial }{\partial X^\beta} \mathcal{O}(\epsilon^2)) \delta X^\beta + \mathcal{O}(\epsilon^2)  ] + \bar{\Delta} \Phi_A
\end{equation}

for variations $ \delta \Phi_A$ and $\bar{\delta} \Phi_A$ we require only the linear portion of this expression with respect to epsilon, therefore the equation reduces to (after neglecting second, third and fourth terms which are $\epsilon^2$ or greater),

\begin{equation}
\delta \Phi_A = \frac{\partial \Phi_A(X^\alpha)}{\partial X^\beta} \delta X^\beta  + \bar{\delta} \Phi_A
\end{equation}

this gives us a relationship between the two different variations of fields which we will commonly abbreviate as,

\begin{equation}
\delta \Phi_A = \frac{\partial \Phi_A}{\partial X^\beta} \delta X^\beta  + \bar{\delta} \Phi_A
\end{equation}

which is similar to the variation taken for Poincare translation in field theory. Typically we will write this more compactly as,

\begin{equation}
\delta \Phi_A = \partial_\beta \Phi_A \delta X^\beta  + \bar{\delta} \Phi_A
\end{equation}

\subsubsection*{Shorter way of calculating the relationship between $ \delta \Phi_A$ and $\bar{\delta} \Phi_A$}

It turns out that the lengthy manipulations used to calculate,

\begin{equation}
\delta \Phi_A = \frac{\partial \Phi_A}{\partial X^\beta} \delta X^\beta  + \bar{\delta} \Phi_A
\end{equation}

in the previous section were not needed. We can calculate this directly,

\begin{equation}
\Delta \Phi_A = \Phi_A^*(X^{*\alpha}) - \Phi_A(X^\alpha) 
\end{equation}

instead of adding and subtracting some term here we can Taylor expand the first term about $X$,

\begin{equation}
\Phi_A^*(X^{*\alpha}) = \Phi_A^*(X^\alpha)  + \frac{\partial \Phi_A^*(X^\alpha)}{\partial X^\beta} \delta X^\beta + \mathcal{O}(\epsilon^2)
\end{equation}

replacing this term we have,

\begin{equation}
\Delta \Phi_A = \Phi_A^*(X^\alpha)  + \frac{\partial \Phi_A^*(X^\alpha)}{\partial X^\beta} \delta X^\beta + \mathcal{O}(\epsilon^2) - \Phi_A(X^\alpha) 
\end{equation}

the first and last terms are exactly,

\begin{equation}
 \bar{\Delta} \Phi_A =  \Phi_A^*(X^\alpha) - \Phi_A(X^\alpha) 
\end{equation}

thus we are left with, 

\begin{equation}
\Delta \Phi_A =  \bar{\Delta} \Phi_A  + \frac{\partial \Phi_A^*(X^\alpha)}{\partial X^\beta} \delta X^\beta + \mathcal{O}(\epsilon^2)
\end{equation}

keeping first order,

\begin{equation}
\delta \Phi_A =  \bar{\delta} \Phi_A  + \frac{\partial \Phi_A}{\partial X^\beta} \delta X^\beta 
\end{equation}

which is exactly what we found via the manipulation method.

\subsubsection*{A relationship between the two derivatives}

Suppose we want to determine the relationship between the derivatives,

\begin{equation}
 \frac{\partial }{\partial X^{*\rho}} \quad , \quad
 \frac{\partial }{\partial X^{\rho}}  
\end{equation}

we can start by noting the relationship between the two coordinates,

\begin{equation}
X^{*\alpha} = X^\alpha + \epsilon g^\alpha (X, \Phi, \partial \Phi, \partial \partial \Phi) + \mathcal{O}(\epsilon^2)
\end{equation}

differentiating both sides of this expression and keeping only terms linear in $\epsilon$, and abbreviating $g$,

\begin{equation}
\frac{\partial X^{*\alpha}}{\partial X^\beta} \approx \frac{\partial X^\alpha}{\partial X^\beta} + \epsilon \frac{\partial g^\alpha (X)}{\partial X^\beta}
\end{equation}

where the first term is just the Kronecker delta,

\begin{equation}
\frac{\partial X^{*\alpha}}{\partial X^\beta} \approx \delta^\alpha_\beta + \epsilon \frac{\partial g^\alpha (X)}{\partial X^\beta}
\end{equation}

to change variable between the two derivatives we have,

\begin{equation}
\frac{\partial}{\partial X^\beta} =  \frac{\partial X^{*\alpha}}{\partial X^\beta} \frac{\partial}{\partial X^{*\alpha}}
\end{equation}

which we can introduce our expression for $\frac{\partial X^{*\alpha}}{\partial X^\beta} $ on the right hand side,

\begin{equation}
\frac{\partial}{\partial X^\beta} = (\delta^\alpha_\beta + \epsilon \frac{\partial g^\alpha (X)}{\partial X^\beta}) \frac{\partial}{\partial X^{*\alpha}}
\end{equation}

our derivative operator can contract with the delta, yielding,

\begin{equation}
\frac{\partial}{\partial X^\beta} =  \frac{\partial}{\partial X^{*\beta}} + \epsilon \frac{\partial g^\alpha (X)}{\partial X^\beta} \frac{\partial}{\partial X^{*\alpha}}
\end{equation}

therefore the difference in the two derivatives is,

\begin{equation}
\frac{\partial}{\partial X^\beta} -  \frac{\partial}{\partial X^{*\beta}} = \epsilon \frac{\partial g^\alpha (X)}{\partial X^\beta} \frac{\partial}{\partial X^{*\alpha}}
\end{equation}

Before turning to the variation of the derivatives of the fields, we will first specify the field variations under coordinate transformations at each rank.

\subsection{Variations of higher ranked $\Phi_A$ under coordinate transformations}

For the first order field theories, we showed how (in the subsection of the same name as this) that under coordinate transformations the substantial variation of the fields can be expressed in terms of the Lie derivative as,

\begin{equation}
\bar{\delta} \Phi_A = -      \mathsterling_{\delta X} \Phi_A 
\end{equation}

more generally for other (e.g., gauge $\delta_g \Phi_A$) transformations this can be defined as,

\begin{equation}
\bar{\delta} \Phi_A = -      \mathsterling_{\delta X} \Phi_A + \delta_g \Phi_A
\end{equation}

however, this result is entirely independent of the order of derivatives in the Lagrangian, so we refer the reader to that earlier section for these calculations.

\subsection{Determining expressions for the variation $\delta (\partial_\rho \Phi_A) $}

\subsubsection*{Introducing the variation $\delta (\partial_\rho \Phi_A) $}

We define the difference $\Delta (\partial_\rho \Phi_A) $ as,

\begin{equation}
\Delta (\partial_\rho \Phi_A)  = \frac{\partial \Phi_A^*(X^{*\alpha})}{\partial X^{*\rho}} - \frac{\partial \Phi_A(X^\alpha)}{\partial X^\rho}
\end{equation}

where the variation is the linear part of the difference between the two derivatives of fields as a function of their respective coordinates (the linear part of the above expression). In order to evaluate this variation we need to add and subtract the following terms just like we did in the previous subsection,

\begin{equation}
\pm \frac{\partial \Phi_A(X^{*\alpha})}{\partial X^{*\rho}} \quad , \quad \pm \frac{\partial \Phi_A(X^{*\alpha})}{\partial X^{\rho}}
\end{equation}

introducing all of these terms,

\begin{equation}
\Delta (\partial_\rho \Phi_A)  = \frac{\partial \Phi_A^*(X^{*\alpha})}{\partial X^{*\rho}} - \frac{\partial \Phi_A(X^\alpha)}{\partial X^\rho} + \frac{\partial \Phi_A(X^{*\alpha})}{\partial X^{*\rho}} - \frac{\partial \Phi_A(X^{*\alpha})}{\partial X^{*\rho}} + \frac{\partial \Phi_A(X^{*\alpha})}{\partial X^{\rho}} - \frac{\partial \Phi_A(X^{*\alpha})}{\partial X^{\rho}}
\end{equation}

and rearranging/ combining,

\begin{equation}
\Delta (\partial_\rho \Phi_A)  = \frac{\partial \Phi_A^*(X^{*\alpha})}{\partial X^{*\rho}} - \frac{\partial \Phi_A(X^{*\alpha})}{\partial X^{*\rho}} 
+ \frac{\partial \Phi_A(X^{*\alpha})}{\partial X^{\rho}}   - \frac{\partial \Phi_A(X^\alpha)}{\partial X^\rho} 
+ \frac{\partial \Phi_A(X^{*\alpha})}{\partial X^{*\rho}} - \frac{\partial \Phi_A(X^{*\alpha})}{\partial X^{\rho}}
\end{equation}

\begin{equation}
\Delta (\partial_\rho \Phi_A)  = \frac{\partial [\Phi_A^*(X^{*\alpha}) - \Phi_A(X^{*\alpha})]}{\partial X^{*\rho}} 
+ \frac{\partial [\Phi_A(X^{*\alpha}) - \Phi_A(X^\alpha)]}{\partial X^{\rho}}  
+ [\frac{\partial }{\partial X^{*\rho}} - \frac{\partial }{\partial X^{\rho}}] \Phi_A(X^{*\alpha})
\end{equation}

each of the 3 terms above deserves separate attention because each involves a fairly lengthy calculation from here. We will start with the first term.

\subsubsection*{$\delta (\partial_\rho \Phi_A) $ First term calculation}

We start with the first term,

\begin{equation}
 \frac{\partial [\Phi_A^*(X^{*\alpha}) - \Phi_A(X^{*\alpha})]}{\partial X^{*\rho}} 
\end{equation}

recall the relationship between the derivatives for the two coordinates,

\begin{equation}
\frac{\partial}{\partial X^\beta} -  \frac{\partial}{\partial X^{*\beta}} = \epsilon \frac{\partial g^\alpha (X)}{\partial X^\beta} \frac{\partial}{\partial X^{*\alpha}}
\end{equation}

thus,

\begin{equation}
\frac{\partial}{\partial X^\rho} -  \frac{\partial}{\partial X^{*\rho}} = \epsilon \frac{\partial g^\alpha (X)}{\partial X^\rho} \frac{\partial}{\partial X^{*\alpha}}
\end{equation}

we can use this to rewrite the derivative,

\begin{equation}
 \frac{\partial [\Phi_A^*(X^{*\alpha}) - \Phi_A(X^{*\alpha})]}{\partial X^{*\rho}}  =  \frac{\partial [\Phi_A^*(X^{*\alpha}) - \Phi_A(X^{*\alpha})]}{\partial X^{\rho}}  -  \epsilon \frac{\partial g^\beta (X)}{\partial X^\rho} \frac{\partial [\Phi_A^*(X^{*\alpha}) - \Phi_A(X^{*\alpha})]}{\partial X^{*\beta}} 
\end{equation}

we can also note that the term in the derivative is related to the variation,

\begin{equation}
\bar{\delta} \Phi_A (X^\alpha) = \Phi_A^*(X^\alpha) - \Phi_A(X^\alpha)
\end{equation}

only here we have this variation as a function of the new coordinates,

\begin{equation}
\bar{\delta} \Phi_A (X^{*\alpha}) = \Phi_A^*(X^{*\alpha}) - \Phi_A(X^{*\alpha}) 
\end{equation}

therefore we can rewrite the first term as,

\begin{equation}
 \frac{\partial [\Phi_A^*(X^{*\alpha}) - \Phi_A(X^{*\alpha})]}{\partial X^{*\rho}}  =  \frac{\partial \bar{\delta} \Phi_A (X^{*\alpha})}{\partial X^{\rho}}  -  \epsilon \frac{\partial g^\beta (X)}{\partial X^\rho} \frac{\partial \bar{\delta} \Phi_A (X^{*\alpha})}{\partial X^{*\beta}} 
\end{equation}

recall that we defined in terms of the barred function,

\begin{equation}
\bar{\delta} \Phi_A(X^\alpha)  = \epsilon \bar{h}_A (X) 
\end{equation}

thus for this case in the new coordinates,

\begin{equation}
\bar{\delta} \Phi_A(X^{*\alpha})  = \epsilon \bar{h}_A (X^*) 
\end{equation}

the first term now reads,

\begin{equation}
 \frac{\partial [\Phi_A^*(X^{*\alpha}) - \Phi_A(X^{*\alpha})]}{\partial X^{*\rho}} =  \frac{\partial \epsilon \bar{h}_A (X^*) }{\partial X^{\rho}}  -  \epsilon \frac{\partial g^\beta (X)}{\partial X^\rho} \frac{\partial \epsilon \bar{h}_A (X^*) }{\partial X^{*\beta}} 
\end{equation}

the second term can be dropped since it is higher order in $\epsilon$, leaving us with only,

\begin{equation}
 \frac{\partial [\Phi_A^*(X^{*\alpha}) - \Phi_A(X^{*\alpha})]}{\partial X^{*\rho}} = \epsilon \frac{\partial  \bar{h}_A (X^*) }{\partial X^{\rho}}
\end{equation}

Taylor expanding $ \bar{h}_A (X^*) $ about $X$,

\begin{equation}
\epsilon \bar{h}_A (X^*) =\epsilon \bar{h}_A (X) + \mathcal{O}(\epsilon^2) 
\end{equation}

again allows only the first term in order $\epsilon$, thus finally we have for the first term,

\begin{equation}
 \frac{\partial [\Phi_A^*(X^{*\alpha}) - \Phi_A(X^{*\alpha})]}{\partial X^{*\rho}} = \epsilon \frac{\partial  \bar{h}_A (X) }{\partial X^{\rho}}
\end{equation}

\subsubsection*{$\delta (\partial_\rho \Phi_A) $ Second term calculation}

Now for the second term,

\begin{equation}
 \frac{\partial [\Phi_A(X^{*\alpha}) - \Phi_A(X^\alpha)]}{\partial X^{\rho}}  
\end{equation}

we can Taylor expand the $\Phi_A(X^{*\alpha})$,

\begin{equation}
\Phi_A(X^{*\alpha}) = \Phi_A(X^\alpha) + \frac{\partial \Phi_A(X^\alpha)}{\partial X^\beta} (X^{*\beta} - X^\beta) + \mathcal{O}(\epsilon^2)
\end{equation}

keeping only the first order in $\epsilon$ and recalling $\delta X^\beta = X^{*\beta} - X^\beta$,

\begin{equation}
\Phi_A(X^{*\alpha}) \approx \Phi_A(X^\alpha) + \frac{\partial \Phi_A(X^\alpha)}{\partial X^\beta} \delta X^\beta
\end{equation}

inserting this expansion into the second term,

\begin{equation}
  \frac{\partial [\Phi_A(X^{*\alpha}) - \Phi_A(X^\alpha)]}{\partial X^{\rho}}  = \frac{\partial [\Phi_A(X^\alpha) + \frac{\partial \Phi_A(X^\alpha)}{\partial X^\beta} \delta X^\beta - \Phi_A(X^\alpha)]}{\partial X^{\rho}}  
\end{equation}

the first and third terms cancel, leaving,

\begin{equation}
  \frac{\partial [\Phi_A(X^{*\alpha}) - \Phi_A(X^\alpha)]}{\partial X^{\rho}}  = \frac{\partial }{\partial X^{\rho}}  [\frac{\partial \Phi_A(X^\alpha)}{\partial X^\beta} \delta X^\beta ]
\end{equation}

we now recall the form of the variation of coordinates in terms of the function $g$,

\begin{equation}
\delta X^\beta = \epsilon g^\beta (X) 
\end{equation}

inserting this,

\begin{equation}
  \frac{\partial [\Phi_A(X^{*\alpha}) - \Phi_A(X^\alpha)]}{\partial X^{\rho}}  = \frac{\partial }{\partial X^{\rho}}  [\frac{\partial \Phi_A(X^\alpha)}{\partial X^\beta}  \epsilon g^\beta (X)  ]
\end{equation}

and using the product rule,

\begin{equation}
  \frac{\partial [\Phi_A(X^{*\alpha}) - \Phi_A(X^\alpha)]}{\partial X^{\rho}}  = \epsilon  [(  \frac{\partial }{\partial X^{\rho}} \frac{\partial \Phi_A(X^\alpha)}{\partial X^\beta})   g^\beta (X) + \frac{\partial \Phi_A(X^\alpha)}{\partial X^\beta}   \frac{\partial  g^\beta (X) }{\partial X^{\rho}} ]
\end{equation}

\begin{equation}
  \frac{\partial [\Phi_A(X^{*\alpha}) - \Phi_A(X^\alpha)]}{\partial X^{\rho}}  = \epsilon  (  \frac{\partial }{\partial X^{\rho}} \frac{\partial \Phi_A(X^\alpha)}{\partial X^\beta})   g^\beta (X) + \epsilon \frac{\partial \Phi_A(X^\alpha)}{\partial X^\beta}   \frac{\partial  g^\beta (X) }{\partial X^{\rho}} 
\end{equation}

we find a first order epsilon expression for the second desired term.

\subsubsection*{$\delta (\partial_\rho \Phi_A) $ Third term calculation}

Finally for the third term,

\begin{equation}
[\frac{\partial }{\partial X^{*\rho}} - \frac{\partial }{\partial X^{\rho}}] \Phi_A(X^{*\alpha})
\end{equation}

we found the relationship between these two derivatives earlier on as,

\begin{equation}
\frac{\partial}{\partial X^\rho} -  \frac{\partial}{\partial X^{*\rho}} = \epsilon \frac{\partial g^\beta (X)}{\partial X^\rho} \frac{\partial}{\partial X^{*\beta}}
\end{equation}

therefore,

\begin{equation}
\frac{\partial}{\partial X^{*\rho}} - \frac{\partial}{\partial X^\rho}   = - \epsilon \frac{\partial g^\beta (X)}{\partial X^\rho} \frac{\partial}{\partial X^{*\beta}}
\end{equation}

and the third term reads,

\begin{equation}
[\frac{\partial }{\partial X^{*\rho}} - \frac{\partial }{\partial X^{\rho}}] \Phi_A(X^{*\alpha}) = - \epsilon \frac{\partial g^\beta (X)}{\partial X^\rho} \frac{\partial \Phi_A(X^{*\alpha}) }{\partial X^{*\beta}} 
\end{equation}

which we can again Taylor expand $\Phi_A(X^{*\alpha})$ and drop higher order terms to basically replace it with $\Phi_A(X^\alpha)$,

\begin{equation}
[\frac{\partial }{\partial X^{*\rho}} - \frac{\partial }{\partial X^{\rho}}] \Phi_A(X^{*\alpha}) = - \epsilon \frac{\partial g^\beta (X)}{\partial X^\rho} \frac{\partial \Phi_A(X^\alpha) }{\partial X^{*\beta}} 
\end{equation}

we can also replace the $\beta$ derivative with,

\begin{equation}
\frac{\partial}{\partial X^\beta} - \frac{\partial}{\partial X^{*\beta}} = \epsilon \frac{\partial g^\sigma (X)}{\partial X^\beta} \frac{\partial}{\partial X^{*\sigma}}
\end{equation}

thus the higher order term again drops and we just get the derivative in the original coordinates,

\begin{equation}
[\frac{\partial }{\partial X^{*\rho}} - \frac{\partial }{\partial X^{\rho}}] \Phi_A(X^{*\alpha}) = - \epsilon \frac{\partial g^\beta (X)}{\partial X^\rho} \frac{\partial \Phi_A(X^\alpha) }{\partial X^\beta} 
\end{equation}

we now have expressions for each of the terms in terms of first order epsilon. It is necessary to put them all together.

\subsubsection*{Combining the 3 terms}

Recall that we wished to determine each of the following 3 terms,

\begin{equation}
\Delta (\partial_\rho \Phi_A)  = \frac{\partial [\Phi_A^*(X^{*\alpha}) - \Phi_A(X^{*\alpha})]}{\partial X^{*\rho}} 
+ \frac{\partial [\Phi_A(X^{*\alpha}) - \Phi_A(X^\alpha)]}{\partial X^{\rho}}  
+ [\frac{\partial }{\partial X^{*\rho}} - \frac{\partial }{\partial X^{\rho}}] \Phi_A(X^{*\alpha})
\end{equation}

with respect to first order in epsilon. We found each of these 3 in the prior sections,

\begin{equation}
 \frac{\partial [\Phi_A^*(X^{*\alpha}) - \Phi_A(X^{*\alpha})]}{\partial X^{*\rho}} = \epsilon \frac{\partial  \bar{h}_A (X) }{\partial X^{\rho}}
\end{equation}

\begin{equation}
  \frac{\partial [\Phi_A(X^{*\alpha}) - \Phi_A(X^\alpha)]}{\partial X^{\rho}}  = \epsilon  (  \frac{\partial }{\partial X^{\rho}} \frac{\partial \Phi_A(X^\alpha)}{\partial X^\beta})   g^\beta (X) + \epsilon \frac{\partial \Phi_A(X^\alpha)}{\partial X^\beta}   \frac{\partial  g^\beta (X) }{\partial X^{\rho}} 
\end{equation}

\begin{equation}
[\frac{\partial }{\partial X^{*\rho}} - \frac{\partial }{\partial X^{\rho}}] \Phi_A(X^{*\alpha}) = - \epsilon \frac{\partial g^\beta (X)}{\partial X^\rho} \frac{\partial \Phi_A(X^\alpha) }{\partial X^\beta} 
\end{equation}

inserting each of them we find,

\begin{equation}
\Delta (\partial_\rho \Phi_A)  = \epsilon \frac{\partial  \bar{h}_A (X) }{\partial X^{\rho}}
+  \epsilon  (  \frac{\partial }{\partial X^{\rho}} \frac{\partial \Phi_A(X^\alpha)}{\partial X^\beta})   g^\beta (X) 
+ \epsilon \frac{\partial \Phi_A(X^\alpha)}{\partial X^\beta}   \frac{\partial  g^\beta (X) }{\partial X^{\rho}} 
- \epsilon \frac{\partial g^\beta (X)}{\partial X^\rho} \frac{\partial \Phi_A(X^\alpha) }{\partial X^\beta} 
\end{equation}

the third and fourth terms exactly cancel!

\begin{equation}
\Delta (\partial_\rho \Phi_A)  = \epsilon \frac{\partial  \bar{h}_A (X) }{\partial X^{\rho}}
+  \epsilon  (  \frac{\partial }{\partial X^{\rho}} \frac{\partial \Phi_A(X^\alpha)}{\partial X^\beta})   g^\beta (X) 
\end{equation}

recalling the variations,

\begin{equation}
\delta X^\beta = \epsilon g^\beta (X) 
\end{equation}

\begin{equation}
\bar{\delta} \Phi_A(X)  = \epsilon \bar{h}_A (X) 
\end{equation}

since we are left with only the linear part we have found the desired variation $\Delta (\partial_\rho \Phi_A)$,

\begin{equation}
\delta (\partial_\rho \Phi_A)  =  \frac{\partial \bar{\delta} \Phi_A(X)}{\partial X^{\rho}} 
+    (  \frac{\partial }{\partial X^{\rho}} \frac{\partial \Phi_A(X^\alpha)}{\partial X^\beta})   \delta X^\beta
\end{equation}

which is the derivative of the bar variation, and a second order contribution due to the fields,

\begin{equation}
\delta (\partial_\rho \Phi_A)  =  \frac{\partial  }{\partial X^{\rho}} \bar{\delta} \Phi_A(X)
+    [ \frac{\partial }{\partial X^{\rho}} \frac{\partial }{\partial X^\beta} \Phi_A(X^\alpha)]   \delta X^\beta
\end{equation}

this is the desired variation which we will be using to simplify the variation of the action from the prior subsection. Typically we will write this more compactly as,

\begin{equation}
\delta (\partial_\rho \Phi_A)  = \partial_\rho \bar{\delta} \Phi_A
+    [\partial_\rho \partial_\beta \Phi_A]   \delta X^\beta
\end{equation}

\subsubsection*{Calculating the variation $\delta (\partial_\rho \Phi_A) $ more directly}

We just calculated the variation,

\begin{equation}
\Delta (\partial_\rho \Phi_A)  = \frac{\partial \Phi_A^*(X^{*\alpha})}{\partial X^{*\rho}} - \frac{\partial \Phi_A(X^\alpha)}{\partial X^\rho}
\end{equation}

by adding and subtracting two terms to this expression. This might seem a bit frustrating --- without knowing what to add it seems like it would be difficult to start this calculation! In fact it can be done in a much more straightforward way, just like we showed for the variation of $\Phi$ in the previous section. Let us start by Taylor expanding the field $\Phi_A^*(X^{*\alpha})$ about $X^\alpha$,

\begin{equation}
\Phi_A^*(X^{*\alpha}) = \Phi_A^*(X^\alpha)  + \frac{\partial \Phi_A^*(X^\alpha)}{\partial X^\beta} \delta X^\beta + \mathcal{O}(\epsilon^2)
\end{equation}

inserting this,

\begin{equation}
\Delta (\partial_\rho \Phi_A)  = \frac{\partial \Phi_A^*(X^\alpha)  + \frac{\partial \Phi_A^*(X^\alpha)}{\partial X^\beta} \delta X^\beta + \mathcal{O}(\epsilon^2)}{\partial X^{*\rho}} - \frac{\partial \Phi_A(X^\alpha)}{\partial X^\rho}
\end{equation}

\begin{equation}
\Delta (\partial_\rho \Phi_A)  = \frac{\partial \Phi_A^*(X^\alpha) }{\partial X^{*\rho}} 
+ \frac{\partial}{\partial X^{*\rho}} [ \frac{\partial \Phi_A^*(X^\alpha)}{\partial X^\beta} \delta X^\beta]
+ \frac{\partial \mathcal{O}(\epsilon^2)}{\partial X^{*\rho}} 
- \frac{\partial \Phi_A(X^\alpha)}{\partial X^\rho}
\end{equation}

also recalling the transformation of the derivative,

\begin{equation}
\frac{\partial}{\partial X^\rho} -  \frac{\partial}{\partial X^{*\rho}} = \epsilon \frac{\partial g^\beta (X)}{\partial X^\rho} \frac{\partial}{\partial X^{*\beta}}
\end{equation}

rewriting the first term,

\begin{equation}
\Delta (\partial_\rho \Phi_A)  = \frac{\partial \Phi_A^*(X^\alpha) }{\partial X^{\rho}} 
-  \epsilon \frac{\partial g^\beta (X)}{\partial X^\rho} \frac{\partial}{\partial X^{*\beta}} \Phi_A^*(X^\alpha) 
+ \frac{\partial}{\partial X^{*\rho}}[  \frac{\partial \Phi_A^*(X^\alpha)}{\partial X^\beta} \delta X^\beta]
+ \frac{\partial \mathcal{O}(\epsilon^2)}{\partial X^{*\rho}} 
- \frac{\partial \Phi_A(X^\alpha)}{\partial X^\rho}
\end{equation}

the first and last term are exactly the derivative of the bar variation,

\begin{equation}
\partial_\rho \bar{\delta} \Phi_A (X^\alpha) = \partial_\rho  [ \Phi_A^*(X^\alpha) - \Phi_A(X^\alpha) ]
\end{equation}

inserting this and keeping only linear terms to introduce the variation,

\begin{equation}
\delta (\partial_\rho \Phi_A)  =\partial_\rho \bar{\delta} \Phi_A (X^\alpha) 
-  \epsilon \frac{\partial g^\beta (X)}{\partial X^\rho} \frac{\partial}{\partial X^\beta} \Phi_A^*(X^\alpha) 
+ \frac{\partial}{\partial X^{\rho}} [ \frac{\partial \Phi_A^*(X^\alpha)}{\partial X^\beta} \delta X^\beta]
\end{equation}

recalling,

\begin{equation}
\delta X^\beta = \epsilon g^\beta (X) 
\end{equation}

and differentiating the third term,

\begin{equation}
\delta (\partial_\rho \Phi_A)  =\partial_\rho \bar{\delta} \Phi_A (X^\alpha) 
-   \frac{\partial \delta X^\beta (X)}{\partial X^\rho} \frac{\partial \Phi_A^*(X^\alpha) }{\partial X^\beta}
+  [( \frac{\partial}{\partial X^{\rho}}\frac{\partial \Phi_A^*(X^\alpha)}{\partial X^\beta}) \delta X^\beta]
+ [ \frac{\partial \Phi_A^*(X^\alpha)}{\partial X^\beta} ( \frac{\partial \delta X^\beta}{\partial X^{\rho}})]
\end{equation}

the second and fourth terms exactly cancel,

\begin{equation}
\delta (\partial_\rho \Phi_A)  =\partial_\rho \bar{\delta} \Phi_A (X^\alpha) 
+  [( \frac{\partial}{\partial X^{\rho}}\frac{\partial \Phi_A^*(X^\alpha)}{\partial X^\beta}) \delta X^\beta]
\end{equation}

the $*$ above drops after writing in terms of the original $\Phi$ and variation, and then dropping higher order, 

\begin{equation}
\delta (\partial_\rho \Phi_A)  =\partial_\rho \bar{\delta} \Phi_A (X^\alpha) 
+  [( \partial_\rho \frac{\partial \Phi_A(X^\alpha)}{\partial X^\beta}) \delta X^\beta]
\end{equation}

finally we are left with the variation,

\begin{equation}
\delta (\partial_\rho \Phi_A)  = \partial_\rho \bar{\delta} \Phi_A
+    [\partial_\rho \partial_\beta \Phi_A]   \delta X^\beta
\end{equation}

which is exactly what we found using the longer method! Therefore for $\delta \Phi$ and $\delta (\partial \Phi)$ the variation can be computed much more directly than what is shown in Gelfand and Fomin.

\subsection{Determining expressions for the variation $\delta (\partial_\omega \partial_\rho \Phi_A) $}

\subsubsection*{Introducing the variation $\delta (\partial_\omega \partial_\rho \Phi_A) $}

We define the difference $\Delta (\partial_\omega \partial_\rho \Phi_A) $ as,

\begin{equation}
\Delta (\partial_\omega \partial_\rho \Phi_A)  = \frac{\partial}{\partial X^{*\omega}} \frac{\partial}{\partial X^{*\rho}} \Phi_A^*(X^{*\alpha}) - \frac{\partial }{\partial X^\omega} \frac{\partial }{\partial X^\rho}\Phi_A(X^\alpha)
\end{equation}

where the variation is the linear part of the difference between the two derivatives of fields as a function of their respective coordinates (the linear part of the above expression). We can write this more compactly as,

\begin{equation}
\Delta (\partial_\omega \partial_\rho \Phi_A)  =\partial_\omega^* \partial_\rho^* \Phi_A^*(X^*) -\partial_\omega \partial_\rho \Phi_A(X)
\end{equation}

recalling the derivative transformation,

\begin{equation}
\frac{\partial}{\partial X^\rho} -  \frac{\partial}{\partial X^{*\rho}} = \epsilon \frac{\partial g^\beta (X)}{\partial X^\rho} \frac{\partial}{\partial X^{*\beta}}
\end{equation}

in compact form,

\begin{equation}
\partial_\rho -  \partial_\rho^* = \epsilon [\partial_\rho  g^\beta (X)] \partial^*_\beta
\end{equation}

\begin{equation}
\partial_\omega -  \partial_\omega^* = \epsilon [\partial_\omega  g^\beta (X)] \partial^*_\beta
\end{equation}

starting by converting $\partial_\rho^*$,

\begin{equation}
\Delta (\partial_\omega \partial_\rho \Phi_A)  =(\partial_\rho - \epsilon [\partial_\rho  g^\beta (X)] \partial^*_\beta)  \partial_\omega^* \Phi_A^*(X^*) -\partial_\omega \partial_\rho \Phi_A(X)
\end{equation}

\begin{equation}
\Delta (\partial_\omega \partial_\rho \Phi_A)  = \partial_\rho \partial_\omega^*  \Phi_A^*(X^*)
-
 \epsilon [\partial_\rho  g^\beta (X)] \partial^*_\beta  \partial_\omega^* \Phi_A^*(X^*)
 -
 \partial_\rho \partial_\omega \Phi_A(X)
\end{equation}

we can combine the first and third terms,

\begin{equation}
\Delta (\partial_\omega \partial_\rho \Phi_A)  = \partial_\rho[ \partial_\omega^*  \Phi_A^*(X^*) - \partial_\omega \Phi_A(X)]
-
 \epsilon [\partial_\rho  g^\beta (X)] \partial^*_\beta  \partial_\omega^* \Phi_A^*(X^*)
\end{equation}

but this is exactly the difference we calculated in the previous section!

\begin{equation}
\Delta (\partial_\omega \Phi_A)  = \frac{\partial \Phi_A^*(X^{*\alpha})}{\partial X^{*\omega}} - \frac{\partial \Phi_A(X^\alpha)}{\partial X^\omega} = \partial_\omega^*  \Phi_A^*(X^*) - \partial_\omega \Phi_A(X)
\end{equation}

with corresponding variation,

\begin{equation}
\delta (\partial_\omega \Phi_A)  = \partial_\omega \bar{\delta} \Phi_A
+    [\partial_\omega \partial_\beta \Phi_A]   \delta X^\beta
\end{equation}

\subsubsection*{Expressing the variation $\delta (\partial_\omega \partial_\rho \Phi_A) $ in terms of $\delta (\partial_\omega \Phi_A)$}

inserting this difference,

\begin{equation}
\Delta (\partial_\omega \partial_\rho \Phi_A)  = \partial_\rho \Delta (\partial_\omega \Phi_A)
-
 \epsilon [\partial_\rho  g^\beta (X)] \partial^*_\beta  \partial_\omega^* \Phi_A^*(X^*)
\end{equation}

Taylor expanding the second term and keeping only linear terms we are left with the same thing (no $*$), thus the variation is,

\begin{equation}
\delta (\partial_\omega \partial_\rho \Phi_A)  = \partial_\rho \delta (\partial_\omega \Phi_A)
-
 \epsilon [\partial_\rho  g^\beta (X)] \partial_\beta  \partial_\omega \Phi_A(X)
\end{equation}

inserting the known variation into the first term,

\begin{equation}
\delta (\partial_\omega \partial_\rho \Phi_A)  = \partial_\rho [\partial_\omega \bar{\delta} \Phi_A
+    (\partial_\omega \partial_\beta \Phi_A)   \delta X^\beta]
- \epsilon [\partial_\rho  g^\beta (X)] \partial_\beta  \partial_\omega \Phi_A
\end{equation}

taking the derivative,

\begin{equation}
\delta (\partial_\omega \partial_\rho \Phi_A)  = \partial_\rho \partial_\omega \bar{\delta} \Phi_A
+    (\partial_\rho \partial_\omega \partial_\beta \Phi_A)   \delta X^\beta
+    (\partial_\omega \partial_\beta \Phi_A)  ( \partial_\rho \delta X^\beta)
- \epsilon [\partial_\rho  g^\beta (X)] \partial_\beta  \partial_\omega \Phi_A
\end{equation}

recall that,

\begin{equation}
\delta X^\beta = \epsilon g^\beta (X) 
\end{equation}

thus,

\begin{equation}
\delta (\partial_\omega \partial_\rho \Phi_A)  = \partial_\rho \partial_\omega \bar{\delta} \Phi_A
+    (\partial_\rho \partial_\omega \partial_\beta \Phi_A)   \delta X^\beta
+    (\partial_\omega \partial_\beta \Phi_A)  ( \partial_\rho \delta X^\beta)
-  [\partial_\rho  \delta X^\beta] \partial_\beta  \partial_\omega \Phi_A
\end{equation}

the third and fourth terms exactly cancel! This leaves only,

\begin{equation}
\delta (\partial_\omega \partial_\rho \Phi_A)  = \partial_\rho \partial_\omega \bar{\delta} \Phi_A
+    (\partial_\rho \partial_\omega \partial_\beta \Phi_A)   \delta X^\beta
\end{equation}

therefore we now have an expression for the variation of the second derivative of the field.

\subsection{Commutation relation of total variation and derivatives}

At this point, you may realize that the field variation relationships all take the same form,

\begin{equation}
\delta \Phi_A = \partial_\beta \Phi_A \delta X^\beta  + \bar{\delta} \Phi_A
\end{equation}

\begin{equation}
\delta (\partial_\rho \Phi_A)  = \partial_\rho \bar{\delta} \Phi_A  +    [\partial_\rho \partial_\beta \Phi_A]   \delta X^\beta
\end{equation}

\begin{equation}
\delta (\partial_\omega \partial_\rho \Phi_A)  = \partial_\rho \partial_\omega \bar{\delta} \Phi_A
+    (\partial_\rho \partial_\omega \partial_\beta \Phi_A)   \delta X^\beta
\end{equation}

it turns out that there is a general commutation relation we can define, which applies to any such function,

\begin{equation}
F = F(X, \Phi, \partial \Phi, \partial \partial \Phi, \dots)
\end{equation}

since the fields and their derivatives depend on $X$ we can write the function as $F(X)$ and transformed as $F^*(X^*)$ such that the total and substantial variations of $F$ are exactly as for the fields,

\begin{equation}
\delta F = F^*(X^{*}) - F(X)
\end{equation}

\begin{equation}
\bar{\delta} F = F^*(X) - F(X)
\end{equation}

keeping only terms linear in $\epsilon$ throughout. Taylor expanding the transformed function about the untransformed coordinates as in the earlier section and using $X^{*\alpha} = X^\alpha + \epsilon g^\alpha + \mathcal{O}(\epsilon^2)$,

\begin{equation}
F^*(X^{*\alpha}) = F^*(X^\alpha) + \epsilon g^\beta \partial_\beta F^*(X^\alpha) + \mathcal{O}(\epsilon^2)
\end{equation}

in the second term $F^*$ can be replaced by $F$ at linear order, since their difference is itself $\mathcal{O}(\epsilon)$ and the term already carries one factor of $\epsilon$,

\begin{equation}
F^*(X^{*\alpha}) = F^*(X^\alpha) + \epsilon g^\beta \partial_\beta F(X^\alpha) + \mathcal{O}(\epsilon^2)
\end{equation}

subtracting $F(X)$ from both sides and using the definitions above together with $\delta X^\beta = \epsilon g^\beta$, we can introduce the total (left) and substantial (right) variations at first order,

\begin{equation}
\delta F = \bar{\delta} F + \delta X^\beta \partial_\beta F
\end{equation}

this holds as an operator identity,

\begin{equation}
\delta = \bar{\delta} + \delta X^\beta \partial_\beta
\end{equation}

therefore we have a general identity valid at any derivative order (compare the three variations summarized at the start of this subsection). Recall that because

\begin{equation}
\bar{\delta} (\partial_\rho F) = \partial_\rho F^*(X) - \partial_\rho F(X) = \partial_\rho [F^*(X) - F(X)] = \partial_\rho \bar{\delta} F
\end{equation}

the substantial variation commutes with partial derivatives,

\begin{equation}
[\bar{\delta}, \partial_\rho] = \bar{\delta} \partial_\rho - \partial_\rho \bar{\delta} = 0
\end{equation}

but not for the total variation since $\delta (\partial_\rho F)$ compares $\partial^*_\rho F^*(X^*)$ to $\partial_\rho F(X)$, derivatives with respect to different coordinates evaluated at different points, which cannot be factored out of the difference. This is the operator statement of the noncommutation $\delta \partial \neq \partial \delta$ emphasized in the introduction. Acting with the operator identity, we re-obtain our three required variations:

\begin{equation}
\delta \Phi_A = \bar{\delta} \Phi_A + \partial_\beta \Phi_A   \delta X^\beta
\end{equation}

\begin{equation}
\delta (\partial_\rho \Phi_A) = \bar{\delta} (\partial_\rho \Phi_A) + \partial_\beta \partial_\rho \Phi_A   \delta X^\beta = \partial_\rho \bar{\delta} \Phi_A + \partial_\rho \partial_\beta \Phi_A   \delta X^\beta
\end{equation}

\begin{equation}
\delta (\partial_\omega \partial_\rho \Phi_A) = \partial_\omega \partial_\rho \bar{\delta} \Phi_A + \partial_\omega \partial_\rho \partial_\beta \Phi_A   \delta X^\beta
\end{equation}

since this pattern continues to any order of derivatives we can directly apply this to derive the Noether identity for Lagrangians of any derivative order.

\subsection{Using the field variations on the variation of the action}

\subsubsection*{Inserting the variations}

Recall that we calculated the variation of the action to be,

\begin{equation}
\delta S =  \int_{R_{X}} [
 \frac{\partial \mathcal{L}}{\partial X^\beta} \delta X^\beta 
+ \frac{\partial \mathcal{L}}{\partial \Phi_A} \delta \Phi_A
+ \frac{\partial \mathcal{L}}{\partial (\partial_\rho \Phi_A)} \delta (\partial_\rho \Phi_A)
+  \frac{\partial \mathcal{L}}{\partial (\partial_\omega \partial_\rho \Phi_A)} \delta (\partial_\omega \partial_\rho \Phi_A)  
+ \epsilon    \mathcal{L}  \frac{\partial g^\alpha}{\partial X^\alpha}  ] dX
\end{equation}

we then found the variations,

\begin{equation}
\delta \Phi_A = \partial_\beta \Phi_A \delta X^\beta  + \bar{\delta} \Phi_A
\end{equation}

\begin{equation}
\delta (\partial_\rho \Phi_A)  = \partial_\rho \bar{\delta} \Phi_A  +    [\partial_\rho \partial_\beta \Phi_A]   \delta X^\beta
\end{equation}

\begin{equation}
\delta (\partial_\omega \partial_\rho \Phi_A)  = \partial_\rho \partial_\omega \bar{\delta} \Phi_A
+    (\partial_\rho \partial_\omega \partial_\beta \Phi_A)   \delta X^\beta
\end{equation}

inserting these variations into the variation of the action,

\begin{multline}
\delta S =  \int_{R_{X}} [
 \frac{\partial \mathcal{L}}{\partial X^\beta} \delta X^\beta 
+ \frac{\partial \mathcal{L}}{\partial \Phi_A} [\partial_\beta \Phi_A \delta X^\beta  + \bar{\delta} \Phi_A]
+ \frac{\partial \mathcal{L}}{\partial (\partial_\rho \Phi_A)} [\partial_\rho \bar{\delta} \Phi_A  +    [\partial_\rho \partial_\beta \Phi_A]   \delta X^\beta]
\\
+  \frac{\partial \mathcal{L}}{\partial (\partial_\omega \partial_\rho \Phi_A)} [\partial_\rho \partial_\omega \bar{\delta} \Phi_A
+    (\partial_\rho \partial_\omega \partial_\beta \Phi_A)   \delta X^\beta]
+ \epsilon    \mathcal{L}  \frac{\partial g^\alpha}{\partial X^\alpha}  ] dX
\end{multline}

we can rewrite the sixth term using,

\begin{equation}
\delta X^\beta = \epsilon g^\beta (X) 
\end{equation}

thus,

\begin{multline}
\delta S =  \int_{R_{X}} [
 \frac{\partial \mathcal{L}}{\partial X^\beta} \delta X^\beta 
+ \frac{\partial \mathcal{L}}{\partial \Phi_A} [\partial_\beta \Phi_A \delta X^\beta  + \bar{\delta} \Phi_A]
+ \frac{\partial \mathcal{L}}{\partial (\partial_\rho \Phi_A)} [\partial_\rho \bar{\delta} \Phi_A  +    [\partial_\rho \partial_\beta \Phi_A]   \delta X^\beta]
\\
+  \frac{\partial \mathcal{L}}{\partial (\partial_\omega \partial_\rho \Phi_A)} [\partial_\rho \partial_\omega \bar{\delta} \Phi_A
+    (\partial_\rho \partial_\omega \partial_\beta \Phi_A)   \delta X^\beta]
+     \mathcal{L}  \partial_\beta \delta X^{\beta}  ] dX
\end{multline}

\subsubsection*{Combining terms in the integrand}

Expanding the terms,

\begin{multline}
\delta S =  \int_{R_{X}} [
 \frac{\partial \mathcal{L}}{\partial X^\beta} \delta X^\beta 
+ \frac{\partial \mathcal{L}}{\partial \Phi_A} (\partial_\beta \Phi_A) \delta X^\beta 
+ \frac{\partial \mathcal{L}}{\partial \Phi_A} ( \bar{\delta} \Phi_A)
+ \frac{\partial \mathcal{L}}{\partial (\partial_\rho \Phi_A)} (\partial_\rho \bar{\delta} \Phi_A )
\\
+ \frac{\partial \mathcal{L}}{\partial (\partial_\rho \Phi_A)}   (\partial_\rho \partial_\beta \Phi_A)   \delta X^\beta
+  \frac{\partial \mathcal{L}}{\partial (\partial_\omega \partial_\rho \Phi_A)} (\partial_\rho \partial_\omega \bar{\delta} \Phi_A)
+  \frac{\partial \mathcal{L}}{\partial (\partial_\omega \partial_\rho \Phi_A)}   (\partial_\rho \partial_\omega \partial_\beta \Phi_A)   \delta X^\beta
+     \mathcal{L}  \partial_\beta \delta X^{\beta}  ] dX
\end{multline}

and rearranging,

\begin{multline}
\delta S =  \int_{R_{X}} [
 \frac{\partial \mathcal{L}}{\partial X^\beta} \delta X^\beta 
+ \frac{\partial \mathcal{L}}{\partial \Phi_A} (\partial_\beta \Phi_A) \delta X^\beta 
+ \frac{\partial \mathcal{L}}{\partial (\partial_\rho \Phi_A)}   (\partial_\rho \partial_\beta \Phi_A)   \delta X^\beta
\\
+  \frac{\partial \mathcal{L}}{\partial (\partial_\omega \partial_\rho \Phi_A)}   (\partial_\rho \partial_\omega \partial_\beta \Phi_A)   \delta X^\beta
+     \mathcal{L}  \partial_\beta \delta X^{\beta} 
+ \frac{\partial \mathcal{L}}{\partial \Phi_A} ( \bar{\delta} \Phi_A)
+ \frac{\partial \mathcal{L}}{\partial (\partial_\rho \Phi_A)} (\partial_\rho \bar{\delta} \Phi_A )
+  \frac{\partial \mathcal{L}}{\partial (\partial_\omega \partial_\rho \Phi_A)} (\partial_\rho \partial_\omega \bar{\delta} \Phi_A)
] dX
\end{multline}

we find that everything from the first two lines can be expressed as a result of the product rule,

\begin{equation}
\partial_\beta (\mathcal{L} \delta X^\beta) = (\partial_\beta \mathcal{L}) \delta X^\beta + \mathcal{L} \partial_\beta \delta X^\beta
\end{equation}

in order to differentiate the multivariable function $\mathcal{L}$ we need to apply the chain rule for a multivariable function,

\begin{equation}
\partial_\beta \mathcal{L} =  \frac{\partial \mathcal{L}}{\partial X^\beta} 
+ \frac{\partial \mathcal{L}}{\partial \Phi_A} \partial_\beta \Phi_A 
 +   \frac{\partial \mathcal{L}}{\partial (\partial_\rho \Phi_A)} [\partial_\beta \partial_\rho \Phi_A]  
+  \frac{\partial \mathcal{L}}{\partial (\partial_\omega \partial_\rho \Phi_A)}   [\partial_\rho \partial_\omega \partial_\beta \Phi_A]
\end{equation}

inserting back into the product rule,

\begin{equation}
\partial_\beta (\mathcal{L} \delta X^\beta) = \frac{\partial \mathcal{L}}{\partial X^\beta} \delta X^\beta 
+ \frac{\partial \mathcal{L}}{\partial \Phi_A} \partial_\beta \Phi_A \delta X^\beta
 +   \frac{\partial \mathcal{L}}{\partial (\partial_\rho \Phi_A)} [\partial_\beta \partial_\rho \Phi_A]   \delta X^\beta  
+ \frac{\partial \mathcal{L}}{\partial (\partial_\omega \partial_\rho \Phi_A)}   [\partial_\rho \partial_\omega \partial_\beta \Phi_A]  \delta X^\beta
+ \mathcal{L} \partial_\beta \delta X^\beta
\end{equation}

but this is exactly what we have in the variation of the action! Therefore, after inserting this back into the action,

\begin{equation}
\delta S =  \int_{R_{X}} [\partial_\beta (\mathcal{L} \delta X^\beta)
+ \frac{\partial \mathcal{L}}{\partial \Phi_A} ( \bar{\delta} \Phi_A)
+ \frac{\partial \mathcal{L}}{\partial (\partial_\rho \Phi_A)} (\partial_\rho \bar{\delta} \Phi_A )
+  \frac{\partial \mathcal{L}}{\partial (\partial_\omega \partial_\rho \Phi_A)} (\partial_\rho \partial_\omega \bar{\delta} \Phi_A)
] dX
\end{equation}

We want to separate the terms into an EOM and a conservation law by writing everything for the EOM in terms of $\bar{\delta} \Phi_A$ and everything else under a divergence. Using the product rule (IBP) we can also expand the third term,

\begin{equation}
 \partial_\rho  [\frac{\partial \mathcal{L}}{\partial (\partial_\rho \Phi_A)}\bar{\delta} \Phi_A]   = [\partial_\rho  \frac{\partial \mathcal{L}}{\partial (\partial_\rho \Phi_A)}] \bar{\delta} \Phi_A   + \frac{\partial \mathcal{L}}{\partial (\partial_\rho \Phi_A)} [\partial_\rho \bar{\delta} \Phi_A  ]
\end{equation}

thus we can reexpress the desired term in our integrand as,

\begin{equation}
 \frac{\partial \mathcal{L}}{\partial (\partial_\rho \Phi_A)} [\partial_\rho \bar{\delta} \Phi_A  ] =  \partial_\rho  [\frac{\partial \mathcal{L}}{\partial (\partial_\rho \Phi_A)}\bar{\delta} \Phi_A] -  [\partial_\rho  \frac{\partial \mathcal{L}}{\partial (\partial_\rho \Phi_A)}] \bar{\delta} \Phi_A
\end{equation}

inserting this back into the action,

\begin{equation}
\delta S =  \int_{R_{X}} [\partial_\beta (\mathcal{L} \delta X^\beta)
+ \frac{\partial \mathcal{L}}{\partial \Phi_A} ( \bar{\delta} \Phi_A)
+ \partial_\rho  [\frac{\partial \mathcal{L}}{\partial (\partial_\rho \Phi_A)}\bar{\delta} \Phi_A] -  [\partial_\rho  \frac{\partial \mathcal{L}}{\partial (\partial_\rho \Phi_A)}] \bar{\delta} \Phi_A
+  \frac{\partial \mathcal{L}}{\partial (\partial_\omega \partial_\rho \Phi_A)} (\partial_\rho \partial_\omega \bar{\delta} \Phi_A)
] dX
\end{equation}

we can again use the product rule on the fifth term,

\begin{equation}
 \partial_\rho [ \frac{\partial \mathcal{L}}{\partial (\partial_\omega \partial_\rho \Phi_A)} (\partial_\omega \bar{\delta} \Phi_A) ] =  
[\partial_\rho \frac{\partial \mathcal{L}}{\partial (\partial_\omega \partial_\rho \Phi_A)}] ( \partial_\omega \bar{\delta} \Phi_A) 
+  \frac{\partial \mathcal{L}}{\partial (\partial_\omega \partial_\rho \Phi_A)} (\partial_\rho \partial_\omega \bar{\delta} \Phi_A)
\end{equation}

thus we can reexpress the desired term in our integrand as,

\begin{equation}
 \frac{\partial \mathcal{L}}{\partial (\partial_\omega \partial_\rho \Phi_A)} (\partial_\rho \partial_\omega \bar{\delta} \Phi_A)
=
 \partial_\rho [ \frac{\partial \mathcal{L}}{\partial (\partial_\omega \partial_\rho \Phi_A)} (\partial_\omega \bar{\delta} \Phi_A) ] 
-
[\partial_\rho \frac{\partial \mathcal{L}}{\partial (\partial_\omega \partial_\rho \Phi_A)}] ( \partial_\omega \bar{\delta} \Phi_A) 
\end{equation}

inserting this back into the action,

\begin{multline}
\delta S =  \int_{R_{X}} [\partial_\beta (\mathcal{L} \delta X^\beta)
+ \frac{\partial \mathcal{L}}{\partial \Phi_A} ( \bar{\delta} \Phi_A)
+ \partial_\rho  [\frac{\partial \mathcal{L}}{\partial (\partial_\rho \Phi_A)}\bar{\delta} \Phi_A] 
-  [\partial_\rho  \frac{\partial \mathcal{L}}{\partial (\partial_\rho \Phi_A)}] \bar{\delta} \Phi_A
\\
+ \partial_\rho [ \frac{\partial \mathcal{L}}{\partial (\partial_\omega \partial_\rho \Phi_A)} (\partial_\omega \bar{\delta} \Phi_A) ] 
-
[\partial_\rho \frac{\partial \mathcal{L}}{\partial (\partial_\omega \partial_\rho \Phi_A)}] ( \partial_\omega \bar{\delta} \Phi_A) 
] dX
\end{multline}

once again we can again use the product rule on the sixth term,

\begin{equation}
\partial_\omega [(\partial_\rho \frac{\partial \mathcal{L}}{\partial (\partial_\omega \partial_\rho \Phi_A)}) \bar{\delta} \Phi_A]
=
[\partial_\omega \partial_\rho \frac{\partial \mathcal{L}}{\partial (\partial_\omega \partial_\rho \Phi_A)}]  \bar{\delta} \Phi_A
+
[\partial_\rho \frac{\partial \mathcal{L}}{\partial (\partial_\omega \partial_\rho \Phi_A)}] ( \partial_\omega \bar{\delta} \Phi_A)
\end{equation}

thus we can reexpress the desired term in our integrand as,

\begin{equation}
[\partial_\rho \frac{\partial \mathcal{L}}{\partial (\partial_\omega \partial_\rho \Phi_A)}] ( \partial_\omega \bar{\delta} \Phi_A)
=
\partial_\omega [(\partial_\rho \frac{\partial \mathcal{L}}{\partial (\partial_\omega \partial_\rho \Phi_A)}) \bar{\delta} \Phi_A]
-
[\partial_\omega \partial_\rho \frac{\partial \mathcal{L}}{\partial (\partial_\omega \partial_\rho \Phi_A)}]  \bar{\delta} \Phi_A
\end{equation}

inserting this back into the action,

\begin{multline}
\delta S =  \int_{R_{X}} [\partial_\beta (\mathcal{L} \delta X^\beta)
+ \frac{\partial \mathcal{L}}{\partial \Phi_A} ( \bar{\delta} \Phi_A)
+ \partial_\rho  [\frac{\partial \mathcal{L}}{\partial (\partial_\rho \Phi_A)}\bar{\delta} \Phi_A] 
-  [\partial_\rho  \frac{\partial \mathcal{L}}{\partial (\partial_\rho \Phi_A)}] \bar{\delta} \Phi_A
\\
+ \partial_\rho [ \frac{\partial \mathcal{L}}{\partial (\partial_\omega \partial_\rho \Phi_A)} (\partial_\omega \bar{\delta} \Phi_A) ] 
-
\partial_\omega [(\partial_\rho \frac{\partial \mathcal{L}}{\partial (\partial_\omega \partial_\rho \Phi_A)}) \bar{\delta} \Phi_A]
+
[\partial_\omega \partial_\rho \frac{\partial \mathcal{L}}{\partial (\partial_\omega \partial_\rho \Phi_A)}]  \bar{\delta} \Phi_A
] dX
\end{multline}

we can separate the EOM from the conservation law,

\begin{multline}
\delta S =  \int_{R_{X}} [
 \frac{\partial \mathcal{L}}{\partial \Phi_A} ( \bar{\delta} \Phi_A)
-  [\partial_\rho  \frac{\partial \mathcal{L}}{\partial (\partial_\rho \Phi_A)}] \bar{\delta} \Phi_A
+
[\partial_\omega \partial_\rho \frac{\partial \mathcal{L}}{\partial (\partial_\omega \partial_\rho \Phi_A)}]  \bar{\delta} \Phi_A
\\
+ \partial_\rho  [\frac{\partial \mathcal{L}}{\partial (\partial_\rho \Phi_A)}\bar{\delta} \Phi_A] 
+ \partial_\rho [ \frac{\partial \mathcal{L}}{\partial (\partial_\omega \partial_\rho \Phi_A)} (\partial_\omega \bar{\delta} \Phi_A) ] 
-\partial_\omega [(\partial_\rho \frac{\partial \mathcal{L}}{\partial (\partial_\omega \partial_\rho \Phi_A)}) \bar{\delta} \Phi_A]
+\partial_\beta (\mathcal{L} \delta X^\beta)
] dX
\end{multline}

combining the first line into the EOM, 

\begin{multline}
\delta S =  \int_{R_{X}} [
\left( \frac{\partial \mathcal{L}}{\partial \Phi_A}
-  \partial_\rho  \frac{\partial \mathcal{L}}{\partial (\partial_\rho \Phi_A)}
+
\partial_\omega \partial_\rho \frac{\partial \mathcal{L}}{\partial (\partial_\omega \partial_\rho \Phi_A)} \right)  \bar{\delta} \Phi_A
\\
+ \partial_\rho  [\frac{\partial \mathcal{L}}{\partial (\partial_\rho \Phi_A)}\bar{\delta} \Phi_A] 
+ \partial_\rho [ \frac{\partial \mathcal{L}}{\partial (\partial_\omega \partial_\rho \Phi_A)} (\partial_\omega \bar{\delta} \Phi_A) ] 
-\partial_\rho [(\partial_\omega \frac{\partial \mathcal{L}}{\partial (\partial_\omega \partial_\rho \Phi_A)}) \bar{\delta} \Phi_A]
+\partial_\rho (\mathcal{L} \delta X^\rho)
] dX
\end{multline}

Here the third boundary term was rewritten using the symmetry of $\frac{\partial \mathcal{L}}{\partial (\partial_\omega \partial_\rho \Phi_A)}$ under $\omega \leftrightarrow \rho$, relabelling the summed dummy indices so that $\partial_\rho$ can be factored out of the whole divergence.

Factoring out the $\partial_\rho$,

\begin{multline}
\delta S =  \int_{R_{X}} [
\left( \frac{\partial \mathcal{L}}{\partial \Phi_A}
-  \partial_\rho  \frac{\partial \mathcal{L}}{\partial (\partial_\rho \Phi_A)}
+
\partial_\omega \partial_\rho \frac{\partial \mathcal{L}}{\partial (\partial_\omega \partial_\rho \Phi_A)} \right)  \bar{\delta} \Phi_A
\\
+ \partial_\rho  \left( \frac{\partial \mathcal{L}}{\partial (\partial_\rho \Phi_A)}\bar{\delta} \Phi_A
+  \frac{\partial \mathcal{L}}{\partial (\partial_\omega \partial_\rho \Phi_A)} (\partial_\omega \bar{\delta} \Phi_A)  
-(\partial_\omega \frac{\partial \mathcal{L}}{\partial (\partial_\omega \partial_\rho \Phi_A)}) \bar{\delta} \Phi_A
+ \mathcal{L} \delta X^\rho \right)
] dX
\end{multline}

We now have an expression for the EOM and conservation law. The required variations for the conservation law are,

\begin{equation}
 \bar{\delta} \Phi_A = \delta \Phi_A -  \partial_\beta \Phi_A \delta X^\beta
\end{equation}

and,

\begin{equation}
\delta X^\beta = X^{*\beta} - X^\beta
\end{equation}

where for coordinate and gauge transformations we express the field transformations as,

\begin{equation}
\bar{\delta} \Phi_A = -      \mathsterling_{\delta X} \Phi_A + \delta_g \Phi_A
\end{equation}

We note that the Lagrangian $\mathcal{L}$ here is the Lagrangian in the original coordinates, dependent on the original coordinates, and the original fields.

\subsubsection*{Noether Identity}

From Noether's paper, the main statement is that an identity exists relating the EOM to the conservation law,

\begin{multline}
\delta S =  \int_{R_{X}} [
\left( \frac{\partial \mathcal{L}}{\partial \Phi_A}
-  \partial_\rho  \frac{\partial \mathcal{L}}{\partial (\partial_\rho \Phi_A)}
+
\partial_\omega \partial_\rho \frac{\partial \mathcal{L}}{\partial (\partial_\omega \partial_\rho \Phi_A)} \right)  \bar{\delta} \Phi_A
\\
+ \partial_\rho  \left( \frac{\partial \mathcal{L}}{\partial (\partial_\rho \Phi_A)}\bar{\delta} \Phi_A
+  \frac{\partial \mathcal{L}}{\partial (\partial_\omega \partial_\rho \Phi_A)} (\partial_\omega \bar{\delta} \Phi_A)  
-(\partial_\omega \frac{\partial \mathcal{L}}{\partial (\partial_\omega \partial_\rho \Phi_A)}) \bar{\delta} \Phi_A
+ \mathcal{L} \delta X^\rho \right)
] dX
\end{multline}

we find this since the action is invariant under the transformations, $\delta S = 0$,

\begin{multline}
0 =  \int_{R_{X}} [
\left( \frac{\partial \mathcal{L}}{\partial \Phi_A}
-  \partial_\rho  \frac{\partial \mathcal{L}}{\partial (\partial_\rho \Phi_A)}
+
\partial_\omega \partial_\rho \frac{\partial \mathcal{L}}{\partial (\partial_\omega \partial_\rho \Phi_A)} \right)  \bar{\delta} \Phi_A
\\
+ \partial_\rho  \left( \frac{\partial \mathcal{L}}{\partial (\partial_\rho \Phi_A)}\bar{\delta} \Phi_A
+  \frac{\partial \mathcal{L}}{\partial (\partial_\omega \partial_\rho \Phi_A)} (\partial_\omega \bar{\delta} \Phi_A)  
-(\partial_\omega \frac{\partial \mathcal{L}}{\partial (\partial_\omega \partial_\rho \Phi_A)}) \bar{\delta} \Phi_A
+ \mathcal{L} \delta X^\rho \right)
] dX
\end{multline}

since this holds for any region of integration, the integrand must vanish, therefore we have the following condition, which we refer to as the Noether identity,

\begin{multline}
0 =  
\left( \frac{\partial \mathcal{L}}{\partial \Phi_A}
-  \partial_\rho  \frac{\partial \mathcal{L}}{\partial (\partial_\rho \Phi_A)}
+
\partial_\omega \partial_\rho \frac{\partial \mathcal{L}}{\partial (\partial_\omega \partial_\rho \Phi_A)} \right)  \bar{\delta} \Phi_A
\\
+ \partial_\rho  \left( \frac{\partial \mathcal{L}}{\partial (\partial_\rho \Phi_A)}\bar{\delta} \Phi_A
+  \frac{\partial \mathcal{L}}{\partial (\partial_\omega \partial_\rho \Phi_A)} (\partial_\omega \bar{\delta} \Phi_A)  
-(\partial_\omega \frac{\partial \mathcal{L}}{\partial (\partial_\omega \partial_\rho \Phi_A)}) \bar{\delta} \Phi_A
+ \mathcal{L} \delta X^\rho \right)
\end{multline}

which can be expressed as,

\begin{multline}  
\left( \frac{\partial \mathcal{L}}{\partial \Phi_A}
-  \partial_\rho  \frac{\partial \mathcal{L}}{\partial (\partial_\rho \Phi_A)}
+
\partial_\omega \partial_\rho \frac{\partial \mathcal{L}}{\partial (\partial_\omega \partial_\rho \Phi_A)} \right)  \bar{\delta} \Phi_A
\\
=
+ \partial_\rho  \left(- \frac{\partial \mathcal{L}}{\partial (\partial_\rho \Phi_A)}\bar{\delta} \Phi_A
-  \frac{\partial \mathcal{L}}{\partial (\partial_\omega \partial_\rho \Phi_A)} (\partial_\omega \bar{\delta} \Phi_A)  
+(\partial_\omega \frac{\partial \mathcal{L}}{\partial (\partial_\omega \partial_\rho \Phi_A)}) \bar{\delta} \Phi_A
- \mathcal{L} \delta X^\rho \right)
\end{multline}

we define the canonical Noether current $J^\rho$ as the quantity under the total divergence when it is written on the same side as the Euler-Lagrange term; when the divergence is moved to the opposite side, as above, an overall minus sign appears. Both conventions occur in the literature.\\

We note that the second order current follows the naive extension of the first order pattern, precisely because the substantial variation commutes with partial derivatives, $\bar{\delta} \partial = \partial \bar{\delta}$.

\subsection{Noether's Theorems}

From the invariance condition $S[\Phi_A (X^\alpha)] = S[\Phi_A^* (X^{*\alpha})]$, i.e. $\delta S = 0$, we found in the previous section the condition,

\begin{multline}
0 =  
\left( \frac{\partial \mathcal{L}}{\partial \Phi_A}
-  \partial_\rho  \frac{\partial \mathcal{L}}{\partial (\partial_\rho \Phi_A)}
+
\partial_\omega \partial_\rho \frac{\partial \mathcal{L}}{\partial (\partial_\omega \partial_\rho \Phi_A)} \right)  \bar{\delta} \Phi_A
\\
+ \partial_\rho  \left( \frac{\partial \mathcal{L}}{\partial (\partial_\rho \Phi_A)}\bar{\delta} \Phi_A
+  \frac{\partial \mathcal{L}}{\partial (\partial_\omega \partial_\rho \Phi_A)} (\partial_\omega \bar{\delta} \Phi_A)  
-(\partial_\omega \frac{\partial \mathcal{L}}{\partial (\partial_\omega \partial_\rho \Phi_A)}) \bar{\delta} \Phi_A
+ \mathcal{L} \delta X^\rho \right)
\end{multline}

This is known as the Noether identity. The first term includes the Euler-Lagrange equation,

\begin{equation}
   E^A =  \frac{\partial \mathcal{L}}{\partial \Phi_A}
-  \partial_\rho  \frac{\partial \mathcal{L}}{\partial (\partial_\rho \Phi_A)}
+
\partial_\omega \partial_\rho \frac{\partial \mathcal{L}}{\partial (\partial_\omega \partial_\rho \Phi_A)}
\end{equation}

The second term has a total divergence of a quantity known as the Noether current,

\begin{equation}
    J^\rho =  \frac{\partial \mathcal{L}}{\partial (\partial_\rho \Phi_A)}\bar{\delta} \Phi_A
+  \frac{\partial \mathcal{L}}{\partial (\partial_\omega \partial_\rho \Phi_A)} (\partial_\omega \bar{\delta} \Phi_A)  
-(\partial_\omega \frac{\partial \mathcal{L}}{\partial (\partial_\omega \partial_\rho \Phi_A)}) \bar{\delta} \Phi_A
+ \mathcal{L} \delta X^\rho
\end{equation}

Therefore the Noether identity can be expressed compactly as $E^A \bar{\delta} \Phi_A + \partial_\rho J^\rho = 0$. It is from this Noether current that conserved objects can be derived.\\

The Noether identity presented above is the starting point for Noether's two theorems.

\subsubsection*{Noether's first theorem}

Noether's first theorem states that every continuous (variational) symmetry of the action with a finite number of parameters yields linearly independent conserved currents, one per parameter. For physical field theories, this means that there is a one to one correspondence between the number of action symmetries (typically the finite-dimensional continuous group of symmetries associated to the spacetime metric, such as Lorentz, Poincar\'e, conformal symmetry groups) and the number of conservation laws derived from these symmetries.\\

Pragmatically, this involves direct use of the Noether current,

\begin{equation}
    J^\rho = \frac{\partial \mathcal{L}}{\partial (\partial_\rho \Phi_A)}\bar{\delta} \Phi_A
+  \frac{\partial \mathcal{L}}{\partial (\partial_\omega \partial_\rho \Phi_A)} (\partial_\omega \bar{\delta} \Phi_A)  
-(\partial_\omega \frac{\partial \mathcal{L}}{\partial (\partial_\omega \partial_\rho \Phi_A)}) \bar{\delta} \Phi_A
+ \mathcal{L} \delta X^\rho
\end{equation}

where one must identify the coordinate symmetries $\delta X^\rho$ (these are the ones associated to the spacetime symmetries, often solved for by solving the Killing equation) and the field symmetries $\bar{\delta} \Phi_A $ (defining these is more nuanced and different methods exist, we will discuss physical applications later in this article). The $\delta X^\rho$ are therefore the Killing vectors $v^\rho$ of a later section, in other words the spacetime symmetry generators, that are solved for in a given metric spacetime (we will give examples later). In applications the infinitesimal parameter $\epsilon$ is absorbed into the transformation parameters (e.g. $a_\beta$, $\omega_{\beta\alpha}$, $S$, $k_\mu$), so that $\delta X^\beta$ and $\bar{\delta} \Phi_A$ are written without an explicit $\epsilon$. We note that internal symmetries, for which $\delta X^\rho = 0$ and $\bar{\delta} \Phi_A \neq 0$, are covered by the same identity; the Killing vector description applies to the spacetime symmetries.\\

More formally, for Noether's first theorem, if $S[\Phi_A (X^\alpha)]$ is invariant, and the equation of motion is satisfied $E^A = 0$, then,

\begin{equation}
 \partial_\rho  \left( \frac{\partial \mathcal{L}}{\partial (\partial_\rho \Phi_A)}\bar{\delta} \Phi_A
+  \frac{\partial \mathcal{L}}{\partial (\partial_\omega \partial_\rho \Phi_A)} (\partial_\omega \bar{\delta} \Phi_A)  
-(\partial_\omega \frac{\partial \mathcal{L}}{\partial (\partial_\omega \partial_\rho \Phi_A)}) \bar{\delta} \Phi_A
+ \mathcal{L} \delta X^\rho \right)= 0
\end{equation}

then for our transforms of coordinates and fields that leave the action invariant, the above quantity will define a conservation law (i.e., a conserved object exists). The currents obtained this way are conserved on shell, and are unique up to identically conserved (improper) currents.

\subsubsection*{Noether's second theorem}

Noether's second theorem is about continuous (infinite) groups of transformations which leave the action invariant. In physics, the common example is the gauge transformations in e.g., electrodynamics. More formally, if $S[\Phi_A (X^\alpha)]$ is invariant under transformations depending on $r$ arbitrary functions, there are $r$ identities connected to the Euler-Lagrange equations. This calculation does not depend on the order of derivatives in the Lagrangian, so it is identical to what we have in the first order section and we won't repeat it here, other than the end result,

\begin{equation}
    E^A a_{Ai} - \partial_\mu (E^A  b_{Ai}^\mu)  + \dots = 0
\end{equation}

the only difference is that $E^A$ now represents the second order Euler-Lagrange expression,

\begin{equation}
   E^A =  \frac{\partial \mathcal{L}}{\partial \Phi_A}
-  \partial_\rho  \frac{\partial \mathcal{L}}{\partial (\partial_\rho \Phi_A)}
+
\partial_\omega \partial_\rho \frac{\partial \mathcal{L}}{\partial (\partial_\omega \partial_\rho \Phi_A)}
\end{equation}

For physics this is a statement that each gauge symmetry implies differential identities of the equations of motion. Examples include $\partial_\mu \partial_\nu F^{\mu\nu} = 0$ in electrodynamics and $\nabla_\mu G^{\mu\nu} = 0$ in general relativity. We will give these examples and their associated calculations later in the article.

\section{Origin of Spacetime Symmetries}

Once the Noether identity is determined, and we have a particular Lagrangian, an important question arises: where do we find the symmetries $\delta X^\alpha$ that we need to substitute into the Noether identity? The way in which the finite group symmetries can be determined is well established, such as in \cite{Olver1993}. For the purpose of this article, where we are focused on flat spacetime (e.g., Minkowski spacetime), the associated spacetime symmetry equations commonly reduce to the Killing equation, where one must solve for the Killing vector associated to the isometries of the given metric. This Killing vector solution is precisely the $\delta X^\alpha$ we will need to insert into the Noether identity. Familiar physical conservation laws such as conservation of energy and momentum are then associated to symmetries of the spacetimes, such as translations of time (energy) and space (momentum).\\

Foundational theories such as electrodynamics also have symmetries known as conformal symmetries which yield additional conservation laws from Noether's theorem. To obtain these, one can solve the conformal Killing equation. The Killing and conformal Killing solutions give finite parameter groups of transformations where each parameter corresponds to a specific conserved quantity. For standard physical field theories in 4D Minkowski spacetime, the breakdown is as follows:

\begin{itemize}
    \item Minkowski metric $\to$ Killing equation $\to$ 10 Parameter Poincar\'e group (associated to conservation of energy-momentum and angular momentum)
    \item Minkowski metric $\to$ conformal Killing equation $\to$ 15 Parameter conformal group (which includes the 10 parameter Poincar\'e group, plus 1 parameter for dilatation and 4 for special conformal transformations)
\end{itemize}

We will detail the way in which each of these transformations can be calculated below. See e.g., \cite{Francesco2012} for some of the related details.

\subsection{Killing equation}

In order to determine the Killing equation, we wish to solve for the vector fields along which the Lie derivative of the metric vanishes; these vector fields generate the isometries (symmetries) of a given metric,

\begin{equation}
\mathsterling_v g_{\mu\nu} = 0
\end{equation}

starting from the partial form of the Lie derivative of the metric,

\begin{equation}
\mathsterling_v g_{\mu\nu} = v^\alpha \partial_\alpha g_{\mu\nu} + g_{\alpha\nu} \partial_\mu v^\alpha + g_{\alpha\mu} \partial_\nu v^\alpha
\end{equation}

we can use the product rule on the second two terms,

\begin{equation}
\mathsterling_v g_{\mu\nu} = v^\alpha \partial_\alpha g_{\mu\nu} + \partial_\mu (g_{\alpha\nu} v^\alpha) - v^\alpha \partial_\mu g_{\alpha\nu} + \partial_\nu (g_{\alpha\mu} v^\alpha) - v^\alpha \partial_\nu g_{\alpha\mu}
\end{equation}

from the first, third and fifth terms we can factor out a $-v^\alpha$,

\begin{equation}
\mathsterling_v g_{\mu\nu} = - v^\alpha (- \partial_\alpha g_{\mu\nu} + \partial_\mu g_{\alpha\nu} +  \partial_\nu g_{\alpha\mu}) + \partial_\mu (g_{\alpha\nu} v^\alpha) + \partial_\nu (g_{\alpha\mu} v^\alpha) 
\end{equation}

but recalling the Christoffel symbols, of first and second kind, respectively,

\begin{equation}
\Gamma_{\alpha \mu \nu}
=
\frac{1}{2} (\partial_\nu g_{\mu \alpha}
+ \partial_\mu g_{\nu\alpha}
- \partial_\alpha  g_{\nu\mu})
\end{equation}

\begin{equation}
\Gamma^\lambda_{ \mu \nu}
=
\frac{1}{2} g^{\alpha\lambda} (\partial_\nu g_{\mu \alpha}
+ \partial_\mu g_{\nu\alpha}
- \partial_\alpha  g_{\nu\mu})
\end{equation}

we recognize the symbol of the first kind inside the brackets of the first term: the combination in brackets is exactly twice $\Gamma_{\alpha \mu \nu}$. Inserting this,

\begin{equation}
\mathsterling_v g_{\mu\nu} = - 2 v^\alpha \Gamma_{\alpha \mu \nu} + \partial_\mu (g_{\alpha\nu} v^\alpha) + \partial_\nu (g_{\alpha\mu} v^\alpha) 
\end{equation}

In the second and third term we can use the metric to lower the index on $v$. In the first term we write $v^\alpha = g^{\alpha\lambda} v_\lambda$, so that the metric combines with the first-kind symbol to produce the second kind,

\begin{equation}
\mathsterling_v g_{\mu\nu} = - 2 v_\lambda g^{\alpha\lambda} \Gamma_{\alpha \mu \nu} + \partial_\mu v_\nu + \partial_\nu v_\mu
\end{equation}

introducing the second kind,

\begin{equation}
\mathsterling_v g_{\mu\nu} = - 2 v_\lambda \Gamma^\lambda_{ \mu \nu} + \partial_\mu v_\nu + \partial_\nu v_\mu
\end{equation}

from here we can notice that the definition of the covariant derivative appears,

\begin{equation}
\nabla_\mu v_\nu = \partial_\mu v_\nu  - \Gamma^\lambda_{ \mu \nu} v_\lambda 
\end{equation}

\begin{equation}
\nabla_\nu v_\mu = \partial_\nu v_\mu  - \Gamma^\lambda_{ \mu \nu} v_\lambda
\end{equation}

where in the second expression we have used the symmetry of the Christoffel symbol in its lower indices. Thus rearranging the Lie derivative,

\begin{equation}
\mathsterling_v g_{\mu\nu} = \partial_\mu v_\nu - v_\lambda \Gamma^\lambda_{ \mu \nu}  + \partial_\nu v_\mu - v_\lambda \Gamma^\lambda_{ \mu \nu}
\end{equation}

we are left with the two covariant derivatives,

\begin{equation}
\mathsterling_v g_{\mu\nu} = \nabla_\mu v_\nu  + \nabla_\nu v_\mu
\end{equation}

the Killing condition is that the metric does not change along $v$,

\begin{equation}
\mathsterling_v g_{\mu\nu} = 0
\end{equation}

thus,

\begin{equation}
\nabla_\mu v_\nu  + \nabla_\nu v_\mu = 0
\end{equation}

\subsection{Killing equation --- Minkowski spacetime}

If we specify the metric to be the Minkowski metric,

\begin{equation}
g_{\mu\nu} = \eta_{\mu\nu}
\end{equation}

then the covariant derivatives are just partial derivatives, since the Christoffel symbols are zero in Minkowski space in Cartesian coordinates, thus the Killing equation simplifies to,

\begin{equation}
\partial_\mu v_\nu  + \partial_\nu v_\mu = 0
\end{equation}

We wish to solve this equation in order to find the symmetries of the Minkowski metric (Poincar\'e transformations). In order to do this we can use a standard index-permutation trick, first differentiating the equation by $\partial_\alpha$,

\begin{equation}
\partial_\alpha \partial_\mu v_\nu  + \partial_\alpha \partial_\nu v_\mu = 0
\end{equation}

cyclically permuting these indices,

\begin{equation}
\partial_\nu \partial_\alpha v_\mu  + \partial_\nu \partial_\mu v_\alpha = 0
\quad , \quad
\partial_\mu \partial_\nu v_\alpha  + \partial_\mu \partial_\alpha v_\nu = 0
\end{equation}

If we subtract the first from the last two we have,

\begin{equation}
\partial_\mu \partial_\nu v_\alpha  + \partial_\mu \partial_\alpha v_\nu + \partial_\nu \partial_\alpha v_\mu  + \partial_\nu \partial_\mu v_\alpha
- \partial_\alpha \partial_\mu v_\nu  - \partial_\alpha \partial_\nu v_\mu = 0
\end{equation}

since the partials commute, the second and fifth terms cancel, as well as the third and sixth terms. The first and fourth combine,

\begin{equation}
2 \partial_\mu \partial_\nu v_\alpha   = 0
\end{equation}

thus we are required to solve the equation,

\begin{equation}
\partial_\mu \partial_\nu v_\alpha   = 0
\end{equation}

this is a linear homogeneous PDE so we can integrate twice to find a solution for $v_\alpha$,

\begin{equation}
\partial_\nu v_\alpha   = \omega_{\alpha\nu}
\end{equation}

\begin{equation}
v_\alpha   = \omega_{\alpha\nu} X^\nu + a_\alpha
\end{equation}

where $\omega$ and $a$ are constant parameters of integration. At this stage $\omega_{\alpha\nu}$ is an arbitrary constant matrix; the Killing equation will constrain it. From the original Killing equation, 

\begin{equation}
\partial_\mu v_\nu  + \partial_\nu v_\mu = 0
\end{equation}

we can sub in this solution, 

\begin{equation}
v_\mu   = \omega_{\mu\alpha} X^\alpha + a_\mu
\quad , \quad
v_\nu   = \omega_{\nu\alpha} X^\alpha + a_\nu
\end{equation}

differentiating, $\partial_\mu v_\nu = \omega_{\nu\alpha} \delta_\mu^\alpha$ and $\partial_\nu v_\mu = \omega_{\mu\alpha} \delta_\nu^\alpha$; the constant $a$ drops out, leaving,

\begin{equation}
\omega_{\nu\alpha} \delta_\mu^\alpha  + \omega_{\mu\alpha} \delta_\nu^\alpha = 0
\end{equation}

\begin{equation}
\omega_{\nu\mu}  + \omega_{\mu\nu}= 0
\end{equation}

thus we have the result that the parameter $\omega$ is antisymmetric,

\begin{equation}
\omega_{\nu\mu}  = - \omega_{\mu\nu}
\end{equation}

we have the familiar result of $v$, the Poincar\'e transformation, with $D(D-1)/2 + D$ parameters ($10$ in $D = 4$),

\begin{equation}
v_\nu   = \omega_{\nu\alpha} X^\alpha + a_\nu
\end{equation}

which, as it is an infinitesimal transformation, is what we often denote as, raising the index with $\eta^{\mu\nu}$ to match the index position used in the Noether identity,

\begin{equation}
\delta X^\nu   = \omega^{\nu}{}_{\alpha} X^\alpha + a^\nu
\end{equation}

\subsection{Conformal Killing equation --- Minkowski spacetime}

The conformal Killing equation comes from modifying the Killing equation,

\begin{equation}
\mathsterling_v g_{\mu\nu} = 0
\end{equation}

to include solutions up to a scaling of the metric,

\begin{equation}
\mathsterling_v g_{\mu\nu} = \lambda(X) g_{\mu\nu}
\end{equation}

where $\lambda$ is some function of the coordinates $X$. Recalling the Lie derivative of the metric,

\begin{equation}
\mathsterling_v g_{\mu\nu} = v^\alpha \partial_\alpha g_{\mu\nu} + g_{\alpha\nu} \partial_\mu v^\alpha + g_{\alpha\mu} \partial_\nu v^\alpha
\end{equation}

 we have,

\begin{equation}
v^\alpha \partial_\alpha g_{\mu\nu} + g_{\alpha\nu} \partial_\mu v^\alpha + g_{\alpha\mu} \partial_\nu v^\alpha = \lambda(X) g_{\mu\nu}
\end{equation}

instead of deriving the covariant derivative form, we will immediately go to Minkowski space because that is all we are interested in at the moment,

\begin{equation}
v^\alpha \partial_\alpha \eta_{\mu\nu} + \eta_{\alpha\nu} \partial_\mu v^\alpha + \eta_{\alpha\mu} \partial_\nu v^\alpha = \lambda(X) \eta_{\mu\nu}
\end{equation}

the first term on the left hand side is just zero (derivative of Minkowski is zero). The second two terms we can lower the $\alpha$ index with Minkowski, yielding,

\begin{equation}
\partial_\mu v_\nu + \partial_\nu v_\mu = \lambda(X) \eta_{\mu\nu}
\end{equation}

We now have a nonhomogeneous equation for $v$ which we must solve in order to determine the conformal transformations.\\

\subsubsection*{Differentiating both sides}

Differentiating both sides like before we have,

\begin{equation}
\partial_\alpha \partial_\mu v_\nu + \partial_\alpha \partial_\nu v_\mu = \eta_{\mu\nu} \partial_\alpha \lambda(X)
\end{equation}

again we need to cyclically permute,

\begin{equation}
\partial_\mu \partial_\nu v_\alpha + \partial_\mu \partial_\alpha v_\nu = \eta_{\nu\alpha} \partial_\mu \lambda(X)
\quad , \quad
\partial_\nu \partial_\alpha v_\mu + \partial_\nu \partial_\mu v_\alpha = \eta_{\alpha\mu} \partial_\nu \lambda(X)
\end{equation}

subtracting the first from the last two, only one unique term on the left hand side remains,

\begin{equation}
2 \partial_\mu \partial_\nu v_\alpha = \eta_{\nu\alpha} \partial_\mu \lambda(X) + \eta_{\alpha\mu} \partial_\nu \lambda(X) - \eta_{\mu\nu} \partial_\alpha \lambda(X)
\end{equation}

contracting both sides with $\eta^{\mu\nu}$, the final term is the dimension D,

\begin{equation}
2 \square v_\alpha = \partial_\alpha \lambda(X) +  \partial_\alpha \lambda(X) - D \partial_\alpha \lambda(X)
\end{equation}

Thus we can combine the right hand side terms,

\begin{equation}
2 \square v_\alpha = (2 - D) \partial_\alpha \lambda(X) 
\end{equation}

\subsubsection*{Differentiating both sides again}

If we differentiate both sides one more time,

\begin{equation}
2 \partial_\beta \square v_\alpha = (2 - D) \partial_\alpha \partial_\beta \lambda(X) 
\end{equation}

we would like to have the equation entirely in terms of $\lambda(X)$. To do this we can write the above expression as,

\begin{equation}
2  \partial^\mu \partial_\mu \partial_\nu  v_\alpha = (2 - D) \partial_\alpha \partial_\nu \lambda(X) 
\end{equation}

we have what we see on the right hand side from the previous section,

\begin{equation}
2 \partial_\mu \partial_\nu v_\alpha = \eta_{\nu\alpha} \partial_\mu \lambda(X) + \eta_{\alpha\mu} \partial_\nu \lambda(X) - \eta_{\mu\nu} \partial_\alpha \lambda(X)
\end{equation}

inserting this,

\begin{equation}
 \partial^\mu [\eta_{\nu\alpha} \partial_\mu \lambda(X) + \eta_{\alpha\mu} \partial_\nu \lambda(X) - \eta_{\mu\nu} \partial_\alpha \lambda(X)] = (2 - D) \partial_\alpha \partial_\nu \lambda(X) 
\end{equation}

expanding,

\begin{equation}
\eta_{\nu\alpha}  \square \lambda(X) + \partial_\alpha  \partial_\nu \lambda(X) -  \partial_\nu  \partial_\alpha \lambda(X) = (2 - D) \partial_\alpha \partial_\nu \lambda(X) 
\end{equation}

the second and third terms on the left hand side cancel each other,

\begin{equation}
\eta_{\nu\alpha}  \square \lambda(X) = (2 - D) \partial_\alpha \partial_\nu \lambda(X) 
\end{equation}

thus we are left with an equation in terms of $\lambda(X)$.

\subsubsection*{Solving the equation for $\lambda(X)$}

We can take the equation for $\lambda(X)$,

\begin{equation}
\eta_{\nu\alpha}  \square \lambda(X) = (2 - D) \partial_\alpha \partial_\nu \lambda(X) 
\end{equation}

and contract both sides by $\eta^{\nu\alpha}$,

\begin{equation}
D  \square \lambda(X) = (2 - D) \square \lambda(X) 
\end{equation}

rearranging,

\begin{equation}
(2 D - 2)  \square \lambda(X) = 0
\end{equation}

or more simply,

\begin{equation}
( D - 1)  \square \lambda(X) = 0
\end{equation}

for $D > 1$ this implies,

\begin{equation}
\square \lambda(X) = 0
\end{equation}

On its own this is a wave equation, whose solutions are not restricted to linear functions. However, inserting $\square \lambda(X) = 0$ back into the uncontracted equation $\eta_{\nu\alpha} \square \lambda(X) = (2 - D) \partial_\alpha \partial_\nu \lambda(X)$, the left hand side vanishes, thus for $D \neq 2$,

\begin{equation}
\partial_\alpha \partial_\nu \lambda(X) = 0
\end{equation}

(The case $D = 2$ is special: there this constraint does not follow, the space of solutions becomes infinite dimensional, and the finite parameter counting below does not apply; we assume $D \geq 3$ from here on.) Solving for $\lambda$, with constant coefficients $A$ and $B_\beta$,

\begin{equation}
 \lambda(X) = A + B_\beta X^\beta
\end{equation}

\subsubsection*{Form of the conformal vector $v_\alpha$}

Since $\lambda(X)$ is linear in the coordinates, its first derivatives are constant. Recalling,

\begin{equation}
2 \partial_\mu \partial_\nu v_\alpha = \eta_{\nu\alpha} \partial_\mu \lambda(X) + \eta_{\alpha\mu} \partial_\nu \lambda(X) - \eta_{\mu\nu} \partial_\alpha \lambda(X)
\end{equation}

the right hand side is therefore constant, so all second derivatives of $v_\alpha$ are constant and all third derivatives of $v_\alpha$ vanish. The solution must then be at most quadratic in the coordinates,

 \begin{equation}
v_\alpha = a_\alpha + b_{\alpha\beta} X^\beta + c_{\alpha\mu\nu} X^\mu X^\nu
\end{equation}

where $a$, $b$ and $c$ are constant parameters. The parameter $c_{\alpha\mu\nu}$ can be taken symmetric in $\mu\nu$: only the part of $c_{\alpha\mu\nu}$ symmetric in $\mu\nu$ survives the contraction with $X^\mu X^\nu$, and indeed from the equation above $c_{\alpha\mu\nu} = \frac{1}{2} \partial_\mu \partial_\nu v_\alpha$, which is symmetric since partial derivatives commute.\\

To determine the parameters it is convenient to first eliminate $\lambda(X)$ in favour of $v_\alpha$.

\subsubsection*{Expressing $\lambda(X)$ in terms of $v_\alpha$}

We can solve for the $\lambda(X)$ in the conformal Killing equation for Minkowski,

\begin{equation}
\partial_\mu v_\nu + \partial_\nu v_\mu = \lambda(X) \eta_{\mu\nu}
\end{equation}

Contracting both sides by $\eta^{\mu\nu}$,

\begin{equation}
\partial^\nu v_\nu + \partial^\nu v_\nu = D \lambda(X)
\end{equation}

thus we have the function,

\begin{equation}
\lambda(X) = \frac{2}{D} \partial_\alpha v^\alpha 
\end{equation}

Subbing this back into the conformal Killing equation,

\begin{equation}
\partial_\mu v_\nu + \partial_\nu v_\mu = \frac{2}{D} \eta_{\mu\nu} \partial_\alpha v^\alpha 
\end{equation}

\subsubsection*{Form of first two parameters due to Killing equation}

Recall we wrote the conformal Killing equation in the form,

\begin{equation}
\partial_\mu v_\nu + \partial_\nu v_\mu = \frac{2}{D} \eta_{\mu\nu} \partial_\alpha v^\alpha 
\end{equation}

If we consider the complete conformal Killing vector solution,

 \begin{equation}
v_\alpha = a_\alpha + b_{\alpha\beta} X^\beta + c_{\alpha\mu\nu} X^\mu X^\nu
\end{equation}

the first term immediately satisfies this equation (constant). The second term,

\begin{equation}
b_{\nu\mu} + b_{\mu\nu} = \frac{2}{D} \eta_{\mu\nu} \eta^{\alpha\gamma} b_{\gamma\alpha}
\end{equation}

this equation is satisfied for,

 \begin{equation}
b_{\alpha\beta} = - b_{\beta\alpha}
\end{equation}

which is the antisymmetric Lorentz parameter $\omega_{\mu\nu} = - \omega_{\nu\mu}$, corresponding to rotations and boosts ($D(D-1)/2$ parameters; $6$ in $D = 4$). We can also solve this equation if $b_{\mu\nu} = S \eta_{\mu\nu}$, in other words any scaling $S$ of the Minkowski metric,

\begin{equation}
S \eta_{\nu\mu} + S \eta_{\mu\nu} = \frac{2}{D} \eta_{\mu\nu} \eta^{\alpha\gamma} S \eta_{\gamma\alpha}
\end{equation}

\begin{equation}
S \eta_{\nu\mu} + S \eta_{\mu\nu} = 2 S \eta_{\mu\nu} 
\end{equation}

therefore LHS = RHS and having $b_{\mu\nu} = S \eta_{\mu\nu}$ also satisfies the Killing equation. In fact these exhaust the solutions: decomposing $b_{\mu\nu}$ into its antisymmetric part, trace part, and symmetric traceless part, the equation forces the symmetric traceless part to vanish. Our conformal Killing vector now reads,

 \begin{equation}
v_\alpha = a_\alpha + (\omega_{\alpha\beta} + S \eta_{\alpha\beta}) X^\beta + c_{\alpha\mu\nu} X^\mu X^\nu
\end{equation}

 \begin{equation}
v_\alpha = a_\alpha + \omega_{\alpha\beta} X^\beta + S X_\alpha + c_{\alpha\mu\nu} X^\mu X^\nu
\end{equation}

\subsubsection*{Form of final (conformal) parameter due to Killing equation}

Finally we need to consider the final term with $c_{\alpha\rho\sigma} X^\rho X^\sigma$, subbing it into the conformal Killing equation,

\begin{equation}
\partial_\mu v_\nu + \partial_\nu v_\mu = \frac{2}{D} \eta_{\mu\nu} \partial_\alpha v^\alpha 
\end{equation}

we get,

\begin{equation}
\partial_\mu (c_{\nu\rho\sigma} X^\rho X^\sigma) + \partial_\nu (c_{\mu\rho\sigma} X^\rho X^\sigma) = \frac{2}{D} \eta_{\mu\nu} \eta^{\alpha\gamma}\partial_\alpha (c_{\gamma\rho\sigma} X^\rho X^\sigma)
\end{equation}

differentiating, the constant parameter is zero,

\begin{equation}
c_{\nu\mu\sigma}  X^\sigma
+ c_{\nu\rho\mu} X^\rho 
+ c_{\mu\nu\sigma}  X^\sigma
+ c_{\mu\rho\nu} X^\rho
= \frac{2}{D} \eta_{\mu\nu} \eta^{\alpha\gamma}c_{\gamma\alpha\sigma} X^\sigma
+ \frac{2}{D} \eta_{\mu\nu} \eta^{\alpha\gamma}c_{\gamma\rho\alpha} X^\rho 
\end{equation}

changing dummies and combining terms,

\begin{equation}
 X^\sigma [c_{\nu\mu\sigma} 
+ c_{\nu\sigma\mu} 
+ c_{\mu\nu\sigma} 
+ c_{\mu\sigma\nu}] 
= \frac{2}{D} \eta_{\mu\nu} \eta^{\alpha\gamma} X^\sigma [c_{\gamma\alpha\sigma} + c_{\gamma\sigma\alpha}]
\end{equation}

using the symmetry property of the parameter $c$ in the last two indices,

\begin{equation}
c_{\nu\mu\sigma} = c_{\nu\sigma\mu}
\end{equation}

several terms combine,

\begin{equation}
 X^\sigma [2 c_{\nu\mu\sigma} 
+ 2 c_{\mu\nu\sigma} ] 
= \frac{4}{D} \eta_{\mu\nu} \eta^{\alpha\gamma} X^\sigma c_{\gamma\alpha\sigma}
\end{equation}

we will define on the right hand side a new vector,

\begin{equation}
k_\sigma = \frac{1}{D} \eta^{\alpha\gamma}  c_{\gamma\alpha\sigma}
\end{equation}

thus,

\begin{equation}
 X^\sigma [2 c_{\nu\mu\sigma} 
+ 2 c_{\mu\nu\sigma} ] 
= 4 \eta_{\mu\nu} k_\sigma X^\sigma 
\end{equation}

\begin{equation}
 c_{\nu\mu\sigma} X^\sigma
+ c_{\mu\nu\sigma} X^\sigma
= 2 \eta_{\mu\nu} k_\sigma X^\sigma 
\end{equation}

rearranging,

\begin{equation}
 [c_{\nu\mu\sigma} 
+ c_{\mu\nu\sigma} 
- 2 \eta_{\mu\nu} k_\sigma] X^\sigma  = 0
\end{equation}

thus,

\begin{equation}
c_{\nu\mu\sigma} 
+ c_{\mu\nu\sigma} 
= 2 \eta_{\mu\nu} k_\sigma
\end{equation}

\subsubsection*{Cyclic permutation to isolate the conformal parameter}

We can cycle the indices,

\begin{equation}
c_{\nu\mu\sigma} 
+ c_{\mu\nu\sigma} 
= 2 \eta_{\mu\nu} k_\sigma
\quad , \quad
c_{\mu\sigma\nu} 
+ c_{\sigma\mu\nu} 
= 2 \eta_{\sigma\mu} k_\nu
\quad , \quad
c_{\sigma\nu\mu} 
+ c_{\nu\sigma\mu} 
= 2 \eta_{\nu\sigma} k_\mu
\end{equation}

Subtracting the third from the first two,

\begin{equation}
2 c_{\mu\nu\sigma} 
= 2 \eta_{\mu\nu} k_\sigma + 2 \eta_{\sigma\mu} k_\nu - 2 \eta_{\nu\sigma} k_\mu
\end{equation}

we are left with,

\begin{equation}
c_{\mu\nu\sigma} = \eta_{\mu\nu} k_\sigma + \eta_{\sigma\mu} k_\nu - \eta_{\nu\sigma} k_\mu
\end{equation}

\subsubsection*{Conformal Killing vector}

Thus if we return to our conformal Killing vector,

 \begin{equation}
v_\alpha = a_\alpha + \omega_{\alpha\beta} X^\beta + S X_\alpha + c_{\alpha\mu\nu} X^\mu X^\nu
\end{equation}

we can sub in our solution for the parameters $c$,

\begin{equation}
c_{\alpha\mu\nu} = \eta_{\alpha\mu} k_\nu + \eta_{\nu\alpha} k_\mu - \eta_{\mu\nu} k_\alpha
\end{equation}

thus the final term in the Killing vector is,

 \begin{equation}
c_{\alpha\mu\nu} X^\mu X^\nu = (\eta_{\alpha\mu} k_\nu + \eta_{\nu\alpha} k_\mu - \eta_{\mu\nu} k_\alpha) X^\mu X^\nu
\end{equation}

contracting Minkowskis,

 \begin{equation}
c_{\alpha\mu\nu} X^\mu X^\nu =  k_\nu X_\alpha X^\nu +  k_\mu X_\alpha X^\mu  - k_\alpha X_\nu X^\nu
\end{equation}

the first two terms combine,

 \begin{equation}
c_{\alpha\mu\nu} X^\mu X^\nu =  2 k_\nu X_\alpha X^\nu  - k_\alpha X_\nu X^\nu
\end{equation}

Altogether we have found the conformal Killing vector to be,

 \begin{equation}
v_\alpha = a_\alpha + \omega_{\alpha\beta} X^\beta + S X_\alpha + 2 k_\nu X_\alpha X^\nu  - k_\alpha X_\nu X^\nu
\end{equation}

the first term corresponds to Poincar\'e translation. The second term to Lorentz rotation. The third to dilatation. The final two to the conformal invariance (special conformal transformations). These are the coordinate variations we sub into Noether to derive conservation laws, raising the index with $\eta^{\alpha\beta}$ to match the index position used in the Noether identity,

 \begin{equation}
\delta X^\alpha = v^\alpha = a^\alpha + \omega^{\alpha}{}_{\beta} X^\beta + S X^\alpha + 2 k_\nu X^\nu X^\alpha  - k^\alpha X_\nu X^\nu
\end{equation}

\section{Applications in Physics}

We will now discuss applications of these ideas to physics. While the coordinate symmetries $\delta X^\beta$ determined in the previous section apply to any field theory on the given spacetime (they are determined by the metric alone, through the Killing and conformal Killing equations), the field symmetries $\bar{\delta} \Phi_A$ must be determined for each theory individually. Typically, the field symmetries consist of two contributions: the Lie derivative part from coordinate symmetries, and possible gauge symmetries of the action. Following Bessel-Hagen \cite{BesselHagen1921,BakerLinnemannSmeenk2021} (see \cite{Chasova2025} for a historical thesis on Bessel-Hagen's work), the proper transformations mix all of these contributions. In this section we apply the Noether identities derived in this article to four examples: for Noether's first theorem, classical electrodynamics (a first order theory) and linearized Gauss-Bonnet gravity (a second order spin-2 higher derivative gravity theory \cite{BK2019,Baker2021DualII}, not to be confused with full Gauss-Bonnet gravity \cite{Lanczos1938}); for Noether's second theorem, classical electrodynamics (a first order theory) and general relativity (a second order theory).\\

Before beginning, we make the note that most of the physics literature has historically focused on deriving the object known as the energy-momentum tensor $T^{\mu\nu}$ \cite{Blaschke2016}. Multiple distinct definitions exist, such as the Hilbert definition $T^{\mu\nu}_H$ (used for the source of Einstein's field equations), the canonical Noether definition $T^{\mu\nu}_C$ (which we will discuss at the end of this article) and the Noether energy-momentum tensor $T^{\mu\nu}_N$ (i.e., the physically accepted energy-momentum tensor derived directly from Noether's first theorem, as we will discuss in this section). Much attention has been placed on the equivalence of these different objects (particularly, using superpotentials and the Belinfante approach \cite{Gotay1993,Pons2011}), however these definitions are not generally equivalent \cite{BKK2021}, raising the question as to which is the foundational definition for e.g., physical field theories. It is our view that the Noether definition is the superior definition which corresponds to this object and therefore is the focus of the following section. We also note that other works, such as \cite{Eriksen1980,Montesinos2006}, were perhaps unaware of Bessel-Hagen's article and thus independently developed comparable direct approaches for obtaining the correct physical conservation laws.

\subsection{Noether's first theorem --- first order theory example (electrodynamics)}

We will now present how one can obtain the 15 conserved tensors associated to the 15 parameter conformal group in 4D classical electrodynamics using the Bessel-Hagen method (as first shown by Bessel-Hagen in 1921 \cite{BesselHagen1921}). The Lagrangian density of classical electrodynamics is,

\begin{equation}
\mathcal{L} = -\frac{1}{4} F_{\mu\nu} F^{\mu\nu}
\end{equation}

where the field strength tensor is defined in terms of the vector potential $A_\mu$,

\begin{equation}
F_{\mu\nu} = \partial_\mu A_\nu - \partial_\nu A_\mu
\end{equation}

this is a first order field theory for the potential $\Phi_A = A_\nu$, thus we recall the first order Noether identity derived in this article,

\begin{equation}
\left(  \frac{\partial \mathcal{L}}{\partial A_\nu}
-  \partial_\rho  \frac{\partial \mathcal{L}}{\partial (\partial_\rho A_\nu)} \right) \bar{\delta} A_\nu 
+ \partial_\rho  \left( \frac{\partial \mathcal{L}}{\partial (\partial_\rho A_\nu)}\bar{\delta} A_\nu 
+ \mathcal{L} \delta X^\rho \right) = 0
\end{equation}

where we will frequently lower the index on the coordinate transformation, $\delta X_\beta = \eta_{\beta\alpha} \delta X^\alpha$, such that $\mathcal{L} \delta X^\rho = \eta^{\rho\beta} \mathcal{L} \delta X_\beta$ \cite{BakerLinnemannSmeenk2021}. We require the derivative of the Lagrangian with respect to the derivative of the potential,

\begin{equation}
\frac{\partial \mathcal{L}}{\partial (\partial_\rho A_\sigma)} = -\frac{1}{2} F^{\mu\nu} \frac{\partial F_{\mu\nu}}{\partial (\partial_\rho A_\sigma)}
\end{equation}

differentiating the field strength tensor,

\begin{equation}
\frac{\partial F_{\mu\nu}}{\partial (\partial_\rho A_\sigma)} = \delta^\rho_\mu \delta^\sigma_\nu - \delta^\rho_\nu \delta^\sigma_\mu
\end{equation}

thus contracting with the antisymmetric $F^{\mu\nu}$,

\begin{equation}
\frac{\partial \mathcal{L}}{\partial (\partial_\rho A_\sigma)} = -\frac{1}{2} (F^{\rho\sigma} - F^{\sigma\rho}) = - F^{\rho\sigma}
\end{equation}

since the Lagrangian does not depend on the undifferentiated potential, $\frac{\partial \mathcal{L}}{\partial A_\nu} = 0$, the Euler-Lagrange expression is,

\begin{equation}
E^\nu = \frac{\partial \mathcal{L}}{\partial A_\nu}
-  \partial_\rho  \frac{\partial \mathcal{L}}{\partial (\partial_\rho A_\nu)} = \partial_\rho F^{\rho\nu}
\end{equation}

which are the sourcefree Maxwell equations. The Noether identity for electrodynamics therefore reads,

\begin{equation}
E^\nu \bar{\delta} A_\nu 
+ \partial_\rho  \left( - F^{\rho\nu} \bar{\delta} A_\nu 
+ \mathcal{L} \delta X^\rho \right) = 0
\end{equation}

\subsubsection*{Identifying the transformations $\bar{\delta} A_\nu$}

We now require the field symmetries $\bar{\delta} A_\nu$. As discussed above, they consist of two parts, the Lie derivative contribution \cite{BakerLinnemannSmeenk2021} (from the canonical part $- \partial^\beta A_\nu \delta X_\beta$, and contragredient \cite{BesselHagen1921} part $- A_\mu \partial_\nu \delta X^\mu$), and a gauge symmetry of the action $\partial_\nu \phi$ following from the gauge transformation $A'_\nu = A_\nu + \partial_\nu \phi$ (where $\phi$ is a scalar). The complete set of transformations is,

\begin{equation}
\bar{\delta} A_\nu = - \mathsterling_{\delta X} A_\nu + \delta_g A_\nu = - \partial^\beta A_\nu \delta X_\beta - A_\mu \partial_\nu \delta X^\mu + \partial_\nu \phi
\end{equation}

Bessel-Hagen solved for the gauge parameter $\phi$ such that the Noether current is gauge invariant \cite{BesselHagen1921,BakerLinnemannSmeenk2021}, obtaining the most trivial scalar combination of the potential and the coordinate transformation,

\begin{equation}
\phi = A_\mu \delta X^\mu
\end{equation}

inserting this and differentiating the third term,

\begin{equation}
\bar{\delta} A_\nu = - \delta X_\beta \partial^\beta A_\nu  - A_\mu \partial_\nu \delta X^\mu + \delta X_\beta \partial_\nu A^\beta + A_\mu \partial_\nu \delta X^\mu
\end{equation}

remarkably the second and fourth terms exactly cancel (those associated to the tensor transformations), leaving,

\begin{equation}
\bar{\delta} A_\nu = \delta X_\beta (\partial_\nu A^\beta - \partial^\beta A_\nu ) = - F^{\beta}_{\ \nu}   \delta X_\beta
\end{equation}

these are the proper (Bessel-Hagen) transformations of electrodynamics \cite{BesselHagen1921,BakerLinnemannSmeenk2021}: the transformation of the potential is the contraction of the field strength tensor with the coordinate transformation. We note that for the proper transformations, the angular momentum current derived below ($M^{\rho\mu\nu} = X^\mu T^{\rho\nu} - X^\nu T^{\rho\mu}$) is built entirely from the symmetric gauge invariant $T^{\rho\beta}$: the spin angular momentum ``superpotential'' contribution that appears in the canonical treatment is not needed, as the correct result comes directly from $\bar{\delta} A_\nu$. This is the content of the Bessel-Hagen improvement, unlike the Belinfante(-Rosenfeld) procedure \cite{Belinfante1940, Rosenfeld1940}, in which the canonical tensor derived from the canonical $\bar{\delta} A_\nu$ is instead ``improved'' by adding the superpotential which is based on the spin angular momentum contribution to correct this. Both routes arrive at the same symmetric and gauge invariant energy-momentum tensor for electrodynamics, however, the Belinfante approach requires on-shell conditions and ad-hoc addition of terms, whereas the Bessel-Hagen result is obtained directly from the symmetries and holds off-shell \cite{BakerLinnemannSmeenk2021}.

\subsubsection*{The energy-momentum tensor from Poincar\'e translation}

Inserting the proper transformations into the Noether identity,

\begin{equation}
E^\nu \bar{\delta} A_\nu 
+ \partial_\rho  \left( (- F^{\rho\nu}) (- F^{\beta}_{\ \nu} \delta X_\beta)
+ \eta^{\rho\beta} \mathcal{L} \delta X_\beta \right) = 0
\end{equation}

thus,

\begin{equation}
E^\nu \bar{\delta} A_\nu 
+ \partial_\rho  \left( \left[ F^{\rho\nu} F^{\beta}_{\ \nu}
- \frac{1}{4} \eta^{\rho\beta} F_{\mu\nu} F^{\mu\nu} \right] \delta X_\beta \right) = 0
\end{equation}

the square brackets contain exactly the accepted physical energy-momentum tensor of classical electrodynamics,

\begin{equation}
T^{\rho\beta} = F^{\rho\nu} F^{\beta}_{\ \nu}
- \frac{1}{4} \eta^{\rho\beta} F_{\mu\nu} F^{\mu\nu}
\end{equation}

which is symmetric, gauge invariant, and traceless (in $D = 4$), thus,

\begin{equation}
E^\nu \bar{\delta} A_\nu 
+ \partial_\rho  \left(T^{\rho\beta}  \delta X_\beta \right) = 0
\end{equation}

\subsubsection*{The 15 conformal conservation laws}

For the 4 parameter Poincar\'e translation $\delta X_\beta = a_\beta$, the constant can be factored out of the total divergence, leaving the identity,

\begin{equation}
E^\nu \bar{\delta} A_\nu 
+ a_\beta   \partial_\rho T^{\rho\beta} = 0
\end{equation}

imposing the equation of motion $E^\nu = 0$ (Noether's first theorem), and using that the 4 parameters $a_\beta$ are arbitrary, we have the conservation law,

\begin{equation}
\partial_\rho T^{\rho\beta} = 0
\end{equation}

The full set of coordinate symmetries for electrodynamics are the 15 parameter conformal transformations derived in the previous section (with lowered index),

\begin{equation}
\delta X_\beta = a_\beta + \omega_{\beta\alpha} X^\alpha + S X_\beta + 2 k_\nu X_\beta X^\nu  - k_\beta X_\nu X^\nu
\end{equation}

for any of these transformations the Noether current is, $J^\rho = T^{\rho\beta} \delta X_\beta$. On shell, the divergence of this current is $
\partial_\rho (T^{\rho\beta} \delta X_\beta) = (\partial_\rho T^{\rho\beta}) \delta X_\beta + T^{\rho\beta} \partial_\rho \delta X_\beta$; the first term vanishes by the conservation law above. For the second term, since $T^{\rho\beta}$ is symmetric only the symmetric part of $\partial_\rho \delta X_\beta$ contributes, which by the conformal Killing equation is proportional to the metric $T^{\rho\beta} \partial_\rho \delta X_\beta = \frac{1}{2} T^{\rho\beta} (\partial_\rho \delta X_\beta + \partial_\beta \delta X_\rho) = \frac{1}{2} \lambda(X)   T^{\rho\beta} \eta_{\rho\beta} = \frac{1}{2} \lambda(X)   T^{\rho}_{\ \rho} = 0$ which vanishes because $T^{\rho\beta}$ is traceless.  See \cite{Baker2021PHD} for analogous calculations to those found in this section for Yang-Mills, Kalb-Ramond and totally antisymmetric fields.\\

In fact the Bessel-Hagen transformations are exact symmetries of the action, which can be verified directly off shell: differentiating the energy-momentum tensor,

\begin{equation}
\partial_\rho T^{\rho\beta} = ( \partial_\rho F^{\rho\nu} ) F^{\beta}_{\ \nu}
+ F^{\rho\nu} \partial_\rho F^{\beta}_{\ \nu}
- \frac{1}{2} F_{\mu\nu} \partial^\beta F^{\mu\nu}
\end{equation}

the last two terms cancel by the Bianchi identity $\partial_{[\alpha} F_{\mu\nu]} = 0$, leaving the identity,

\begin{equation}
\partial_\rho T^{\rho\beta} = F^{\beta}_{\ \nu}   \partial_\rho F^{\rho\nu}
\end{equation}

which holds true on-shell since we are left with a term proportional to the Euler-Lagrange (Maxwell) equation,

\begin{equation}
\partial_\rho T^{\rho\beta} = F^{\beta}_{\ \nu} E^\nu
\end{equation}

the Noether identity with the Bessel-Hagen transformations is therefore satisfied identically for all 15 parameters, confirming that the transformations are symmetries. Therefore all 15 parameters yield conserved currents \cite{BesselHagen1921,BakerLinnemannSmeenk2021}; four from the divergence of the energy-momentum tensor,

\begin{equation}
T^{\rho\beta} = F^{\rho\nu} F^{\beta}_{\ \nu}
- \frac{1}{4} \eta^{\rho\beta} F_{\mu\nu} F^{\mu\nu}
\end{equation}

six from the divergence of the angular momentum tensor, obtained by factoring the antisymmetric parameter $\omega_{\mu\nu}$ out of the current,

\begin{equation}
M^{\rho\mu\nu} = X^\mu T^{\rho\nu} - X^\nu T^{\rho\mu}
\end{equation}

one from the divergence of the dilatation tensor,

\begin{equation}
D^\rho = T^{\rho\beta} X_\beta
\end{equation}

and four from the divergence of the conformal tensor,

\begin{equation}
C^{\rho\alpha} = T^{\rho\beta} (2 X_\beta X^\alpha - \delta^\alpha_\beta X_\sigma X^\sigma)
\end{equation}

these are the 15 conformal conservation laws of classical electrodynamics, first derived from Noether's first theorem by Bessel-Hagen \cite{BesselHagen1921}. The conformal invariance of Maxwell's equations themselves was established earlier by Cunningham and Bateman \cite{Cunningham1910,Bateman1909, Bateman1910}.\\

 We note that, as discussed in \cite{BakerLinnemannSmeenk2021}, Noether proved both her theorems and their converses (see also \cite{Rosen1980}). Therefore while we focus on action symmetries being used to determine conserved tensors here, the existence of these conserved tensors imply the existence of an action symmetry (in the converse direction). As discussed in \cite{BakerLinnemannSmeenk2021} and later in this article, much of the historical literature has approached this from the so-called ``canonical Noether'' energy-momentum tensor, which does not correspond to the physical energy-momentum tensor above. Convention has been to ``improve'' this object in order to obtain the known result, a longstanding issue \cite{Forger2004} that is not the focus of our present article. We also note that the above approach is the most conventional way to define electrodynamics and its physical objects covariantly, but others exist, such as defining a second potential such that all of Maxwell's equations follow from the Euler-Lagrange equation (see, e.g., \cite{Cameron2012}).

\subsection{Noether's first theorem --- second order theory example (linearized Gauss-Bonnet gravity)}

\subsubsection*{Linearized geometry and the Lagrangian}

We now consider a second order theory for a symmetric second rank potential $\Phi_A = h_{\mu\nu}$, following the energy-momentum tensor derivation in \cite{BK2019} (see \cite{Bak1994} for a more general discussion of energy-momentum in gravity theories). Linearized higher order gravity models are the ideal candidate for second order theories with a rank 2 tensor potential. The ``field strength'' analogies to electrodynamics are the linearized Riemann tensor,

\begin{equation}
R^{\mu\nu\alpha\beta} = \frac{1}{2}(\partial^{\mu}\partial^{\beta}h^{\nu\alpha} + \partial^{\nu}\partial^{\alpha}h^{\mu\beta} - \partial^{\mu}\partial^{\alpha}h^{\nu\beta} - \partial^{\nu}\partial^{\beta}h^{\mu\alpha})
\end{equation}

the linearized Ricci tensor,

\begin{equation}
R^{\nu\beta} = \eta_{\mu\alpha} R^{\mu\nu\alpha\beta} = \frac{1}{2}(\partial^{\beta}\partial^{\alpha}h^{\nu}_{\ \alpha} + \partial^{\nu}\partial^{\alpha}h^{\beta}_{\ \alpha} - \square h^{\nu\beta} - \partial^{\nu}\partial^{\beta}h)
\end{equation}

and the linearized Ricci scalar,

\begin{equation}
R = \eta_{\nu\beta} R^{\nu\beta} = \partial_\mu \partial_\nu h^{\mu\nu} - \square h
\end{equation}

where $h = h^\mu_{\ \mu}$ and $\square = \partial_\mu \partial^\mu$. The most general Lagrangian density quadratic in second derivatives of $h_{\mu\nu}$ which is invariant under the spin-2 gauge transformation $h'_{\mu\nu} = h_{\mu\nu} + \partial_\mu \zeta_\nu + \partial_\nu \zeta_\mu$ (that is, linearized diffeomorphisms, the same gauge transformation as Fierz-Pauli theory \cite{Fierz1939}) can be written as \cite{BK2019},

\begin{equation}
\mathcal{L} = \tilde{a} R_{\mu\nu\alpha\beta} R^{\mu\nu\alpha\beta} + \tilde{b} R_{\mu\nu} R^{\mu\nu} + \tilde{c} R^2
\end{equation}

with free coefficients $\tilde{a}$, $\tilde{b}$, $\tilde{c}$. For this problem we require the second order Noether identity derived earlier,

\begin{multline}
0 =  
\left( \frac{\partial \mathcal{L}}{\partial \Phi_A}
-  \partial_\rho  \frac{\partial \mathcal{L}}{\partial (\partial_\rho \Phi_A)}
+
\partial_\omega \partial_\rho \frac{\partial \mathcal{L}}{\partial (\partial_\omega \partial_\rho \Phi_A)} \right)  \bar{\delta} \Phi_A
\\
+ \partial_\rho  \left( \frac{\partial \mathcal{L}}{\partial (\partial_\rho \Phi_A)}\bar{\delta} \Phi_A
+  \frac{\partial \mathcal{L}}{\partial (\partial_\omega \partial_\rho \Phi_A)} (\partial_\omega \bar{\delta} \Phi_A)  
-(\partial_\omega \frac{\partial \mathcal{L}}{\partial (\partial_\omega \partial_\rho \Phi_A)}) \bar{\delta} \Phi_A
+ \mathcal{L} \delta X^\rho \right)
\end{multline}

where we will replace the fields as $\Phi_A \to h_{\rho\sigma}$. Since this Lagrangian depends on neither the undifferentiated potential nor its first derivatives, we have $\frac{\partial \mathcal{L}}{\partial h_{\rho\sigma}} = 0$ and $\frac{\partial \mathcal{L}}{\partial (\partial_\lambda h_{\rho\sigma})} = 0$, and the second order Noether identity derived in this article reduces to,

\begin{equation}
E^{\rho\sigma} \bar{\delta} h_{\rho\sigma}
+ \partial_\omega \left( \eta^{\omega\nu} \mathcal{L} \delta X_\nu 
+ \frac{\partial \mathcal{L}}{\partial (\partial_\omega \partial_\lambda h_{\rho\sigma})} \partial_\lambda \bar{\delta} h_{\rho\sigma}
- \left[ \partial_\lambda \frac{\partial \mathcal{L}}{\partial (\partial_\omega \partial_\lambda h_{\rho\sigma})} \right] \bar{\delta} h_{\rho\sigma}
\right) = 0
\end{equation}

with the Euler-Lagrange expression,

\begin{equation}
E^{\rho\sigma} = \partial_\omega \partial_\lambda \frac{\partial \mathcal{L}}{\partial (\partial_\omega \partial_\lambda h_{\rho\sigma})}
\end{equation}

\subsubsection*{Identifying the transformation $\bar{\delta} h_{\rho\sigma}$}

For the 4 parameter Poincar\'e translation $\delta X^\beta = a^\beta$ (eliminating the contragredient part of the Lie derivative), the field transformations consist of the canonical part and the spin-2 gauge symmetry with gauge vector $\zeta_\sigma$,

\begin{equation}
\bar{\delta} h_{\rho\sigma} = - \partial_\beta h_{\rho\sigma} \delta X^\beta + \partial_\rho \zeta_\sigma + \partial_\sigma \zeta_\rho
\end{equation}

applying the Bessel-Hagen method with the most general vector built from the potential and the coordinate transformation, $\zeta_\sigma = \tilde{A} h_{\sigma\beta} \delta X^\beta + \tilde{B} h \eta_{\sigma\nu} \delta X^\nu$, gauge invariance of the current selects $\tilde{A} = 1$, $\tilde{B} = 0$ \cite{BK2019}. Inserting $\zeta_\sigma = h_{\sigma\beta} \delta X^\beta$ with constant $\delta X^\beta$,

\begin{equation}
\bar{\delta} h_{\rho\sigma} = [- \partial_\beta h_{\rho\sigma} + \partial_\rho h_{\sigma\beta} + \partial_\sigma h_{\rho\beta}] \delta X^\beta
\end{equation}

remarkably this combination is exactly proportional to the linearized Christoffel symbol,

\begin{equation}
\bar{\delta} h_{\rho\sigma} = - 2 \Gamma^\nu_{\ \rho\sigma} \delta X_\nu
\quad , \quad
\Gamma^\nu_{\ \rho\sigma} = \frac{1}{2} (\partial^\nu h_{\rho\sigma} - \partial_\rho h^\nu_{\ \sigma} - \partial_\sigma h^\nu_{\ \rho})
\end{equation}

note that this $\Gamma^\nu_{\ \rho\sigma}$ is minus the linearization of the standard Christoffel symbol of general relativity. This is the second order analogue of the proper transformations $\bar{\delta} A_\nu = - F^\beta_{\ \nu} \delta X_\beta$ of electrodynamics: in both cases the transformation of the potential is the contraction of a well known ``first-derivative'' relation of the theory with the coordinate transformation \cite{BK2019}.

\subsubsection*{Evaluating the Noether current}

From the Lagrangian we calculate,

\begin{multline}
\frac{\partial \mathcal{L}}{\partial (\partial_\omega \partial_\lambda h_{\rho\sigma})} = 2 \tilde{a} [R^{\rho\omega\lambda\sigma} + R^{\lambda\rho\sigma\omega}]
+ \tilde{b} [ - \eta^{\omega\lambda} R^{\rho\sigma} - \eta^{\rho\sigma} R^{\omega\lambda} + \frac{1}{2}(\eta^{\lambda\sigma} R^{\rho\omega} + \eta^{\lambda\rho} R^{\sigma\omega} + \eta^{\omega\sigma} R^{\rho\lambda} + \eta^{\omega\rho} R^{\sigma\lambda})]
\\
+ 2 \tilde{c} R [ - \eta^{\omega\lambda} \eta^{\rho\sigma} + \frac{1}{2}(\eta^{\omega\rho} \eta^{\lambda\sigma} + \eta^{\omega\sigma} \eta^{\lambda\rho})]
\end{multline}

two identities equivalent to the linearized Bianchi identities are used throughout,

\begin{equation}
\partial_\omega R^{\lambda\rho\omega\sigma} = \partial^\lambda R^{\rho\sigma} - \partial^\rho R^{\lambda\sigma}
\quad , \quad
\partial^\rho R = 2 \partial_\omega R^{\omega\rho}
\end{equation}

inserting the transformations $\bar{\delta} h_{\rho\sigma} = - 2 \Gamma^\nu_{\ \rho\sigma} \delta X_\nu$, the second term in the current evaluates to \cite{BK2019},

\begin{equation}
\frac{\partial \mathcal{L}}{\partial (\partial_\omega \partial_\lambda h_{\rho\sigma})} \partial_\lambda \bar{\delta} h_{\rho\sigma}
= ( - 4 \tilde{a} R^{\omega\rho\lambda\sigma} R^{\nu}_{\ \rho\lambda\sigma} - 2 \tilde{b} R_{\rho\sigma} R^{\omega\rho\nu\sigma} - 2 \tilde{b} R^{\omega\lambda} R^{\nu}_{\ \lambda} - 4 \tilde{c} R R^{\nu\omega} ) \delta X_\nu
\end{equation}

which is manifestly gauge invariant for any coefficients. The third term in the current evaluates to \cite{BK2019},

\begin{multline}
\left[ \partial_\lambda \frac{\partial \mathcal{L}}{\partial (\partial_\omega \partial_\lambda h_{\rho\sigma})} \right] \bar{\delta} h_{\rho\sigma}
= \frac{1}{2} (4 \tilde{c} + \tilde{b}) [2 \partial^\omega R   \eta^{\rho\sigma} - \partial^\sigma R   \eta^{\rho\omega} - \partial^\rho R   \eta^{\sigma\omega}] \Gamma^\nu_{\ \rho\sigma} \delta X_\nu
\\
+ (4 \tilde{a} + \tilde{b}) [2 \partial^\omega R^{\rho\sigma} - \partial^\sigma R^{\rho\omega} - \partial^\rho R^{\sigma\omega}] \Gamma^\nu_{\ \rho\sigma} \delta X_\nu
\end{multline}

this term is not gauge invariant (it depends on the undifferentiated linearized Christoffel symbol), thus a gauge invariant energy-momentum tensor requires that it vanishes identically. This fixes the coefficients up to an overall normalization, which we choose as,

\begin{equation}
\tilde{a} = \frac{1}{4} \quad , \quad \tilde{b} = -1 \quad , \quad \tilde{c} = \frac{1}{4}
\end{equation}

which corresponds to exactly the linearized Gauss-Bonnet Lagrangian \cite{BK2019},

\begin{equation}
\mathcal{L} = \frac{1}{4} (R_{\mu\nu\alpha\beta} R^{\mu\nu\alpha\beta} - 4 R_{\mu\nu} R^{\mu\nu} + R^2)
\end{equation}

\subsubsection*{The energy-momentum tensor}

Collecting the first and second terms of the current with these coefficients, the Noether identity reads,

\begin{equation}
E^{\rho\sigma} \bar{\delta} h_{\rho\sigma} + \partial_\omega ( T^{\omega\nu} \delta X_\nu ) = 0
\end{equation}

with the energy-momentum tensor of linearized Gauss-Bonnet gravity \cite{BK2019},

\begin{equation}
T^{\omega\nu} = - R^{\omega\rho\lambda\sigma} R^{\nu}_{\ \rho\lambda\sigma} + 2 R_{\rho\sigma} R^{\omega\rho\nu\sigma} + 2 R^{\omega\lambda} R^{\nu}_{\ \lambda} - R R^{\nu\omega} + \frac{1}{4} \eta^{\omega\nu} (R_{\mu\lambda\alpha\beta} R^{\mu\lambda\alpha\beta} - 4 R_{\mu\lambda} R^{\mu\lambda} + R^2)
\end{equation}

which is symmetric, gauge invariant, and built entirely from the linearized curvature tensors, in analogy with the electrodynamic energy-momentum tensor built from $F_{\mu\nu}$. For the Poincare translation $\delta X_\nu = a_\nu$ we have the conservation law,

\begin{equation}
\partial_\omega T^{\omega\nu} = 0
\end{equation}

It is important to note that this holds off-shell, since the equation of motion term is identically zero (a Bianchi identity) and does not need to be solved for a specific on-shell condition (it is a topological theory, no dynamics). This is in distinction with standard dynamical field theory, but it is the requirement to have the second order linearized gravity theory yield a gauge invariant energy-momentum tensor.  \\

We note that this model was extended to dual formalisms in \cite{Baker2021DualII}, and this technique was used to develop higher spin gauge theory Lagrangians in \cite{Baker2021CJP}.

\subsection{Noether's second theorem --- first order theory example (electrodynamics)}

For Noether's second theorem we return to the Noether identity of electrodynamics,

\begin{equation}
E^\nu \bar{\delta} A_\nu 
+ \partial_\rho  \left( - F^{\rho\nu} \bar{\delta} A_\nu 
+ \mathcal{L} \delta X^\rho \right) = 0
\end{equation}

and consider the gauge symmetry of the action alone: the coordinates are not transformed, and the fields transform under the gauge transformation $A'_\nu = A_\nu + \partial_\nu \phi$ where $\phi(X)$ is an arbitrary function,

\begin{equation}
\bar{\delta} A_\nu = \partial_\nu \phi
\end{equation}

the Noether identity is defined at all points; to apply Noether's second theorem we integrate it (in the action condition),

\begin{equation}
\int \left( E^\nu \bar{\delta} A_\nu 
+ \partial_\rho  \left( - F^{\rho\nu} \bar{\delta} A_\nu 
+ \mathcal{L} \delta X^\rho \right) \right) dX  = 0
\end{equation}

and apply boundary conditions such that all total divergence terms integrate to zero, leaving just the Euler-Lagrange term,

\begin{equation}
\int E^\nu   \bar{\delta} A_\nu   dX  = 0
\end{equation}

we then restrict consideration of the field variation to the gauge function as $\bar{\delta} A_\nu  = \partial_\nu \phi$ such that, after also inserting $E^\nu = \partial_\rho F^{\rho\nu}$

\begin{equation}
 \int \partial_\rho F^{\rho\nu}   \partial_\nu \phi   dX = 0
\end{equation}

integrating by parts

\begin{equation}
\int  \partial_\rho F^{\rho\nu}   \partial_\nu \phi   dX
= \int  \partial_\nu ( \partial_\rho F^{\rho\nu}   \phi )   dX
- \int  \partial_\nu \partial_\rho F^{\rho\nu}   \phi   dX
\end{equation}

the first term on the right hand side is again a boundary term which vanishes for our choice of $\phi$, thus,

\begin{equation}
\int  \partial_\nu \partial_\rho F^{\rho\nu}   \phi   dX = 0
\end{equation}

since $\phi(X)$ is an arbitrary function, we have the differential identity of Noether's second theorem for electrodynamics \cite{BakerLinnemannSmeenk2021},

\begin{equation}
\partial_\nu \partial_\rho F^{\rho\nu} = 0
\end{equation}

this identity holds off-shell (it is automatic by the antisymmetry of $F^{\rho\nu}$).

\subsection{Noether's second theorem --- second order theory example (general relativity)}

Finally we consider general relativity, with the Einstein-Hilbert Lagrangian density,

\begin{equation}
\mathcal{L} = \sqrt{-g}   g^{\mu\nu} R_{\mu\nu}
\end{equation}

where here $R_{\mu\nu}$ is the full (not linearized) Ricci tensor. We make the disclaimer here that while the article is focused on flat spacetimes, the general relativity example is a prototypical one for Noether's second theorem (and some of the original motivation \cite{Noether1918,Klein1918}), so we will break from this focus of the article in this subsection (and this subsection alone).\\

Since the Ricci tensor contains second derivatives of the metric, this is a second order field theory for the potential $\Phi_A = g_{\mu\nu}$, and the second order Noether identity of this article applies. The Euler-Lagrange expression of the Einstein-Hilbert Lagrangian is proportional to the Einstein tensor \cite{Carmeli2001},

\begin{equation}
E^{\mu\nu} 
= - \sqrt{-g}   G^{\mu\nu}
\quad , \quad
G^{\mu\nu} = R^{\mu\nu} - \frac{1}{2} g^{\mu\nu} R
\end{equation}

The ``gauge'' symmetry of general relativity is the diffeomorphism invariance of the action: the coordinate transformation $\delta X^\mu$ is now an arbitrary function of the coordinates (not a Killing vector), and the metric transforms through (minus) its Lie derivative,

\begin{equation}
\bar{\delta} g_{\mu\nu} = - \mathsterling_{\delta X}   g_{\mu\nu} = - \delta X^\lambda \partial_\lambda g_{\mu\nu} - g_{\mu\lambda} \partial_\nu \delta X^\lambda - g_{\nu\lambda} \partial_\mu \delta X^\lambda
\end{equation}

which, as in the calculation of the Killing equation section, can be written covariantly as,

\begin{equation}
\bar{\delta} g_{\mu\nu} = - (\nabla_\mu \delta X_\nu + \nabla_\nu \delta X_\mu)
\end{equation}

the Noether identity is defined at all points; to apply Noether's second theorem we integrate it (in the action condition), and apply boundary conditions such that all total divergence terms integrate to zero, leaving just the Euler-Lagrange term,

\begin{equation}
\int \sqrt{-g}   G^{\mu\nu} ( \nabla_\mu \delta X_\nu + \nabla_\nu \delta X_\mu )   dX = 0
\end{equation}

using the symmetry of the Einstein tensor and changing dummies the two terms combine,

\begin{equation}
2 \int \sqrt{-g}   G^{\mu\nu} \nabla_\mu \delta X_\nu   dX = 0
\end{equation}

integrating by parts inside the integral with the covariant derivative,

\begin{equation}
G^{\mu\nu} \nabla_\mu \delta X_\nu = \nabla_\mu ( G^{\mu\nu} \delta X_\nu ) - ( \nabla_\mu G^{\mu\nu} ) \delta X_\nu
\end{equation}

and recalling that for any vector $V^\mu$ the covariant divergence satisfies $\sqrt{-g}   \nabla_\mu V^\mu = \partial_\mu ( \sqrt{-g}   V^\mu )$, the first term is a boundary term which vanishes, leaving only,

\begin{equation}
\int \sqrt{-g}   ( \nabla_\mu G^{\mu\nu} )   \delta X_\nu   dX = 0
\end{equation}

since $\delta X_\nu$ is an arbitrary function, we have the differential identity of Noether's second theorem for general relativity which is the contracted Bianchi identity,

\begin{equation}
\nabla_\mu G^{\mu\nu} = 0
\end{equation}

which holds off-shell, following from diffeomorphism invariance analogous to $\partial_\nu \partial_\rho F^{\rho\nu} = 0$ following from the gauge invariance of electrodynamics.\\

We note a key point about how the first and second theorems differ: when the coordinate transformation $\delta X^\mu$ is restricted to the Killing or conformal Killing vectors of the previous section, Noether's first theorem yields conserved currents; when $\delta X^\mu$ is left as an arbitrary function, Noether's second theorem yields off-shell differential identities of the equations of motion.

\section{Discussion}

Noether's theorems originate in Noether's 1918 article \cite{Noether1918}, which contains both theorems in full generality. The application of the first theorem to the 15 conformal symmetries of electrodynamics followed three years later in the work of Bessel-Hagen \cite{BesselHagen1921}, who also introduced the notion of invariance up to a surface term. Much of the physics literature descends from simplified presentations such as that of Hill \cite{Hill1951}, who presented only a special case of Noether's results (although he did cite Noether \cite{Noether1918} and Bessel-Hagen \cite{BesselHagen1921}). A more rigorous presentation was later given in \cite{Trautman1967}. In fact, it was not until 1971 that an English translation of Noether's original work appeared \cite{Noether1918}. As Lanczos, one of the top physicists working with variational techniques in the twentieth century, wrote: \textit{``Every theoretical physicist is familiar with the expression ``Noether’s theorem'' or ``Noether’s principle,'' although none of them actually reads Emmy Noether’s original paper''} \cite{Lanczos1973}. Later, modern generalizations were developed such as Olver \cite{Olver1993} and Martinez Alonso \cite{Martinez1979}, with primary focus on the mathematics associated (rather than physical application in mind). Some physics articles focus on generalized approaches, e.g., \cite{Agostini2007,Bravetti2021}, see also \cite{Hydon2011} for extension of the second theorem. More recently the history of the theorems and their reception was documented by Kosmann-Schwarzbach \cite{KosmannSchwarzbach2011} (which includes an English translation of Noether's original paper \cite{Noether1918}). Various other books have been dedicated to Noether's theorems and symmetries in recent decades \cite{Gieres1997,Sardanashvily2016,Neuenschwander2017}, which cover to various degrees topics we discuss in this article. Of course, application and discussion of Noether's results are widespread in physics \cite{Bluman1988,Ibragimov1998,Sniatycki1998,Garcia2000,Rosenhaus2002,Sanyal2002,Struckmeier2002,Rudowicz2003,Hanc2004,Kara2006,Cicogna2007,Sarlet1981b,Feroze2011,Tsamparlis2011,Tsamparlis2011A}, a summary of which we will include at the end of this section. Our focus in the present article was not on application (although some included), but to give a presentation using rigorous methods of the calculus of variations for how the Noether identity and theorems themselves can be obtained in a language familiar to those working with fields in physics: particularly in flat spacetimes (for our applications). An analogous presentation for curved spacetimes requires more extensive generalization of the initial action condition and calculations that may be the subject of future work. The focus of this final section is to briefly discuss key points in this area such as the canonical Noether energy-momentum tensor, Bessel-Hagen's divergence symmetry generalization, and some of the many places these ideas have been applied in the literature. 

\subsection{The canonical Noether energy-momentum tensor}

Suppose we start from the Noether identity, e.g., for first order theories,

\begin{equation}
\left( \frac{\partial \mathcal{L}}{\partial \Phi_A}
- \partial_\rho \frac{\partial \mathcal{L}}{\partial (\partial_\rho \Phi_A)} \right) \bar{\delta} \Phi_A
+ \partial_\rho \left( \frac{\partial \mathcal{L}}{\partial (\partial_\rho \Phi_A)} \bar{\delta} \Phi_A
+ \mathcal{L} \delta X^\rho  \right) = 0
\end{equation}

The most standard conserved tensor we could wish to derive for a physical field theory is the energy-momentum tensor $T^{\mu\nu}$. For translation $\delta X^\beta = a^\beta$ one has $h_A = 0$ for tensor fields (since $a^\beta$ is a constant the derivative is zero), i.e., $\delta \Phi_A = 0$ and $\bar{\delta} \Phi_A = - a^\beta \partial_\beta \Phi_A$, and the current reduces to $J^\rho = -a_\beta   T^{\rho\beta}_C$ where $T^{\rho\beta}_C$ is the so-called canonical Noether energy-momentum tensor,

\begin{equation}
    T^{\rho\beta}_C
 = \frac{\partial \mathcal{L}}{\partial (\partial_\rho \Phi_A)}   \partial^\beta \Phi_A
- \eta^{\rho\beta}   \mathcal{L}
\end{equation}

This tensor does not correspond to the well known physical energy-momentum tensor for standard physical theories, a decades old problem \cite{Forger2004} that continues to be discussed in the literature. This is the object which the Belinfante ``improvement'' is used to obtain the desired physical result \cite{Belinfante1940,BakerLinnemannSmeenk2021}; we refer the reader to \cite{Baker2021PHD} and \cite{Forger2004} for discussion of this historical issue. As demonstrated in this article, using the Bessel-Hagen method one can directly derive the physical energy-momentum tensor from the Noether identity without the need for any ad-hoc improvement procedure. While these improvements continue to be used in the literature, more recent approaches have emphasized the methods which can be used to obtain physical conserved tensors directly, therefore we will not dedicate time in this article to the technical details behind the (arguably now mostly resolved) ``improvement'' issue.\\

It is also important to emphasize that this ``canonical Noether'' object, which features Noether's name, was neither proposed nor endorsed by Noether. Therefore any connection of this result to something which is ``not working'' to derive a physically conserved object should not be attributed to Emmy Noether herself, rather later authors who attributed this object to her due to the connection to her theorems.

\subsection{Extensions of Noether's work}

Several authors have extended Noether's work to various degrees \cite{Martinez1979,Olver1993}. In fact, Bessel-Hagen \cite{BesselHagen1921}, in the same article as presenting the method for deriving physical conservation laws we used earlier, also discussed actions invariant up to a boundary term, for example,

\begin{equation}
\delta S = \int \partial_\rho C^\rho  dX
\end{equation}

for some functions $C^\rho (X, \Phi, \partial \Phi)$ (a divergence symmetry, using Bessel-Hagen's notation ``C'' for this contribution). Basically, instead of the action variation being identically zero, it is set equal to the above contribution such that the result Noether identity is,

\begin{equation}
\left( \frac{\partial \mathcal{L}}{\partial \Phi_A}
- \partial_\rho \frac{\partial \mathcal{L}}{\partial (\partial_\rho \Phi_A)} \right) \bar{\delta} \Phi_A
+ \partial_\rho \left( \frac{\partial \mathcal{L}}{\partial (\partial_\rho \Phi_A)} \bar{\delta} \Phi_A
+ \mathcal{L} \delta X^\rho  \right) =  \partial_\rho C^\rho
\end{equation}

where on the left hand side we have the same identity as before (for first order theories). Combining this extra contribution with the Noether current gives,

\begin{equation}
\left( \frac{\partial \mathcal{L}}{\partial \Phi_A}
- \partial_\rho \frac{\partial \mathcal{L}}{\partial (\partial_\rho \Phi_A)} \right) \bar{\delta} \Phi_A
+ \partial_\rho \left( \frac{\partial \mathcal{L}}{\partial (\partial_\rho \Phi_A)} \bar{\delta} \Phi_A
+ \mathcal{L} \delta X^\rho - C^\rho \right) = 0
\end{equation}

so that the conserved current is modified as $J^\rho - C^\rho$.\\

Boundary modifications of this form are common with certain types of ``improvements'', such as for scalar field theory \cite{CallanColemanJackiw1970,Coleman1971,KuzminMcKeon2001}. See also \cite{Gieres2022} for a recent article on comparing ``improvements'' to total divergence addition to the Lagrangian (and related discussion in \cite{Katz2008}). For electrodynamics with the Bessel-Hagen transformations strict invariance holds ($C^\rho = 0$), which is why the treatment of the 15 conformal currents in the applications section is complete for that case. The same generalization applies in exactly the same way to the second order identity, with $C^\rho$ subtracted from the second order current. Further extensions of the ideas of divergence symmetries can be found in \cite{Martinez1979,Olver1993}.

\subsection{Other Noether applications, results and discussion in physics literature}

In this final section we will do a brief (non-exhaustive) review of the physics literature regarding applications of Noether's theorems. It all started with Noether's original article, which did not focus on the derivation of physical conserved tensors, rather the general theorems \cite{Noether1918}. Bessel-Hagen then introduced the notion of invariance up to a surface term as well as was the first to derive the 15 physical conservation laws of electrodynamics directly from Noether's first theorem \cite{BesselHagen1921}. Later, perhaps unaware of Bessel-Hagen's result (which was not translated until recently \cite{BesselHagen1921}), authors began to ``improve'' the so-called ``canonical Noether'' energy-momentum tensor which did not correspond to the physical energy-momentum tensor, perhaps most famously the ``symmetrization improvement'' using the spin contribution by Belinfante and Rosenfeld \cite{Belinfante1940, Rosenfeld1940}. Other famous reasons to improve this canonical object included for conformal transformations in scalar field theory \cite{CallanColemanJackiw1970}, which can equivalently be determined by modifying the Lagrangian with a total divergence term \cite{KuzminMcKeon2001} (see \cite{Gieres2022} for more discussion on this). More recently authors have continued the discussion of energy-momentum \cite{Forger2004,Blaschke2016,BKK2021,Baker2021PHD,Gieres2022}, with some lingering issues needing to be resolved about the general nature of these objects.\\

While energy-momentum and other conserved tensors are relevant to the current article (focused primarily on classical field theories \cite{Rosen1972} in flat spacetimes), Noether's theorems have been applied and discussed for various other branches of physics. Bessel-Hagen also introduced results related to mechanics \cite{BesselHagen1921}, which continues to be an important point of discussion in the physics literature \cite{Sarlet1981,Brading2005,Badin2017,Song2018,Hecht2026}. The Ward-Takahashi identities used in quantum field theory give quantum counterparts of the classical Noether identities \cite{Ward1950, Takahashi1957}. These were later generalized to the non-Abelian case (the Slavnov-Taylor identities \cite{Taylor1971, Slavnov1972}). Since then, many articles have focused on Noether's theorems with respect to quantum mechanics \cite{Brown2004,Albeverio2006,Marvian2014} and quantum field theory \cite{Buchholz1986,Avery2016,Benedetti2022,Freese2022,Freese2026}. Even connections to statistical mechanics have been explored \cite{Hermann2022}. Of course, Noether was a mathematician, and much of the interest in these results remains in the mathematics literature \cite{Bluman1989,Dorodnitsyn2001,Cotter2013,Zhang2015,Zhang2016,Olver2021,Peng2022}. There is also continual interest in other areas such as the philosophy of physics community \cite{Byers1998,Brading2002,Brading2003,BrownHolland2004,Romero2015,Read2022}. While this summary is by no means exhaustive of the various interest in Noether's theorems in the literature, it covers some of the main branches of interest which continue to push forward Noether's ideas, and applications of her results.\\

Noether's theorems are some of the most important mathematical results which are used in physics. While the machinery itself is solid, actual application has issues in certain models such as linearized gravity \cite{Babak1999,Magnano2002,Butcher2008,Bicak2016,Baker2021CQG,TaylorBaker2024}, where a long standing non-uniqueness issue is problematic for cases where unique expressions are required (see, e.g., the Padmanabhan-Deser debate \cite{Padmanabhan2008,Deser2010,Barcelo2014,LinnemannSmeenkBaker2023}). There exists a related longstanding but distinct issue in full (nonlinear) general relativity regarding defining energy-momentum at all \cite{Moller1958,Moller1961,MTW1973,Cooperstock1978,Cooperstock1992,Chang1999,Szabados2009}. Beyond this issue, there exists a large body of literature focused on the topic of Noether's work with respect to gravitational physics \cite{Brading2005GR,Basilakos2011,Dimakis2017,Sk2017,Deser2019,Bajardi2022,Aoki2022,DeHaro2022,Cresto2026}. It is our view that application of Noether's ideas to various gravity theories is an area with a literature gap that has the potential to address various tensions between theory and observation in astrophysics. In the physics literature more broadly, recent work has continued to explore novel applications, issues, and discussions surrounding the use of these theorems in physics \cite{Rosenhaus2016,Brauner2020,Balondo2021,Kourkoulou2022,Rignon2023,Houchmandzadeh2025,Rivas2025}; what we wish to emphasize is that Noether's results are not merely a tool which physicists use peripherally to their core area of research, but that themselves are of interest of investigation for physicists in the contemporary physics community. The power of Noether's ideas are universally acclaimed at this point, but there remain issues (such as non-uniqueness of conserved tensors \cite{TaylorBaker2024} or methodological approaches \cite{BakerLinnemannSmeenk2021}) which still have not been conclusively addressed. It is our view that addressing these issues can help push forward some of the major unsolved physics problems of this era.

\bibliographystyle{unsrt}
\bibliography{NoetherReferences}

\end{document}